\documentclass[11pt]{article}

\usepackage{enumerate}
\usepackage{graphicx}
\usepackage{amssymb}
\usepackage{cancel}

\usepackage{algorithm}
\usepackage{algpseudocode}
\usepackage{siunitx}

\usepackage{amsmath}
\usepackage{setspace}
\usepackage{soul}
\usepackage{color}
\def\boxit#1{\vbox{\hrule\hbox{\vrule\kern6pt\vbox{\kern6pt#1\kern6pt}\kern6pt\vrule}\hrule}}

\newcommand{\RNum}[1]{\uppercase\expandafter{\romannumeral #1\relax}}

\usepackage{geometry}


\usepackage{setspace}
\usepackage[utf8]{inputenc}
\usepackage{amsthm}
\newtheorem{theorem}{Theorem}

\newtheorem{proposition}{Proposition}

\newtheorem{lemma}{Lemma}
\theoremstyle{definition}

\usepackage{smile}   
\usepackage{bbm}
\usepackage[colorlinks,
linkcolor=red,
anchorcolor=blue,
citecolor=blue
]{hyperref}
\usepackage{enumitem}

\usepackage{fullpage}

\usepackage{tikz}

\graphicspath{ {./plots/} }
\usepackage{authblk}
\providecommand{\keywords}[1]{\textit{Keywords:} #1}

\begin{document}
	\title{funFMC: Overlapping Clustering for Functional Data}
	\author[1]{Abhiti Mishra}
\author[1]{Kean Ming Tan}
\author[1]{Tailen Hsing}
\affil[1]{Department of Statistics, University of Michigan}
    \date{}
    \maketitle

\begin{abstract}
In applications such as neuroscience and environmental science, data are naturally modeled as multivariate functional data and often exhibit overlapping cluster structure.  
Existing clustering methods for functional data typically impose mutually exclusive memberships and therefore fail to capture such structure.  
We propose a latent factor model based approach with functional factors and a real-valued loading matrix that encodes potentially overlapping cluster memberships. 
Under mild conditions, we establish identifiability of the loading matrix up to permutation, ensuring that the overlapping cluster structure is recoverable up to label switching. 
We develop a procedure for estimating both the number of clusters and the associated cluster memberships. 
This involves solving an infinite-dimensional regression problem in operators, whose solution is characterized using the inner product on the space of Hilbert–Schmidt operators and expressed in terms of real-valued matrices. 
This formulation enables rigorous asymptotic analysis, and we establish a central limit theorem to facilitate statistical inference on overlapping cluster memberships. 
We demonstrate the performance of our method using numerical studies and an application to functional magnetic resonance imaging data.
\end{abstract}



\section{Introduction}

Functional data analysis is the branch of statistics in which data are modeled as smooth functions \citep{fdabook2, fdabook}.
With the availability of increasingly high-dimensional data, functional data techniques are relevant in a wide range of scientific disciplines.
When data contain multiple curves across different samples, a natural first step is to cluster homogeneous curves together. 
Such unsupervised techniques can reveal hidden similarity between curves.
For example, time series extracted from functional magnetic resonance imaging (fMRI) data can be clustered to reveal regions (parcellations) of the brain that show similar patterns of activation.

Clustering methods for functional data are largely inspired by their counterparts for scalar and vector-valued data. 
The $k$-means algorithm has been adapted to functional data in several ways.
Some adaptations involve extraction of finite-dimensional coefficients using an appropriate basis expansion \citep{Abraham2003, k-means-fpca, Delaigle2019, Denis2020RegularizedFunctionalClustering, func_proj_kmeans}, while others propose notions of distance between infinite-dimensional curves \citep{Ieva2013MultivariateFunctionalClustering} or covariance operators \citep{mult_func_clus_martino}.
Examples of hierarchical clustering algorithms include recursive construction of binary trees \citep{GOLOVKINE2022binarytrees}, and agglomerative clustering techniques combined with quantile estimation \citep{Gaetan2016quantile}.
Latent mixture models have been developed for clustering independent \citep{mixedeffects, adaptive_mixture_model, Bouveyron2015funFEM,  funHDDC-multivariate} as well as dependent \citep{Liang2021} functional data.
\cite{wtkmeans} formulate the sparse clustering of functional data as a variational problem.
We refer the reader to surveys on functional clustering for a thorough review \citep{old_funcclreview,funcclusreview}.
Most of the existing techniques are incapable of estimating overlapping cluster memberships, which is relevant when curves are related to multiple clusters.

The limited work on identification of overlapping clusters includes the functional extensions of fuzzy $k$-means \citep{mult_func_clus_tokushige} and fuzzy $k$-mediods \citep{Giordani2020FuzzyClustering}. 
\cite{nguyen2011dirichlet} use a Bayesian mixture model in which each curve can belong to multiple groups, and the allocations follow a Dirichlet process prior.
More recently, \cite{masarotto2024} consider the problem of overlapping clustering of covariance operators using the Wasserstein-Procrustes metric.
Closer in spirit to our work, \cite{bayesian_cov_mult_func_clus} propose a nonparametric Bayesian sparse latent factor model for multivariate functional clustering, and \cite{Bouveyron2022CoClustering} use a functional latent block model for co-clustering. 
However, neither provide frequentist theoretical guarantees for overlapping membership recovery.

Recently, \cite{bing} introduced a latent factor model for identifying overlapping cluster structure among the components of $X\in \mathbb{R}^p$. 
They posit a low-dimensional representation $X = AZ + E$, where $Z \in \mathbb{R}^K$ contains the latent factors, $A \in \mathbb{R}^{p \times K}$ the loading matrix, and $E$ is noise. Each factor $Z_k$ is associated with a cluster, and the $k$-th column of $A$ encodes the membership weights of the variables in that cluster, naturally accommodating overlapping assignments. 
The estimation of $A$ 
relies on the covariance matrix of $X$, denoted by $\Sigma \in \RR^{p \times p}$.
In this paper, we extend the approach of \cite{bing} to the setting where $X$ is assumed to be a $p$-dimensional functional variable where each element $X_i$ is in some function space $\HH$. 
The covariance in this case is a linear mapping from $\HH^p$ to $\HH^p$.  
The methodology in \cite{bing} is not directly applicable in the functional setting. 
For example, comparing the magnitudes of the absolute covariances $|\Sigma_{ij}|$ is a key element in estimating $A$ in \cite{bing}. 
However, there is no natural ordering of the cross covariance operators in the infinite-dimensional setting. 
Thus, we need to consider a fundamentally different way to compare the $\Sigma_{ij}$ so as to capture the underlying structure of the functional latent factor model.

In this paper, we develop a covariance-based clustering procedure that is capable of estimating overlapping cluster memberships.
We propose a latent factor model with functional factors and a real-valued loading matrix that encodes cluster memberships.
We assume the presence of \emph{pure variables}, which are variables depending only on one latent factor.
Under mild scaling assumptions on $A$ and an assumption on separation among latent factors, we establish identifiability of the loading matrix up to post-multiplication by a signed permutation matrix.
This ensures recovery of cluster memberships up to label-switching and a change in the direction of dependence.
We develop an estimation procedure for recovering the number of clusters as well as the cluster memberships.
It is a two-stage process where the pure variables are estimated first, followed by the estimation of the non-pure variables that belong to multiple clusters.
The second step is based on a regression problem involving infinite-dimensional operators. 
We express the solution of this problem in terms of real-valued matrices by leveraging Hilbert-Schmidt inner products.

We establish rigorous statistical guarantees showing that the number of clusters is estimated correctly with high probability.
We also characterize false positives in pure variable estimation in terms of the model parameters.
Additionally, we present a novel asymptotic analysis for the part of $A$ corresponding to non-pure variables, which facilitates inference on the overlapping cluster memberships.
Our results are different from well-known asymptotic results for hard clustering using $k$-means \citep{hartigan1978, Pollard1982strongconsistency_kmeans, SerinkoBabu1992}, as well as their extensions to functional data \citep{ Delaigle2019}.
To the best of our knowledge, our results are the first to show the asymptotic distribution of the degree of overlap for soft clustering.

The rest of the paper is arranged as follows: in Section \ref{sec:prelim_and_setup}, we introduce standard notation for functional data as well as the underlying factor model. 
We also establish the identifiability of the loading matrix under the model assumptions. 
Section \ref{sec:estimation} describes the estimation procedure. 
Section \ref{sec:th-guar} provides theoretical guarantees for estimation of the number of clusters and the loading matrix, which in turn determines the group memberships. 
Section \ref{sec:num-exp} shows the effectiveness of the proposed method in simulation studies.
In Section~\ref{sec:fMRI}, we apply our method to open-source, task-based fMRI data.


\section{Preliminaries and Problem Setup} \label{sec:prelim_and_setup}
\subsection{Preliminaries}\label{sec:prelim}

We start with a brief overview of multivariate functional data and refer the reader to \cite{mult_longi_data} and \cite{mult_func_data} for a more comprehensive review. 
Functional data is modeled using random variables that take values in a separable Hilbert space $\mathbb{H}$. 
This space is equipped with an inner product $\langle \cdot,\cdot \rangle_{\HH}$ and a norm $\|\cdot\|_{\HH} = \langle \cdot,\cdot \rangle_{\HH}^{1/2}$. 
We drop the subscript $\HH$, but we will reintroduce it for distinguishing between different Hilbert spaces when necessary.
An example of a Hilbert space is $L^2([0,1], \mathbb{R})$ that consists of square-integrable functions from the closed interval $[0,1]$ to $\RR$. 
For any $f \in L^2([0, 1], \RR)$,  $\int_0^1 f^2(\tau) \,d\tau < \infty$. 
The inner product is given by $\langle f,g \rangle = \int_0^1 f(\tau)g(\tau) \,d\tau$ for all $f,g \in L^2([0, 1], \RR)$. 
Another example is the space $\WW_q[0,1]$ for some $q \geq 1$. 
This space consists of functions from $[0,1]$ to $\RR$ that are $q-1$ times differentiable. 
Let $f^{(q-1)}$ denote the $(q-1)^{th}$ derivative of $f$. 
This space consists of functions that have $f^{(q-1)}$ absolutely continuous and a derivative $f^{(q)}$ almost everywhere with $\int_{0}^1 \left(f^{(q)}(\tau)\right)^2 \,d\tau < \infty$. 
The inner product is defined as $\langle f,g\rangle = \int_0^1 f(\tau)g(\tau)\,d\tau + \int_0^1 f^{(q)}(\tau)g^{(q)}(\tau)\,d\tau$.

Multivariate functional data are modeled using vectors in the space $\HH^p$. 
For any $x,y \in \HH^p$, $\langle x,y\rangle_{\HH^p} = \sum_{i=1}^p \langle x_i, y_i\rangle_{\HH}$, where $x_i$ and $y_i$ denote the $i$th components of $x$ and $y$, respectively.  
We consider the centered random vector $X = (X_1, \ldots, X_p) \in \mathbb{H}^p$.  
Let $\mathcal{L}(\mathbb{H})$ denote the space of bounded linear operators from $\mathbb{H}$ to $\mathbb{H}$. 
The covariance operator of $X$ lies in $\mathcal{L}(\mathbb{H}^p)$ and is given by $\Sigma = \mathbb{E}[X \otimes X]$. 
This operator takes an element $y \in \mathbb{H}^p$ to $\mathbb{E}[\langle X, y\rangle_{\HH^p} X]$. 
We can represent $\Sigma$ as a block covariance operator of the form
\begin{equation} \label{eqn:block-structure}
    \Sigma = \begin{bmatrix} 
    \Sigma_{11} & \Sigma_{12} & \ldots & \Sigma_{1p} \\
    \vdots & \vdots & \ddots & \vdots \\
    \Sigma_{p1}& \Sigma_{p2} & \ldots & \Sigma_{pp}
\end{bmatrix},
\end{equation}
where $\Sigma_{ij} \in \mathcal{L}(\mathbb{H})$ is the cross-covariance operator between $X_{i}$ and $X_j$. 
Let $(\Omega, \mathcal{F}, \mathbb{P})$ be the probability space on $\mathbb{H}$. The operator $\Sigma_{ij}$ is given by $\Sigma_{ij} = \int_{\Omega} X_j \otimes X_i \,d\mathbb{P},$
where the integral is a Bochner integral (Definition 2.6.2.\ in \citealp{fdabook}) and the tensor product, $X_j\otimes X_i$, is the mapping that takes any element $y \in \mathbb{H}$ to $\langle X_j, y\rangle X_i$, a random element in $\HH$. 
In order to leverage the matrix structure of $\Sigma$, we state the conventions for addition and multiplication of an operator with a scalar. 
Let $\Psi_1, \Psi_2 \in \mathcal{L}(\mathbb{H})$. 
Their sum is defined as $(\Psi_1 + \Psi_2)y = \Psi_1y + \Psi_2y$ for all $y \in \mathbb{H}$. Let $a \in \mathbb{R}$. Then $(a\Psi_1 ) y = a(\Psi_1 y)$ for all elements $y \in \mathbb{H}$. 
Let $A \in \mathbb{R}^{p \times K}$, $\Sigma \in \mathcal{L}(\mathbb{H}^p)$ and $\Psi \in \mathcal{L}(\mathbb{H}^K)$. 
Following conventions, if $\Sigma = A\Psi A^\top$, then $\Sigma_{ij} = \sum_{k_1=1}^K \sum_{k_2 = 1}^K  A_{ik_1}A_{jk_2}\Psi_{k_1k_2}$.  

The space $\mathcal{L}(\mathbb{H})$ can be equipped with the operator norm, which measures how much an operator stretches an element of unit norm. 
The operator norm is defined as $\|\Psi_1\|_{op} = \sup_{\|x\| = 1} \|\Psi_1 x\|$ for any $\Psi_1 \in \mathcal{L}(\HH)$. 
For simplicity, we drop the operator norm notation, so $\|\Psi\|$ for an operator should be interpreted as the operator norm, unless stated otherwise.
The operator norm is not associated with a natural inner product. 
To this end, we introduce a smaller class of operators which can be equipped with an inner product. 
Let $\{e_k\}_{k=1}^{\infty}$ be an orthonormal basis in $\HH$.
An operator $\Psi \in \mathcal{L}(\HH)$ is called a Hilbert-Schmidt operator if  
$\sum_{k=1}^{\infty}\|\Psi e_k\|^2 < \infty$. 
This definition is not dependent on the choice of the orthonormal basis $\{e_k\}$. 
We denote the space of Hilbert-Schmidt operators from $\HH$ to $\HH$ by $\TT_{1,1}$. 
This is a Hilbert space where the inner product of $\Psi_1, \Psi_2 \in \TT_{1,1}$ is defined as
$\langle \Psi_1, \Psi_2 \rangle_{\TT_{1,1}} = \sum_{k=1}^{\infty} \langle \Psi_1 e_k, \Psi_2 e_k\rangle_{\HH}$, where $\{e_k\}$ is an orthonormal basis in $\HH$ \citep[cf.\ Section 4.4 of][]{fdabook}. Covariance operators as well as cross-covariance operators in $\cL(\HH)$ belong to $\TT_{1,1}$. 

Finally, we introduce adjoint operators that play an important role in functional regression problems. Let $\HH_1$ and $\HH_2$ be two separable Hilbert spaces and let $\mathcal{L}(\HH_1, \HH_2)$ denote the space of bounded linear operators from $\HH_1$ to $\HH_2$. For every $\Psi \in \mathcal{L}(\HH_1, \HH_2)$, there exists a unique element $\Psi^* \in \mathcal{L}(\HH_2, \HH_1)$ such that $\langle \Psi x, y\rangle_{\HH_1} = \langle x, \Psi^*y\rangle_{\HH_2}$ for all $x \in \HH_1$ and $y \in \HH_2$. The operator $\Psi^*$ is called the adjoint of the operator $\Psi$. 


\subsection{Factor Model} \label{sec:factor_model}
Let $X \in \HH^{p}$ be a multivariate functional random variable that has a low-dimensional representation given by
\begin{equation} \label{eqn:model}
    X = AZ + E,
\end{equation}
where $Z \in \mathbb{H}^K$ and $E \in \mathbb{H}^p$ are the vector of multivariate functional latent factors and random noise, respectively. 
Here, $A \in \mathbb{R}^{p \times K}$ is the loading matrix that encodes the dependence of the variables $X$ on the latent factors $Z$. 
We assume that the random noise are uncorrelated, but we place no such restriction on the latent factors. 
The variables $A, Z$ and $E$ are all unobserved in our model. 

In the context of clustering, each latent factor  can be interpreted as a cluster. 
The loading matrix $A$ encodes the cluster memberships of the random variables: the $i$th random variable $X_i$ depends on a latent factor $Z_a$ if and only if $A_{ia} \neq 0$. 
We allow each variable to belong to multiple clusters simultaneously since the $i$th variable can have multiple $A_{ia} \neq 0$ for $a \in [K]$.
We define $G_a$ to be the set of $X_i$'s that depend on $Z_a$:
\begin{equation} \label{eqn:Ga}
    G_a = \{i \in [p]: A_{ia} \neq 0\}.
\end{equation}
If $i, j \in G_a$, then $X_i$ and $X_j$ are correlated through their dependence on $Z_a$ and are interpreted to belong to the same cluster. 

However, under Model \eqref{eqn:model}, the loading matrix $A$ is not identifiable in general. 
Specifically, Model \eqref{eqn:model} can be rewritten as $X = (AQ)(Q^{-1}Z) + E$ where $Q$ is an invertible matrix. 
The problem of identifiability in factor models has been studied extensively \citep{identif-factor-models}. 
We now state some assumptions to ensure that the loading matrix $A$ is uniformly scaled and identifiable.
\begin{assumption} \label{assump1}
$\sum_{a=1}^K |A_{ja}| \leq 1$ for all $1\leq j \leq p$.
\end{assumption}
This assumption ensures uniform scaling for each row of $A$. 
Although this assumption is not a requirement for the identifiability of the model, it rectifies scaling disparities among observed variables. 
It is also flexible enough to allow an observed variable to be related to none of the clusters, which occurs when $\sum_{a=1}^K|A_{ja}| = 0$. 

The next assumption requires each cluster to have two variables that belong exclusively to that cluster, referred to as the \emph{pure variables}.   
That is, for each pure variable $X_j$, we have $|A_{ja}| =1$ for some $a$ and $A_{jb} = 0$ for all $b \neq a$.
The pure variables can be interpreted as observable proxies of the latent factors since $X_j = \pm Z_a + E_j$. 
Throughout the manuscript, let $I_a$ be the set of indices $j \in [p]$ for pure variables that belong to the $a$th cluster.  
The set of all pure variables is denoted by $I = \cup_{a=1}^K I_a$.
The rest of the variables are \emph{non-pure variables}, and their indices are denoted by the set $J = [p]\backslash I$. 
For each $j\in J$, $A_{ja}$ is non-zero for at least two values of $a \in [K]$. 

\begin{assumption} \label{assump2}
$|I_a| \geq 2$ for all $a \in [K]$.
\end{assumption}

Assumption~\ref{assump2} ensures that each latent factor has at least two observable proxies to help in identifying its cluster membership. If the number of latent factors $K$ is known, then Assumption \ref{assump2} is sufficient to guarantee the identifiability of Model \eqref{eqn:model} \citep{identif-factor-models-econ,identif-factor-models}. However, when $K$ is unknown, additional assumptions are needed on the latent factors. We now revisit the correlation structure of the latent factors. Let $C$ denote the covariance operator of $Z$. 

\begin{assumption} \label{assump3}
The minimum separation $\Delta(C) = \min_{a\neq b} \left(\|C_{aa}\| \wedge \|C_{bb}\| - \|C_{ab}\| \right) > 0$ and $C$ has independent columns.
\end{assumption}

If $Z_a = Z_b$ or $Z_a = -Z_b$, then $\|C_{aa}\| = \|C_{bb}\| = \|C_{ab}\|$ and the minimum separation $\Delta(C)$ is zero.
Therefore, Assumption~\ref{assump3} ensures that no pair of latent factors are equal in norm with probability one.

We discuss the two parts of Assumption~\ref{assump3} separately. 
The identifiability conditions are motivated by \cite{bing}, which considered the case where $X$ and $Z$ are multivariate random variables in $\RR^p$ and $\RR^K$, respectively, so $C \in \RR^{K \times K}$. 
In this case, the minimum separation assumption $C_{aa} \wedge C_{bb} - |C_{ab}|>0$ ensures that all $2\times 2$ minors of $C$ have non-zero determinant, which guarantees that no two columns are proportional. 
However, three or more columns of $C$ may be linearly dependent and $C$ may not have full rank.
Analogously, in the functional case, the minimum separation condition prevents any two columns from being dependent, but places no restriction on three or more columns of $C$.

The second part of Assumption~\ref{assump3}, requiring that $C$ have independent columns, is not implied by the first part and merits separate discussion. 
In the scalar case, independence of columns of $C \in \RR^{K \times K}$ is equivalent to invertibility. 
In the functional case, however, this equivalence breaks down. 
A simple counterexample is a block diagonal operator $C$ in which at least one diagonal block is non-invertible; such an operator has independent columns but fails to be invertible.
We refer the reader to \cite{bing} for an in-depth comparison of Assumptions~\ref{assump1}--\ref{assump3} with other identifiability assumptions in the literature.

In Theorem \ref{thm2:existence}, we show that under Assumptions~\ref{assump1}--\ref{assump3}, the loading matrix $A$ is identifiable.  To formalize this, we utilize the covariance structure of the variables in the model. As introduced above, $C \in \mathcal{L}(\HH^K)$ is the covariance operator of $Z$. Let $\Sigma, \Gamma \in \mathcal{L}(\HH^p)$ be the covariance operators of $X$ and $E$, respectively.
Then
\begin{equation} \label{eqn:cov_decomp}
    \Sigma = ACA^\top + \Gamma,\qquad \Sigma_{ij} = \sum_{k_1=1}^K \sum_{k_2=1}^K A_{ik_1}A_{jk_2}C_{k_1k_2} + \Gamma_{ij}
\end{equation}
using the block matrix structure of the covariance operators. Since the errors are assumed to be uncorrelated,  $\Gamma_{ij} = \mathbf{0}$ for $i \neq j$ where $\mathbf{0}$ is the zero operator.  

Recall that a signed permutation matrix is a square matrix that has exactly one nonzero entry in each row and each column, where each nonzero entry is either $+1$ or $-1$. 
Observe that if $A$ and $C$ satisfy \eqref{eqn:cov_decomp}, then, for any signed permutation matrix $Q \in \RR^{K \times K}$, $\widetilde A:= AQ$ and $\widetilde C := Q^\top C Q$ satisfy \eqref{eqn:cov_decomp} as well.

\begin{theorem} \label{thm2:existence}
    (Identifiability) Under Model \eqref{eqn:model} and Assumptions \ref{assump1}--\ref{assump3}, the loading matrix $A$ and covariance operator $C$ can be identified from $\Sigma$ up to a signed permutation matrix.
\end{theorem}
Theorem~\ref{thm2:existence} implies that the associated overlapping clusters $G_a$ for $1 \leq a \leq K$ are identifiable up to label switching. 
Moreover, $\Gamma$ is also identifiable. 
In Section~\ref{sec:identify-a}, we propose an algorithm to select a representative version of $A$ from the collection $\{AQ: Q \in \RR^{K \times K} \text{ is a signed}$  $\text{permutation matrix}\}.$


\section{Estimation} \label{sec:estimation}
Recall from Section \ref{sec:factor_model} that $I_a$ is the set of pure variables corresponding to the latent factor $Z_a$. 
The set of all pure variables is denoted by $I = \cup_{a=1}^K I_a$, and the set of non-pure variables is denoted by $J = [p] \backslash I$. 
The partition of $I$ used to track the assignment of pure variables to different clusters is denoted by $\mathcal{I} = \{I_1, \ldots, I_K\}$.  
We denote the cardinality of a set $S$ by $|S|$. 
For any matrix $M\in \RR^{m_1 \times m_2}$ and subsets $I \subseteq [m_1], J \subseteq [m_2]$, $M_I$ represents the $|I| \times m_2$ sub-matrix of $M$ containing elements $M_{ij}$ such that $i \in I, j\in [m_2]$, and $M_{IJ}$ represents the $|I| \times |J|$ matrix containing elements $M_{ij}$ such that $i \in I, j \in J$. 
Similarly, for a block operator $\Phi \in \HH^{m_1 \times m_2}$, $\Phi_{IJ}$ denotes the matrix containing blocks $\Phi_{ij}$ such that $i \in I, j\in J$.

Let  $X^{(1)}, \ldots, X^{(n)}$ be $n$ 
realizations of the random vector  $X \in \HH^p$.  
Given  $X^{(1)}, \ldots, X^{(n)}$, the goal is to estimate the loading matrix $A$ in Model \eqref{eqn:model}. 
We propose to estimate the corresponding components for the pure variables $A_I$ and non-pure variables $A_J$ via a two-step procedure. 
In Section~\ref{sec:identify-a}, we propose an algorithm to identify $A$ when $\Sigma$ is known.
We develop an estimation procedure for $A$ in Section \ref{sec:est-sample}.

\subsection{Identifying $A$} \label{sec:identify-a}
The key observation underlying the identification and estimation of $A_I$ is that pure variables induce a distinctive structure on the operator norms of the blocks of $\Sigma$. 
To see this, suppose that $i \in I_a$. 
Since $X_i$ is a pure variable, $A_{ik_1} = 0$ for all $k_1 \neq a$ and $A_{ia} \in \{+1, -1\}$. 
For any $j \neq i$, \eqref{eqn:cov_decomp} reduces to $\Sigma_{ij} = A_{ia}\sum_{k_2 = 1}^K A_{jk_2}C_{ak_2}$, so $\Sigma_{ij}$ is a linear combination of the $a$th row of $C$, with the weights given by the loadings of the $j$th variable. 
When $j \in I_a$, this expression further simplifies to $\Sigma_{ij} = A_{ia}A_{ja}C_{aa}$. 
In this case, the cross-covariance between  the $i$th and $j$th variables is equal to the covariance of the latent factor $Z_a$, up to a sign change. 
Therefore, $\|\Sigma_{ij}\| = \|C_{aa}\|$. 

For any $X_j$, Assumption \ref{assump1} ensures that the sum of loadings $\sum_{k_2=1}^K |A_{jk_2}| \le 1$. 
Furthermore, Assumption~\ref{assump3} ensures that $\|C_{aa}\|$ is the maximum of the operator norms in the $a$th row of $C$. 
Consequently, $\|\Sigma_{ij}\| \leq \|C_{aa}\|$. 
Equality is attained only when the weight on the $a$th component $|A_{ja}|$ is equal to $1$, which occurs when $j \in I_a$. 
We conclude that if $i \in I_a$, the remaining elements of $I_a$ are the column indices of the $i$th row at which the maximum norm is achieved. 

To formalize this argument, we introduce two key quantities. Let
\begin{equation} \label{eqn:Mi}
    M_i = \max_{j \in [p] \backslash \{i\}}  \|\Sigma_{ij}\|
\end{equation}
be the maximum of the operator norms of covariance operators in the $i$th row of $\Sigma$, for all $1\leq i \leq p$. Let
\begin{equation} \label{eqn:Si}
    S_i = \{j \in [p] \backslash \{i\} : \|\Sigma_{ij}\| = M_i\}
\end{equation}
be the set of column indices at which $M_i$ is attained by the operator norms of covariance operators in the $i$th row. 

\begin{theorem} \label{thm:pure_var}
Under Model \eqref{eqn:model} and Assumptions \ref{assump1}--\ref{assump3}, the following holds:
    \begin{enumerate} [label=(\alph*)]
        \item $i \in I \Longleftrightarrow M_i = M_j$ for all $j \in S_i$.
        \item The pure variable set $I$ can be uniquely determined from $\Sigma$. The partition $\mathcal{I}$ is unique and can be determined from $\Sigma$ up to permutations. 
    \end{enumerate}
\end{theorem} 

Given the covariance operator $\Sigma$, Theorem \ref{thm:pure_var} allows us to devise an algorithm for identifying the pure variable set $I$ and its partition $\cI$. 
For each index $i \in [p]$, we compute $M_i$ and $S_i$. 
If the condition in Theorem \ref{thm:pure_var} is satisfied for index $i$, then $X_i$ is declared as a pure variable and we define the associated cluster as $I^{i} = S_i \cup \{i\}$. 
We repeat this procedure until all indices in $[p]$ have been exhausted.
Let $K$ denote the number of distinct sets $I^i$ identified in the process. 
We then set the partition $\cI = \{I^{i_1}, \ldots, I^{i_K}\}$ and the set of pure variables $I = \cup_{k=1}^K I^{i_k}$. 
Finally, the set of indices for the non-pure variables is $J = [p] \backslash I$. 

To construct $A_I$ given $\cI$, let  $(Z_1, \ldots, Z_K)$ be a set of latent variables that correspond to $I^{i_1}, \ldots,I^{i_k}$. 
However, $(c_1Z_{\pi(1)}, \ldots, c_kZ_{\pi(K)})$, where $\pi$ is a permutation of the set $[K]$ and $c_k \in \{-1, +1\}$, is also a valid set of latent variables. 
This corresponds to the identifiability of $A$ up to a signed permutation matrix, as shown in Theorem~\ref{thm2:existence}. 
For the rest of this paper, we will focus on a representative version of $A$ that serves as a definitive target during inference. 
Let $s_k = \min\{i:i \in I_k\}$ denote the smallest indices in the pure variable sets, and let $r_j$ be the inverse rank of $s_j$ in the set $\{s_k: 1\le k\le K\}$. For example, if $s_{j_0}$ is the smallest among the $s_k$, then $r_{j_0} = 1$.
We set $A_{s_jr_j} = 1$ and for all indices $i$ that are in the same group as $s_j$, set $A_{i r_j} =  1$ if $\|\Sigma_{s_js_j} + \Sigma_{s_ji}\|> \|\Sigma_{s_js_j} - \Sigma_{s_ji}\|$ and $-1$ otherwise. 
This algorithm uniquely determines $A_I$ from $\cI$, and we re-label $I_a = I^{i_k}$ for $k$ such that $r_k=a$.
The matrix $A_I$ has the properties that the leading non-zero element of each column is $1$, and location of the leading non-zero element is strictly increasing with the column number.

With $A_I$ determined, we proceed with the construction of the rest of the loading matrix $A_J$. 
To this end, we compute $C$, the covariance operator of the latent factors, using only $A_I$ and the non-diagonal elements of $\Sigma_{II}$.
Once $C$ is available, $A_J$ can be constructed as the solution of a regression problem in operators involving both $\Sigma$ and $C$. 
The resulting optimization can be simplified to a finite-dimensional problem. 
In the following, we provide the details for the construction of $A_J$.

The block structure of $\Sigma$ allows \eqref{eqn:cov_decomp} to be decomposed as 
\begin{equation} \label{eqn:compos-main}
    \Sigma_{II} = A_ICA_I^\top + \Gamma_{II},\qquad \Sigma_{IJ} = A_ICA_J^\top,\qquad \Sigma_{JJ} = A_JCA_J^\top + \Gamma_{JJ}.
\end{equation}
By~\eqref{eqn:compos-main}, it can be shown that $\Sigma_{i_1i_2} = A_{i_1a}A_{i_2a}C_{aa}$ for all $i_1,i_2 \in I_a$ such that $i_1 \neq i_2$.
Averaging over all such indices leads us to the formula
\begin{equation} \label{eqn:c_aa}
    C_{aa} = \frac{1}{|I_a|(|I_a|-1)} \sum_{i_1,i_2 \in I_a, i_1 \neq i_2} A_{i_1a}A_{i_2a}\Sigma_{i_1i_2},
\end{equation}
since $|A_{ia}| =1 $ for all $i \in I_a$. Let $W =A_I^\top A_I$, which is a diagonal matrix with entries $W_{aa} = |I_a|$ for all $1\leq a \leq K$.
To estimate $C_{ab}$ for $a \neq b$, it can be shown by~\eqref{eqn:compos-main} that
\begin{equation} \label{eqn:I-rows}
    A_I^\top \Sigma_{II} A_I = WCW + A_I^\top \Gamma_{II} A_I.
\end{equation}
When $a \neq b$, the columns of $A_I$  at indices $a$ and $b$ do not have any overlapping non-zero entries. 
 Since $\Gamma_{II}$ is a diagonal matrix, $(A_I^\top \Gamma_{II} A_I)_{ab}$ $ = 0$. 
 The $(a,b)$th element of the matrices in \eqref{eqn:I-rows} yields the equality
$|I_a\|I_b|C_{ab} = \sum_{i_1,i_2 \in I} A_{i_1a}\Sigma_{i_1i_2}A_{i_2b}.$ 
The elements $\{A_{ia}:i \in I\}$ are non-zero if and only if $i \in I_a$. Therefore,
\begin{equation}\label{eqn:c_ab}
    C_{ab} = \frac{1}{|I_a\|I_b|} \sum_{i_1 \in I_a,i_2 \in I_b} A_{i_1a}A_{i_2b}\Sigma_{i_1i_2}.
\end{equation}

Recall the expression for $\Sigma_{IJ}$ from \eqref{eqn:compos-main} and pre-multiply both sides by $W^{-1}A_I^\top$ to derive
\begin{equation} \label{eqn:aj-main} 
    W^{-1} A_I^\top \Sigma_{IJ} = C A_J^\top,
\end{equation}  
where all terms other than $A_J$ have been computed. 
Then \eqref{eqn:aj-main} is a regression problem in operators and it uniquely determines $A_J^\top$ because $C$ has linearly independent columns by Assumption~\ref{assump3}.
Despite having independent columns, $C$ may not be invertible since it is an infinite-dimensional compact operator.
We change this into a tractable problem by using the inner product on the space $\TT_{1,1}$, introduced in Section \ref{sec:prelim}. 
Recall that for any two operators $\Psi_1, \Psi_2 \in \TT_{1,1}$, $\langle \Psi_1, \Psi_2 \rangle_{\TT_{1,1}} = \sum_{k=1}^{\infty} \langle \Psi_1 e_k, \Psi_2 e_k \rangle_{\HH}$, where $e_k$ is an orthonormal basis in $\HH$. 
Let $\TT_{|J|,K}$ be the space of Hilbert-Schmidt operators from $\HH^{|J|}$ to $\HH^K$.
Any $\Psi^1 \in \TT_{|J|, K}$ has a block operator form given by $(\Psi^1_{k,j}, k\in [K], j \in [|J|])$.
The space $\TT_{|J|,K}$ is equipped with the inner product $\langle\Psi^1, \Psi^2\rangle_{\TT_{|J|,K}} = \sum_{j=1}^{|J|} \sum_{k=1}^K\langle \Psi^1_{k,j}, \Psi^1_{k,j}\rangle_{\TT_{1,1}} $ for any $\Psi^1, \Psi^2 \in \TT_{|J|,K}$.
The operators $CA_J^\top$ and $W^{-1}A_I^\top \Sigma_{IJ}$ belong to $\TT_{|J|,K}$.
We express \eqref{eqn:aj-main} as a minimization problem in the Hilbert-Schmidt norm:
\begin{equation} \label{eqn:optim-aj}
    A_J = \argmin_{Y \in \mathbb{R}^{|J| \times K} } \left\|W^{-1} A_I^\top \Sigma_{IJ} - CY^\top \right\|_{\TT_{|J|,K}}.
\end{equation}
The following theorem  (see, for instance, Theorem 3.5.10 in \citealp{fdabook}) shows that the optimization in \eqref{eqn:optim-aj} can be solved by inverting a matrix of scalars that is constructed using the inner product on $\TT_{|J|,K}$.

\begin{theorem} \label{thm:op_to_matrix}
Let $\Psi:\RR^{|J|\times K} \to \TT_{|J|,K}$ be the mapping that acts on a matrix $Y \in \RR^{|J|\times K}$ as
$$ \Psi( Y) = \mathscr{C} Y^\top = \left(\sum_{s=1}^K Y_{j,s} \mathscr{C}_{k,s}, k \in [K], j\in [|J|]\right). $$
Let the operator $\CU \in \TT_{|J|,K}$ be given and consider the optimization problem
$$\argmin_{Y\in\RR^{|J|\times K}} \| \CU-\Psi Y\|_{\TT_{|J|,K}}.$$
Assume that $\CU \in \mathrm{Dom}(\Psi^\dagger)$, where $\mathrm{Dom}$ refers to the domain of an operator and $M^\dagger$ denotes the Moore-Penrose inverse of $M$.
The solution to this optimization problem is given by\\ $Y^{\mathrm{optim}}  = (\Psi^*\Psi)^{\dagger} \Psi^*\CU $, which admits the representation 
\begin{equation}\label{eqn:optim_scalar}
Y^{\mathrm{optim}}=\left\{\sum_{k=1}^K\left\langle\mathscr{C}_{k, s}, \mathscr{C}_{k, s^{\prime}}\right\rangle_{\mathbb{T}_{1,1}}\right\}_{s, s^{\prime} \in[K]}^{\dagger}\left\{\sum_{k=1}^K\left\langle\mathscr{C}_{k, s}, \mathscr{U}_{k, j}\right\rangle_{\mathbb{T}_{1,1}}\right\}_{s \in[K], j \in [|J|]}.
\end{equation}
\end{theorem}

Both the matrices in \eqref{eqn:optim_scalar} are real-valued. 
The expression for $Y^{\mathrm{optim}}$ reduces to the familiar least squares solution if $\HH=\RR$. 
To solve \eqref{eqn:optim-aj}, we set $\CU = W^{-1}A_I^\top \Sigma_{IJ}$ and $\CCC = C$ in Theorem \ref{thm:op_to_matrix}. 
We note that $\CU \in \text{Dom}(\Psi^\dagger)$ for these specifications.


\subsection{Estimating $A$} \label{sec:est-sample}

In practice, we are given $n$ realizations of the random vector $X \in \HH^p$, which we denote as $X^{(1)}, \ldots, X^{(n)}$. We estimate the covariance operator $\Sigma$ using the sample covariance operator $\widehat \Sigma$, defined as follows: 
\begin{equation} \label{eqn:sample-cov}
    \hat{\Sigma}u = n^{-1} \sum_{\ell=1}^n\langle X^{(\ell)}, u\rangle X^{(\ell)} \mathrm{~~for~all~} u \in \HH^p.
\end{equation} 
To estimate $A_I$, we modify the procedure in Section \ref{sec:identify-a} to incorporate sampling error. 
We compute the maximum operator norm in the $i$th row as $\widehat M_i = \max_{j \in [p] \backslash \{i\}} \left\|\widehat{\Sigma}_{ij}\right\|$. The set $\widehat S_i = \left\{ j\in [p] \backslash \{i\} : \widehat M_i- \left\|\widehat{\Sigma}_{ij}\right\|   \leq 2\delta \right\}$ captures the column indices at which the operator norm is within $2\delta$ distance of $\widehat M_i$. 
We also relax the pure variable criterion in Theorem \ref{thm:pure_var} so that the $i$th variable is classified as pure if $|\widehat M_j -  \|\widehat \Sigma_{ij}\|| \leq 2\delta$ for all $j \in \hat S_i$.
The tuning parameter $\delta$ is based on how well the sample covariance operator estimates the true covariance operator. 
In particular, $\delta$ is the maximum distance between an element of $\widehat \Sigma$ and the corresponding element of $\Sigma$ in the operator norm.
We provide statistical guarantees for our method based on $\delta$ in Section \ref{sec:th-guar} and propose a  cross-validation procedure when the value of $\delta$ is unknown in Section \ref{sec:num-exp}. 
Motivated by Theorem \ref{thm:pure_var}, we devise an algorithm to estimate the pure variable set $I$ and its partition $\cI$ in Algorithm \ref{algo:pure_set}.

\begin{algorithm}
\caption{Estimating the pure variable set $I$ and its partition $\mathcal{I}$.} \label{algo:pure_set}
Input the  sample covariance $\widehat \Sigma$ and a tuning parameter $\delta >0$.
\begin{enumerate}
    \item Initialize $\widehat I^{i} = \varnothing$ for all $1\leq i \leq p$. Repeat the following steps for all $i \in \{1, \ldots, p\}$:
    \begin{enumerate}
        \item Calculate the maximum operator norm $\widehat M_i = \max_{\ell \in [p] \backslash \{i\}} \left\|\widehat{\Sigma}_{i\ell}\right\|$.
        \item Identify the set $\widehat S_i = \left\{ \ell \in [p] \backslash \{i\} : \widehat M_i- \left\|\widehat{\Sigma}_{i\ell}\right\|   \leq 2\delta \right\}$.
        \item Calculate $\widehat M_j = \max_{\ell \in [p] \backslash \{i\}} \left\|\widehat{\Sigma}_{j\ell}\right\|$ for all $j \in \widehat S_i$.
        \item If $\left| \widehat M_j - \left\|\widehat{\Sigma}_{ij}\right\|  \right|  \leq 2 \delta$ for all $j \in \widehat S_i$, then set $I^i = \widehat S_i ~\cup~ \{i\}$.
    \end{enumerate}
    \item Let $\widehat \cI = \{\widehat I^i: \widehat I^i \neq \varnothing \}$. If $\widehat I^{j} \cap \widehat I^{k} \neq \varnothing$, then replace $\widehat I^{j}$ and $\widehat I^{k}$ by $\widehat I^{j} \cap \widehat I^{k}$.
    \item Set $\widehat K$ to be the number of distinct sets in $\widehat \cI$. Relabel the elements in $\widehat \cI$ such that $\widehat \cI = \{\widehat I^a:a \in [\widehat K]\}$.
\end{enumerate}
\end{algorithm}

Once $\widehat \cI$ is estimated, we define $\widehat A_{\wh I}$ using the algorithm described in Section~\ref{sec:identify-a}.
Recall that this algorithm ensures that the leading non-zero element of each column is $1$, and the location of the leading non-zero element is strictly increasing with the column number.
In the rest of the manuscript, we denote $\widehat A_{\wh I}$ by $\wh A_I$ for notational ease.
We estimate the covariance operator $C$ and then solve the sample version of the regression problem in \eqref{eqn:optim-aj} to estimate $A_J$. 
We estimate $C$ using the sample versions of \eqref{eqn:c_aa} and \eqref{eqn:c_ab} as
$$\widehat{C}_{aa} = \frac{1}{\left|\hat{I}_a \right| \left(\left|\hat{I}_a \right|-1\right)} \sum_{i_1, i_2 \in \hat{I}_a, i_1 \neq i_2} \widehat A_{i_1a}\widehat A_{i_2a}\widehat{\Sigma}_{i_1i_2},\quad \widehat{C}_{ab} = \frac{1}{\left|\hat{I}_a \right| \left|\hat{I}_b \right|} \sum_{i_1 \in \hat{I}_a, i_2 \in \hat{I}_b} \widehat{A}_{i_1a}\widehat{A}_{i_2b}\widehat{\Sigma}_{i_1i_2}.$$
Let $\widehat{\mathscr{U}} = \hat{W}^{-1} \hat{A}_I^\top \hat{\Sigma}_{IJ}$, where $\widehat W = \widehat A_I^\top \widehat A_I$. 
We substitute $\CU$ with $\widehat \CU$ and $\mathscr{C}$ with $\widehat{C}$ in Theorem \ref{thm:op_to_matrix} to obtain
$\widehat A_J^\top =  \left\{\sum_{k=1}^{\hat K}\left\langle \widehat{C}_{k,s},  \widehat{C}_{k,s'}\right\rangle_{\TT_{1,1}}\right\}_{s,s'\in [\widehat K]} ^{-1} \left\{\sum_{k=1}^{\hat K} \langle \widehat{C}_{k,s}, \hat{\CU}_{k,j}\rangle_{\TT_{1,1}}\right\}_{s\in [\widehat K], j\in \widehat J}$. 
These steps are summarized in Algorithm \ref{algo:2}. 
We refer to the proposed method for estimation of overlapping clusters as funFMC, a shorthand for functional Factor Model-based Clustering.

\begin{algorithm}
\caption{Estimating the loadings of the non-pure variables $A_J$.} \label{algo:2}
Input the sample covariance $\widehat \Sigma$, the estimated pure variable partition $\widehat \cI$ and the estimated loadings corresponding to pure variables $\widehat A_I$.
\begin{enumerate}
    \item Compute the covariance of the latent factors $\widehat C$ as 
    \begin{enumerate}
        \item $\widehat C_{aa} = \frac{1}{|\hat I_a|(|\hat I_a| - 1)} \sum_{i_1, i_2 \in \hat I_a, i_1 \neq i_2} \widehat A_{i_1a} \widehat A_{i_2a}\widehat \Sigma_{i_1i_2}$ for all $a \in [\widehat K]$,
        \item $\hat C_{ab} = \frac{1}{|\hat I_a||\hat I_b| } \sum_{i_1 \in \hat I_a, i_2 \in \hat I_b} \hat A_{i_1a} \hat A_{i_2b} \hat \Sigma_{i_1i_2} $ for all $a,b \in [\widehat K]$ such that $a\neq b$.
    \end{enumerate}
    \item Calculate the operator $\widehat\CU = \widehat W^{-1} \widehat A_I^\top \widehat \Sigma_{IJ} $, where $\widehat W = \widehat A_I^\top \widehat A_I$.
    \item Calculate $\widehat A_J^\top =  \left\{\sum_{k=1}^{\hat K}\left\langle \widehat{C}_{k,s},  \widehat{C}_{k,s'}\right\rangle_{\TT_{1,1}}\right\}_{s,s'\in [\widehat K]} ^{-1} \left\{\sum_{k=1}^{\hat K} \langle \widehat{C}_{k,s}, \hat{\CU}_{k,j}\rangle_{\TT_{1,1}}\right\}_{s\in [\widehat K], j\in \widehat J}$.
\end{enumerate}
\end{algorithm}

We discuss the estimation of $\widehat \Sigma$ from discretely-observed samples in Section \ref{sec:num-exp}.


\section{Statistical Guarantees} \label{sec:th-guar}
This section provides statistical guarantees for cluster recovery using the proposed method. 
We establish (i) consistency of estimation of the number of clusters $K$, (ii) recovery of the pure variable set $I$, and (iii) asymptotic results for the estimator of $A_J$. 
The consistency of $\widehat A_J$ follows from the convergence of $\sqrt{n}(\widehat A_J - A_J)$ to a zero-mean Gaussian random variable.
Section \ref{sec:pure-var-guar} establishes consistency of the estimators of $K$ and $I$ using accuracy of $\widehat \Sigma$ in the operator norm. 
Section \ref{sec:aj-guar} derives asymptotic normality for the estimator of $A_J$ using the Hilbert-Schmidt inner product on covariance operators.
 
\subsection{Statistical Guarantees for Estimation of Pure Variables}\label{sec:pure-var-guar}

In this section, we establish consistency of estimation of the number of clusters $K$ and the pure variable set $I$ using Algorithm \ref{algo:pure_set}. 
We first derive a concentration inequality for the sample covariance operator $\widehat \Sigma$, defined in \eqref{eqn:sample-cov}, using the operator norm. 
We then quantify which variables are selected as pure variables with high probability in Algorithm \ref{algo:pure_set}.

The sample covariance operator $\widehat \Sigma \in \cL(\HH^p)$ has a block operator structure, with its $(i, j)$th block being the sample covariance operator $\widehat \Sigma_{ij} \in \cL(\HH)$.
This block is defined as $\widehat \Sigma_{ij}y = n^{-1} \sum_{\ell=1}^n \langle X^{(\ell)}_{j}, y\rangle_{\mathbb{H}} X^{(\ell)}_{i}$ for all $y \in \HH$, where $X_j^{(\ell)}$ denotes the $j$th component of the $\ell$th sample vector.
We derive a concentration inequality for $\widehat \Sigma$ in the operator norm by applying existing results on its block operators.
Such results have been proved for subgaussian and pregaussian random variables, which we define next.

Let $Y \in \HH$ be a centered, weakly integrable random variable with covariance operator $\Sigma_Y$ and sample covariance operator $\widehat \Sigma_Y$. 
The variable $Y$ is called subgaussian if for all $y \in \HH$, $\|\langle Y, y \rangle_{\HH} \|_{\psi_2} \leq \|\langle Y, y \rangle_{\HH} \|_{L_2(\PP)}$, where $\|\cdot\|_{\psi_2}$ denotes the Orlicz norm.
Theorem 9 in \cite{conc} states a concentration bound for $\|\widehat \Sigma_Y - \Sigma_Y\|$ in terms of the quantity $r(\Sigma_Y) = (\EE\|Y\|_{\HH})^2/\|\Sigma_Y\|$.
We derive a bound for $\|\widehat \Sigma - \Sigma\|$ by extending this result to the multivariate case.

\begin{proposition} \label{prop:event}
Let $X = (X_1, \ldots, X_p)^\top \in \HH^p$ be a centered, subgaussian 
random variable with covariance operator $\Sigma$. 
Define $\Xi$ as
\begin{equation*}
\Xi_{ij} = \begin{cases} 
\Sigma_{ii}
    & \text{if } i=j \\
\Psi_{ij} 
    & \text{if } i\neq j
\end{cases},
\end{equation*}
where $\Psi_{ij}$ is the covariance operator of $(X_i, X_j) \in \HH^2$. 
For any $t\ge 1$, define
\begin{enumerate}[label=(\alph*)]
    \item $\delta_a = c_0 \max_{1 \leq i, j \leq p} \|\Xi_{ij}\| \left( \sqrt{\frac{r(\Xi_{ij})}{n}} \bigvee \frac{r(\Xi_{ij})}{n} \bigvee \sqrt{ \frac{t}{n}} \bigvee \frac{t}{n} \right),$
    \item $\delta_b = c_0 \max_{1 \leq i , j \leq p} (\|\Sigma_{ii}\| + \|\Sigma_{jj}\| ) \left( \sqrt{\frac{2 ( r(\Sigma_{ii})+ r(\Sigma_{jj}))}{n}} \bigvee \frac{2 ( r(\Sigma_{ii})+ r(\Sigma_{jj}))}{n} \bigvee \sqrt{ \frac{t}{n}} \bigvee \frac{t}{n} \right),$
\end{enumerate}
where $c_0>0$ is a constant. 
Then, the event 
    \begin{equation} \label{eqn:eventepsilon}
    \mathcal{E}(\delta)= \left\{\max_{1 \leq i,j \leq p} \|\hat{\Sigma}_{ij} - \Sigma_{ij}\| \leq \delta \right\}
\end{equation}
occurs with probability at least $1- p^2 e^{-t}$ for $\delta \in \{\delta_a, \delta_b\}$.
\end{proposition}

Note that Theorem 9 of \cite{conc} requires that $X$ is also pregaussian, meaning that there exists a Gaussian random variable in $\HH$ with the same covariance operator as $X$. However, that is automatic when $\HH^p$ is a Hilbert space.

Proposition~\ref{prop:event} states that the event $\cE(\delta)$ in
\eqref{eqn:eventepsilon}, where each element of $\Sigma$ is estimated well, occurs with high probability. 
The estimation gap $\delta$ can be defined either in terms of the two-dimensional block operators $\Psi_{ij}\in \cL(\HH^2)$ by using $\delta_a$ or in terms of $\Sigma_{ij} \in \cL(\HH)$ by using $\delta_b$.
The use of $\delta_a$ results in a tighter bound, but the computation of $\|\Psi_{ij}\|$ is more  expensive compared to $\|\Sigma_{ij}\|$.

In order to obtain a high-dimensional bound, we set $t = 3 \log{p}$ in Proposition~\ref{prop:event} and derive $\PP[\cE(\delta)] \geq 1-1/p.$
When $p$ is fixed but the sample size $n$ increases, we use the the notation $\delta_n$ to highlight the dependence of $\delta$ on $n$.
In this case, we choose $t = \gamma \log n$ for any $\gamma >0$.
Then, as $n \to \infty$, $\delta_n \to 0$ and $\PP[\cE(\delta_n)] = 1-p^2n^{-\gamma} \to 1$.
Hence, for large enough $n$, the estimation error $\delta_n$ decreases with increasing $n$ and the probability of the event $\cE(\delta_n)$ increases.

We now return to the original goal of proving consistency of $\widehat I$.
We show that on the event $\cE(\delta)$, variables with indices in $I$ are always estimated as pure variables.
However, there are also false discoveries. 
The set $\widehat I ~\backslash~ I$ consists of non-pure variables that have a strong association with one latent factor.
We quantify such false discoveries in terms of the sampling error $\delta$ and the separation between the latent factors.

Firstly, we consider the indices $i, j_1, j_2 \in I_a$. 
Recall from Section \ref{sec:identify-a} that $\|\Sigma_{ij_1}\| = \|\Sigma_{ij_2}\| = \|C_{aa}\|$. 
On the event $\cE(\delta)$, since $\widehat \Sigma_{ij_1}$ and $\widehat \Sigma_{ij_2}$ are at most $\delta$ distance away from $\Sigma_{ij_1}$ and $\Sigma_{ij_2}$ respectively, the triangle inequality leads to 
\begin{equation} \label{eqn:est-ai-2delta}
   | \|\widehat \Sigma_{ij_1}\|-\|\widehat \Sigma_{ij_2}\| | \leq 2\delta.
\end{equation}
This motivates the condition for pure-variable estimation in Algorithm \ref{algo:pure_set}.

However, \eqref{eqn:est-ai-2delta} is also satisfied if $j_2 \notin I_a$, but $X_{j_2}$ has a strong association with $Z_a$.
Let $\nu = \Delta(C) >0$, defined in Assumption~\ref{assump3}, denote the separation of the blocks of $C$ in the operator norm.
As before, let the indices $i, j_1 \in I_a$. Then $\|\Sigma_{ij_1}\| =\|C_{aa}\|$.
The term $\|\Sigma_{ij_2}\| = \|\sum_{k_2=1}^K A_{j_2k_2}C_{ak_2}\| \leq |A_{j_2a}|\|C_{aa}\| + (1-|A_{j_2a}|)(\|C_{aa}\|-\nu)$, where the second step uses the triangle inequality and definition of $\nu$. 
Consequently, $\|\Sigma_{ij_2}\| \leq \|C_{aa}\| - (1-|A_{j_2a}|)\nu$. 
Taking into consideration $\delta$-perturbations on the event $\cE(\delta)$,
\begin{equation} \label{eqn:quasi-pure-moti}
    \|\widehat \Sigma_{ij_1}\| - \|\widehat \Sigma_{ij_2}\| \geq (1-|A_{j_2a}|)\nu - 2\delta.
\end{equation}
If the variable $X_{j_2}$ is weakly associated with $Z_a$, then $(1-|A_{j_2a}|)$ is large and the sample covariances in \eqref{eqn:quasi-pure-moti} are well-separated. 
As a result, Algorithm \ref{algo:pure_set} does not identify $X_{j_2}$ as a pure variable. 
However, if $X_{j_2}$ is strongly associated with $Z_a$, then it might be identified as a pure variable. 
In particular, \eqref{eqn:est-ai-2delta} may be satisfied if $(1-|A_{j_2a}|)\nu \leq 4\delta$.

We call $X_j$ a \emph{quasi-pure} variable if it is a non-pure variable and $|A_{ja}|\geq 1-4\delta/\nu$ for some $a \in [K]$. 
We denote the indices of quasi-pure variables strongly associated with the latent factor $Z_a$ by $J_1^a = \{j \in J : |A_{ja}| \geq 1 - 4\delta/\nu\}.$ 
The set of all quasi-pure variables is denoted by $J_1 = \cup_{a=1}^K J_1^a$. 
If $j \in J_1^a$, then Assumption \ref{assump2} implies that $|A_{jb}| \leq 4\delta/\nu$ for all $b \neq a$. 
Lastly, let
$\|C\|_{\infty} = \max_{1\leq i , j\leq K} \|C_{ij}\|.$ 

\begin{theorem} \label{thm:pureindexguaran}
Suppose that Model \eqref{eqn:model} and Assumptions \ref{assump1}--\ref{assump3} hold. 
Also assume that \\$\nu > 2 \max (2 \delta, \sqrt{2 \|C\|_{\infty} \delta}).$
Then on the event $\cE(\delta)$,
\begin{enumerate} [label=(\alph*)]
    \item $\hat{K} = K$;
    \item 
there exists a label permutation $\pi$ of the set $\{1, \ldots, K\}$ such that $\hat{\cI}$ satisfies:
$$I_{\pi(a)} \subseteq \hat{I}_a \subseteq \left(I_{\pi(a)} \cup J_1^{\pi(a)} \right);$$
\item there exists a signed permutation matrix $Q$ such that
$\bar A= AQ$ satisfies the property: $\mathrm{sign}(
\bar A_{ia}) = \mathrm{sign}(\hat A_{ia})$ for
all $i \in \hat I_a, a \in [K]$.
\end{enumerate}
 \end{theorem}

The first two parts of Theorem \ref{thm:pureindexguaran} show that the labels of $\hat A_I$ can be aligned with those of $A_I$. 
The third part states that the signs of these matrices can also be aligned using the signed permutation matrix $Q$.
An immediate consequence of part $(b)$ is that with high probability, $I \subseteq \hat{I} \subseteq (I \cup J_1)$.
Therefore $X_i$ is estimated as a pure variable for all indices $i \in I$ and all non-pure indices in $\widehat I$ lie in $J_1$. 
The pure variable sets are identifiable up to label switching, which agrees with Theorem \ref{thm:pure_var}. 

When $J_1 = \varnothing,$ pure variables are estimated perfectly. 
Then the algorithm for the construction of $\widehat A_I$ given $\widehat \cI$ ensures that under the assumptions of Theorem~\ref{thm:pureindexguaran}, $ \widehat I_a = I_a$ in part (b), and $Q = I$ in part (c).
The set $J_1 = \varnothing$ when no entries of the loading matrix corresponding to non-pure variables are greater than $1-4\delta/\nu$ in absolute value.
This may occur when $\nu$ is very large, that is, the latent factors are well-separated.
This may also occur for any value of $\delta$ and $\nu$, when non-pure variables are not highly associated with any latent factor.
In particular, when $p$ and $\nu$ are fixed, but $n$ increases, we know that $\delta_n \to 0$, hence there are no quasi-pure variables for large enough $n$.


\subsection{Statistical Guarantees for Estimation of Non-Pure Variables}\label{sec:aj-guar}

In this section, we establish asymptotic normality of $\widehat{A}_J$, which in particular implies consistency.
The main idea is that $\widehat{A}_J$ is a smooth function of $\widehat{\Sigma}$, and since the asymptotics of $\widehat{\Sigma}$ are known, those of $\widehat{A}_J$ can be computed.
We show that $\sqrt{n}(\widehat{A}_J^\top - A_J^\top)$ converges in distribution to a zero-mean Gaussian random matrix $\mathcal{G}(\mathfrak{Z})$, where $\mathfrak{Z}$ is the limiting Gaussian for $\sqrt{n}(\widehat{\Sigma}-\Sigma)$ and $\mathcal{G}$ is a bounded linear map defined explicitly below.
The covariance structure of the limit is fully determined by $\mathcal{G}$ and the covariance of $\mathfrak{Z}$.

We begin by recalling the asymptotics of the sample covariance operator $\widehat{\Sigma}$ \citep[Chapter~8]{fdabook}.
If $\mathbb{E}\|X\|^4 < \infty$, then $\sqrt{n}(\widehat{\Sigma} - \Sigma) \xrightarrow{d} \mathfrak{Z}$ in $\mathbb{T}_{p,p}$ as $n \to \infty$, where $\mathfrak{Z}$ is a zero-mean Gaussian random variable with covariance operator $\mathscr{S} = \mathbb{E}[(X \otimes X-\Sigma) \otimes_{\mathrm{HS}}(X \otimes X-\Sigma)]$.
Here $\mathscr{S}: \mathbb{T}_{p,p} \to \mathbb{T}_{p,p}$ is a bounded positive semidefinite operator.

From Algorithm~\ref{algo:2}, we can see that $\widehat A_J$ depends on $\widehat \Sigma$ through the operators $\widehat C$ and $\widehat \CU$, and their inner products.
To formalize this dependence, we define the matrices $V, \widehat V \in \RR^{K \times K}$ as $V = \left\{\sum_{k=1}^K\left\langle C_{k,s},  C_{k,s'}\right\rangle_{\TT_{1,1}}\right\}_{s,s'\in [ K]}$ and $\widehat V = \left\{\sum_{k=1}^K\left\langle \widehat{C}_{k,s},  \widehat{C}_{k,s'}\right\rangle_{\TT_{1,1}}\right\}_{s,s'\in [\widehat K]}$. 
The map $\cF: \TT_{p,p} \to \TT_{|J|,K}$ is defined by
$\cF(B) = W^{-1}A_I^\top B_{IJ}$.
In particular, $\cF(\Sigma) = \CU$.
The map $\mathcal{T}: \TT_{p,p} \to \TT_{K,K}$ is a linear map defined using \eqref{eqn:c_aa} and \eqref{eqn:c_ab} such that $\mathcal{T}(\Sigma) = C$.

We decompose $\widehat A_J - A_J$ into a sum of linear and non-linear terms as
\begin{equation} \label{eqn:ajhat-aj}
    \widehat A_J^\top - A_J^\top = \widehat V^{-1}(\widehat V\widehat A_J^\top - VA_J^\top) + (\widehat V^{-1}-V^{-1}) V A_J^\top.
\end{equation}
The map $\cG: \TT_{p,p} \to \RR^{K \times K}$ is the sum of three terms:
\begin{equation}\label{eqn:entire-map}
    \cG(B) = V^{-1}\bigl[\mathcal{G}_1(B) + \mathcal{G}_2(B)- \mathcal{G}_3(B)A_J^\top\bigr]
\end{equation}
for any $B \in \TT_{p,p}$.
The maps $\mathcal{G}_1$ and $\mathcal{G}_2$ arise from the linear terms of $\widehat{V}\widehat{A}_J^\top - VA_J^\top$, and are defined as $[\mathcal{G}_1(B)]_{sj} = \sum_{k=1}^K \langle C_{ks},\, \cF(B)_{kj} \rangle_{\mathbb{T}_{1,1}},$ and  
$[\mathcal{G}_2(B)]_{sj} = \sum_{k=1}^K \langle \mathcal{T}(B)_{ks},\, \CU_{kj}\rangle_{\mathbb{T}_{1,1}}$ for any $s \in [K], j \in J$. 
To simplify the second term of \eqref{eqn:ajhat-aj}, we write 
$\widehat V^{-1}-V^{-1} = -\widehat V^{-1}(\wh V-V) V^{-1}.$
The map $\cG_3$ captures the linear terms of $\wh V-V$ and we refer the reader to Section~\ref{appx:proofs-asymp-normality} in the supplementary material for a precise definition.

\begin{theorem}\label{thm:asymp_aj} 
Assume that $\mathbb{E}\|X\|^4 < \infty$, the matrices $V$ and $\wh V$ are invertible, and $\widehat I=I$.
Let $\mathscr{S}: \mathbb{T}_{p,p} \to \mathbb{T}_{p,p}$ be the
covariance operator of $\mathfrak{Z}$, and let $\mathcal{G}$ be as defined in \eqref{eqn:entire-map}.
For fixed $p$ and $\nu$, as $n \to \infty$,
  \[
    \sqrt{n}(\widehat{A}_J^\top - A_J^\top)
    \xrightarrow{d} \mathcal{G}(\mathfrak{Z})
    \quad \text{in } \quad \mathbb{R}^{K\times |J|},
  \]
where $\cG(\mathfrak{Z})$ a zero-mean Gaussian random matrix with covariance
\begin{equation}\label{eqn:cov_full}
\mathrm{Cov}\bigl[\cG(\mathfrak{Z})_{s_1,j_1},\,\cG(\mathfrak{Z})_{s_2,j_2}\bigr]= [V^{-1}\,\mathscr{T}\, V^{-1}]_{s_1j_1,\,s_2j_2},\end{equation}
where $\mathscr{T} \in \mathbb{R}^{K|J| \times K|J|}$ is a symmetric positive semidefinite matrix. 
The product is defined as $[V^{-1}\,\mathscr{T}\, V^{-1}]_{s_1j_1,\,s_2j_2} = \sum_{k_1, k_2 = 1}^K V^{-1}_{s_1k_1}\mathscr{T}_{k_1j_1, k_2j_2} V^{-1}_{s_2k_2}$.
For $s_1,s_2 \in [K]$ and $j_1,j_2 \in J$, $\mathscr{T}_{s_1j_1,\,s_2j_2}$ is defined as
{\allowdisplaybreaks
\begin{align} 
    & \underbrace{\sum_{k_1,k_2 \in [K]}
       \langle (\cF\mathscr{S}\cF^*)_{k_1j_1, k_2j_2}\,
       C_{k_2s_2},\, C_{k_1s_1}
       \rangle}_{\text{(I): from }\mathcal{G}_1}
    \quad + \underbrace{\sum_{k_1,k_2 \in [K]}
       \langle (\mathcal{T}\mathscr{S}\mathcal{T}^*)_{k_1s_1,k_2s_2}\,
       \CU_{k_2j_2},\, \CU_{k_1j_1}
       \rangle}_{\text{(II): from }\mathcal{G}_2}
    \nonumber  \\
    & + \underbrace{\sum_{k_1,k_2 \in [K]}
       \Bigl(
       \langle (\cF\mathscr{S}\mathcal{T}^*)_{k_1j_1,k_2s_2}\,
       \CU_{k_2j_2},\, C_{k_1s_1}\rangle +
       \langle (\cF\mathscr{S}\mathcal{T}^*)_{k_1j_2,k_2s_1}\,
       \CU_{k_2j_1},\, C_{k_1s_2}\rangle
       \Bigr)}_{\text{(III): cross covariance }\mathcal{G}_1,\mathcal{G}_2}
    \nonumber \\
    & - \underbrace{2\sum_{s\in[K]}(A_J)_{j_2s}
       \sum_{k_1,k_2\in[K]}
       \Bigl(
       \langle(\cF\mathscr{S}\mathcal{T}^*)_{k_1j_1,k_2s_2}\,
       C_{k_2s},\,C_{k_1s_1}\rangle +
       \langle(\cF\mathscr{S}\mathcal{T}^*)_{k_1j_1,k_2s}\,
       C_{k_2s_2},\,C_{k_1s_1}\rangle
       \Bigr)}_{\text{(IV): cross term }\mathcal{G}_1,\mathcal{G}_3A_J^\top}
    \nonumber \\
    & - \underbrace{2\sum_{s\in[K]}(A_J)_{j_2s}
       \sum_{k_1,k_2\in[K]}
       \Bigl(
       \langle(\mathcal{T}\mathscr{S}\mathcal{T}^*)_{k_1s_1,k_2s_2}\,
       \CU_{k_2s},\,\CU_{k_1j_1}\rangle
       +
       \langle(\mathcal{T}\mathscr{S}\mathcal{T}^*)_{k_1s_1,k_2s}\,
       \CU_{k_2s_2},\,\CU_{k_1j_1}\rangle \Bigr)}_{\text{(V): cross term }\mathcal{G}_2,\mathcal{G}_3A_J^\top}
    \nonumber \\
    & + \underbrace{\sum_{s,s'\in[K]}(A_J)_{j_1s}(A_J)_{j_2s'}
       \sum_{k_1,k_2\in[K]}
       \Bigl(
       \langle(\mathcal{T}\mathscr{S}\mathcal{T}^*)_{k_1s_1,k_2s_2}\,
       C_{k_2s'},\,C_{k_1s}\rangle +\langle(\mathcal{T}\mathscr{S}\mathcal{T}^*)_{k_1s_1,k_2s'}\,
       C_{k_2s_2},\,C_{k_1s}\rangle}_{\text{(VI): from }\mathcal{G}_3A_J^\top}
    \nonumber \\
    &\qquad\qquad\qquad \qquad \qquad \quad \underbrace{
       +\, \langle(\mathcal{T}\mathscr{S}\mathcal{T}^*)_{k_1s,k_2s_2}\,
       C_{k_2s'},\,C_{k_1s_1}\rangle +\,\langle(\mathcal{T}\mathscr{S}\mathcal{T}^*)_{k_1s,k_2s'}\,
       C_{k_2s_2},\,C_{k_1s_1}\rangle
       \Bigr).}_{\text{(VI) cont.}}\nonumber
\end{align}}
In particular, 
$\widehat{A}_J \xrightarrow{p} A_J$.
\end{theorem}

\begin{remark}
    \begin{enumerate}[label=(\alph*)]
        \item The assumption $\wh I = I$ in Theorem~\ref{thm:asymp_aj} ensures that $\wh A_J$ and $A_J$ have the same dimension. 
        Recall from the discussion after Theorem~\ref{thm:pureindexguaran} that for fixed $p$ and $\nu$, there are no quasi-pure variables for large enough $n$.
        Then $\wh I=I$ eventually on the events $\cE(\delta_n)$ whose probability $\mathbb{P}[\cE(\delta_n)] \to 1$ as $n \to \infty$.
        Hence, the asymptotics of $\sqrt{n}(\wh A_J^\top - A_J)$ are the same as those of $\sqrt{n}(\wh A_J^\top - A_J)\mathbf{1}(\wh I = I)$.
        \item In the special case that the latent factors are uncorrelated, $C_{k_1k_2} = \mathbf{0}$ for $k_1 \neq k_2$.
        This simplifies some of the summations involving $C$ in the definition of $\mathscr{T}$.
        Setting $\HH = \RR$ recovers the asymptotics for the latent factor model in \cite{bing}.
        In this case, $\cF:\RR^{p \times p} \to \RR^{|J|\times K}$ and $\cT:\RR^{p \times p} \to \RR^{K \times K}$ are maps between real matrices and can be represented as 4-tensors.
        The inner products correspond to multiplication between numbers in $\RR$.
        If we additionally assume that $X$ is a zero-mean Gaussian, the fourth moment is $\mathscr{S}_{i_1j_1,i_2j_2}=\Sigma_{i_1i_2}\Sigma_{j_1j_2}+\Sigma_{i_1j_2}\Sigma_{j_1i_2}$ for any $i_1, i_2, j_1, j_2 \in [p]$ using Isserlis' Theorem.
        Then the elements of $V^{-1}\mathscr{T}V^{-1}$ can be computed efficiently.
    \end{enumerate}
\end{remark}

Theorem~\ref{thm:asymp_aj} specifies the covariance structure of the limiting normal random variable, hence it can be used to conduct inference on the overlapping cluster assignments in $A_J$.
In applications, the practical significance of the $k$th cluster may be inferred from the corresponding pure variables.
For any indices $j \in J, k \in [K]$, suppose we want check whether the $j$th variable lies in the $k$th cluster. 
Then we test $H_{0, jk}:A_{jk}=0$ versus $H_{1, jk}: A_{jk} \neq 0$.
We know that $\sqrt{n}(\widehat A_{jk} -A_{jk}) \to \cN(0, \sigma_{jk}^2)$, where $\sigma_{jk}^2 = [V^{-1}\mathscr{T} V^{-1}]_{kj, kj}$.
We obtain a $p$-value by using the plug-in estimators: $\frac{\widehat A_{jk}}{ \sqrt{\widehat \sigma^2_{jk}/n}}$ and compare it to the null distribution of $\cN(0,1)$.
The tests are only asymptotically valid and we have not considered a second-order correction based on the finite-sample estimator $\widehat \sigma^2_{jk}$. 

Sometimes, multiple hypothesis tests may be more useful.
In particular, if $A_{j_1k}$ and $A_{j_2k}$ are both found to be non-zero for a fixed $k$ and distinct $j_1, j_2 \in J$, then  $X_{j_1}$ and $X_{j_2}$ are correlated through their dependence on $Z_k$.
When such tests are conducted simultaneously over many indices $j \in J$ and $k \in [K]$, multiple hypothesis testing corrections should be applied.
We refer the reader to the vast literature on this topic.
In particular, covariance-aware methods such as those proposed by \cite{Fan2012} may be used since the dependence among the $p$-values under the null hypothesis is characterized by $V^{-1}\mathscr{T}V^{-1}$.


\section{Numerical Experiments} \label{sec:num-exp}

In this section, we validate the empirical performance of our method, funFMC, under various simulation settings.
We discuss computational details for discretely-observed function samples. 
We also describe a cross-validation procedure for selecting the tuning parameter $\delta$.
In Section~\ref{sec:eval-metric}, we introduce evaluation metrics and in Sections \ref{sec:non-overlap} and \ref{sec:overlap}, we present the accuracy of our method for identifying overlapping and non-overlapping clusters, respectively.
For non-overlapping clusters, we compare our method to a latent mixture model-based technique \citep{funHDDC-multivariate} and functional $k$-means \citep{horvath2012inference}.
We provide details for the competing methods in Section~\ref{sec:non-overlap}.
To the best of our knowledge, there are no existing methods that perform covariance-based overlapping clustering for multivariate functional data.

Recall that the proposed method relies on accurate estimation of the covariance operator $\Sigma$. 
Throughout this section, we assume $\HH = L^2[0,1]$.
Since infinite-dimensional functions are never observed fully, we assume that each univariate function sample is observed on $m$ equidistant points in $[0,1]$.
This corresponds to time series extracted from fMRI data, which is observed at regular time intervals.
Then $X^{(\ell)} \in \RR^{p \times m}$ and the $i$th function $X^{(\ell)}_i$ is observed on the grid $\{s/(m-1):0\le s\le m-1\}$ for all indices $1\le i\le p$ and all samples $1\le \ell \le n$.
We use the \texttt{scikit-fda} package \citep{scikit-fda} in python to utilize both the basis and the discrete representation of functions.
We represent each $X_i^{(\ell)}$ in terms of a univariate $B$-spline basis of order four with 10 basis functions. 
We approximate $X^{(\ell)}_i \approx \sum_{k=1}^{10}b^{(\ell)}_{i,k}\phi_{k}$ where $\{\phi_k\}_{k=1}^{10}$ are the basis functions and $b^{(\ell)}_{i,k}$ are the coefficients. 

To compute $\widehat \Sigma_{ij}$, we construct $D_i, D_j \in \RR^{n \times 10}$, which store the coefficients $b^{(\ell)}_{i,k}$ and $b^{(\ell)}_{j,k}$, respectively, for $1\le \ell \le n$ and $1\le k \le 10$.
Let $F_{ij} \in \RR^{10 \times 10}$ be the sample covariance matrix between $D_i$ and $D_j$.
Then $\widehat \Sigma_{ij} \approx \sum_{k_1=1}^{10} \sum_{k_2=1}^{10} [F_{ij}]_{k_1, k_2} (\phi_{k_1} \otimes \phi_{k_2}),$ where  $\{\phi_{k_1} \otimes \phi_{k_2}\}_{k_1,k_2=1}^{10}$ is the tensor product of $B$-splines.
Let $\widehat M_{ij} \in \RR^{m \times m}$ be the discrete representation of $\widehat \Sigma_{ij}$, computed by evaluating the tensor $B$-splines on the grid $\left\{\left(\frac{s}{(m-1)},\frac{t}{(m-1)}\right): 0\le s,t \le m-1\right\}$.
All covariance operators in $L^2[0,1]$ are integral operators, hence they are associated with a kernel.
This bi-variate, symmetric kernel function is called the covariance function.
Let $\psi_{ij}$ be the covariance function associated with $\widehat \Sigma_{ij}$.
Then the $(s,t)$th element of $\widehat M_{ij}$ approximates $\psi_{ij}\left(\frac{s}{(m-1)},\frac{t}{(m-1)}\right)$.
We compute $\|\Sigma_{ij}\| \approx \|\widehat M_{ij}\|_{op}/(m-1)$ and $\|\Sigma_{ij}\|_{\TT_{1,1}} \approx  \|\widehat M_{ij}\|_{F}/(m-1) $, where $\|\cdot\|_{op}$ and $\|\cdot\|_F$ denote the spectral norm and the Frobenius norm on matrices, respectively.
We provide motivation for the computational methods in Section~\ref{appx:num-methods} in the supplementary material.

Recall that Algorithm~\ref{algo:pure_set} involves selecting a tuning parameter $\delta$, which we choose using a cross-validation type method.
Specifically, inspired by Proposition~\ref{prop:event}, we select $\delta$ from the grid $\delta_q = c_q \sqrt{\log{p}/n}$ with varying constants $c_q$ using cross-validation.
For each cross-validation step, the data is split into two independent parts of equal size. 
Using the first part, we compute $\widehat \Sigma^1$.
The second part is used to compute $\widehat \Sigma^2$, which is input into Algorithm~\ref{algo:pure_set} to obtain $\widehat K(q)$ and $\widehat I(q)$. 
Then $\widehat A_{\widehat I(q)}$ is used to compute $\widehat C(q)$ using Algorithm~\ref{algo:2} and we set $\widehat \cV(q) = \widehat A_{\widehat I(q)} \widehat C(q) \widehat A_{\widehat I(q)}^\top$. 
We choose $\delta = \delta_{q^*}$, where
$$q^* = \argmin_q \frac{ \sqrt{\sum_{i_1, i_2 \in \widehat I(q), i_1\neq i_2} (\|\widehat \Sigma^1_{i_1i_2}\| - \|[\widehat V(q)]_{i_1i_2}\|)^2 }}{\sqrt{|\widehat I(q)|(|\widehat I(q)| -1 )}}.$$
In practice, the range of the constants $c_q$ is scaled by the first quantile of the set $\{\|\Sigma_{ij}\|\}_{i,j \in [p]}$.


\subsection{Evaluation Metrics} \label{sec:eval-metric}

To compare results across the different methods, we use the pairwise comparison approach described in \cite{pairwise}. 
Recall the set $G_a$ from \eqref{eqn:Ga} that consists of the indices $i$ such that $X_i$ is associated with $Z_a$. 
Let $\widehat G_a$ denote its estimated counterpart for all $1\le a \le K$.
For any pair $1\le i<j \le p$, define
\begin{align}\label{eqn:tp-ij}
& TP_{ij} = \mathbf{1}\left\{ i,j \in G_a \mathrm{~and~} i, j \in \widehat G_b \mathrm{~for~some~} a \in [K], b \in [\widehat K] \right\}, \nonumber \\
&TN_{ij} = \mathbf{1}\left\{ i,j \notin G_a \mathrm{~and~} i, j \notin \widehat G_b \mathrm{~for~all~} a \in [K], b \in [\widehat K] \right\},\nonumber \\
&FP_{ij} = \mathbf{1}\left\{ i,j \notin G_a \mathrm{~for~all~} a \in [K] \mathrm{~and~} i, j \in \widehat G_b   \mathrm{~for~some~} b \in [\widehat K] \right\},\nonumber \\
&FN_{ij} = \mathbf{1}\left\{ i,j \in G_a \mathrm{~for~some~} a \in [K] \mathrm{~and~} i, j \notin \widehat G_b \mathrm{~for~all~}  b \in [\widehat K] \right\}.
\end{align}
The overall true positives (TP), true negatives (TN), false positives (FP) and false negatives (FN) are calculated by summing over all pairs $i, j \in [p], i<j$. We then compute the specificity (SP), sensitivity (SN) and Rand index (RI) as follows:
$$SP = \frac{TN}{TN+FP},\qquad SN=\frac{TP}{TP+FN},\qquad RI = \frac{TN+TP}{TN+TP+FN+FP} .$$
Specificity penalizes false positives, sensitivity penalizes false negatives and Rand index penalizes all incorrect labels.
We report these metrics for both non-overlapping and overlapping clustering.
In the case of non-overlapping clustering, we threshold $\widehat A$ to ensure fair comparison with competing methods that perform hard clustering.
In the overlapping case, all methods return a membership matrix which is thresholded before determining $\widehat G_a$.
We provide details in the following sections.


\subsection{Non-Overlapping Clusters} \label{sec:non-overlap}

In this section, we present two scenarios with non-overlapping clusters.
We compare our method to the following approaches for clustering multivariate functional data: 
(i) latent mixture models in group-specific subspaces \citep{funHDDC, funHDDC-multivariate}, implemented using the \texttt{R} package \texttt{funHDDC}; and (ii) functional $k$-means clustering using the $L^2$ distance \citep{horvath2012inference}. The $L^2$ distance between any two samples $X^{(\ell_1)}$ and $X^{(\ell_2)}$ is defined as \begin{equation} \label{eqn:kmeans_dist}
    d(X^{(\ell_1)},X^{(\ell_2)}) = \sqrt{\sum_{i=1}^p \int_0^1 \left(X^{(\ell_1)}_i(t) - X^{(\ell_2)}_i(t)\right)^2 \,dt},
\end{equation}
which is computed using numerical integration. 
Both \texttt{funHDDC} and functional $k$-means assign non-overlapping labels to the variables. 
In contrast, our proposed method allows overlapping cluster memberships.
We apply a thresholding step to $\widehat A$: any element with an absolute value greater than $0.9$ is set to $1$ or $-1$, depending on its sign, while all remaining elements in the row are set to zero. 
Both competing methods require the true number of clusters $K$ to be entered as input.
In contrast, our method does not require $K$ to be known in advance.
The number of clusters is estimated by our method in a data-driven manner and we allow $\widehat K \neq K$.  

In Scenario (\RNum{1}), we generate data using Model \eqref{eqn:model}. 
We set $K=3$ and $p=100$ and use Matern kernels to specify $C$.
A Matern kernel with smoothness parameter $\nu$ and scale parameter $\alpha$ is defined as
\begin{equation}\label{eqn:matern}
    K(x,x') = \frac{2^{1-\nu}}{\Gamma(\nu)} \left(\frac{\sqrt{2\nu} |x-x'|}{\alpha} \right)^{\nu} K_{\nu}\left(\frac{\sqrt{2\nu} |x-x'|}{\alpha} \right),
\end{equation}
where $K_{\nu}$ is the Bessel function. 
The latent factors are drawn independently from zero-mean Gaussian processes with Matern kernels with scale parameters $5,1$ and $0.1$, respectively. 
The latent factors share a common $\nu = 1.5$.
We generate the loading matrix $A$ by randomly partitioning its rows into three clusters of sizes $34, 33$ and $33$. 
We first randomly select $34$ out of the $100$ rows and assign them to cluster one.
We randomly choose $33$ of the remaining rows and assign them to cluster two; the rest of the rows belong to cluster three.
If the row $i$ is assigned to cluster $k$, then we set $A_{ik}=1$ and $A_{ij}=0$ for all $j \neq k$.
Let $E^{(\ell)}_i \in \RR^m$ denote the discretized version of the $i$th noise function of the $\ell$th sample.
For each index $i \in [p]$, we draw $n$ independent samples from $\cN_m(\mathbf{0}, \sigma_{i}^2I_m)$, where $\sigma^2_{i}$ is drawn independently from a uniform distribution on $[1,3]$.
The noise samples are stored in $E_i^{(\ell)}$ for $1\le \ell \le n$.

In Scenario (\RNum{2}), each sample $X^{(\ell)}$ is generated so that $\{X^{(\ell)}_i : 1 \leq i \leq 25\}$ and $\{X^{(\ell)}_i : 26 \leq i \leq 50\}$ belong to two distinct clusters \citep{mult_func_clus_martino}.
These two collections, differing in both mean and covariance, are defined as follows:
\begin{align*}
    & X^{(\ell)}_i(t) = m_1(t) + \sum_{r=1}^R \xi^{(\ell)}_{i,r} \sqrt{\rho_{r}} \psi_{r}(t) \text{~~for~~} i = 1, \ldots, 25. \\
    & X^{(\ell)}_i(t) = m_2(t) + \sum_{r=1}^R \xi^{(\ell)}_{i,r} \sqrt{\rho_{r}} \psi_{r}(t) \text{~~for~~} i = 26, \ldots, 50.
\end{align*}
The functions $\psi_{r}(t)$ are the Fourier basis functions. 
The mean functions are $m_1(t) = t(1-t), m_2(t) =m_1(t) + \sum_{r=1}^3 \sqrt{\rho_{r}}\psi_{r}(t)$. 
We set $R = 100$. 
The scores $\xi^{(\ell)}_{i,r}$ are dependent across the index $i$, which ensures that unlike Scenario (I), the two clusters are dependent. 
For each $\ell \in [n], r \in [R]$, $\left\{\xi^{(\ell)}_{i,r}\right\}_{i=1}^p$ is drawn from a multivariate normal with mean zero and covariance matrix $C'$. 
We set $C'_{ii} = 1$ for all $i \in \{1, \ldots, 50\}$, $C'_{ij} = 0.8$ for $i,j \in \{1, \ldots 25\}, i \neq j$ and $i, j \in \{26, \ldots, 50\}, i\neq j$, and the rest of the entries of $C'$ are zero. 
Across samples $\ell$ and directions $r$, the scores are drawn independently. 
The deterministic sequence 
\begin{equation*}
  \rho_{r} =
    \begin{cases}
      1/(r+1) & \text{for~~} r \in \{1,2,3\}\\
      1/(r+1)^2 & \text{for~~} r \geq 4
    \end{cases}
\end{equation*}
ensures that the first three Fourier functions explain most of the variance. 

For both the scenarios, we vary $n$ and report clustering metrics averaged over $50$ simulations. The results for Scenarios (I) and (II) are presented in Table \ref{table:compare_matern} and Table \ref{table:compare_kmeans2}, respectively. 
In Scenario (\RNum{1}), funFMC performs the best, followed by \texttt{funHDDC} and $k$-means since the clusters differ in their covariance structures.
In Scenario (\RNum{2}), funFMC competes with $k$-means and \texttt{funHDDC}, and performs better as $n$ increases. 
In both scenarios, the accuracy of $k$-means does not necessarily increase with $n$ because it clusters $p$ multivariate objects each represented by an $n \times m $ matrix.
For a particular model specification, \texttt{funHDDC} diverged sometimes and did not produce a result.
We used four specifications ($a_{kj}b_kQ_kD_k$, $a_kb_kQ_kD_k$, $ab_kQ_kD_k$, $a_{kj}bQ_kD_k$) and for each simulation, out of the models that converged, selected the one with the lowest BIC.

\begin{table}[ht]
\centering
\caption{Clustering metrics for the three competing methods in Scenario (\RNum{1}). funFMC estimated the true number of clusters correctly in at least 90\% of simulations for each sample size.}
\label{table:compare_matern}
\sisetup{table-format=1.4}
\begin{tabular}{lccccccccc}
\toprule
&
\multicolumn{3}{c}{$n=50$} &
\multicolumn{3}{c}{$n=100$} &
\multicolumn{3}{c}{$n=250$} \\
\cmidrule(lr){2-4}
\cmidrule(lr){5-7}
\cmidrule(lr){8-10}
Method &
{SP} & {SN} & {RI} &
{SP} & {SN} & {RI} &
{SP} & {SN} & {RI} \\
\midrule
$k$-means
& \msd{0.7018}{0.0339}         
& \msd{0.9761}{0.0039} 
& \msd{0.7914}{0.0235}
& \msd{0.7145}{0.0312}
& \msd{0.9799}{0.0032} 
& \msd{0.8012}{0.0216}
& \msd{0.7003}{0.0302} 
& \msd{0.9821}{0.0016} 
& \msd{0.7923}{0.0208} \\

funHDDC
& \msd{0.9044}{0.0210} 
& \msd{0.9704}{0.0068} 
& \msd{0.9259}{0.0160}
& \msd{0.8777}{0.0219} 
& \msd{0.9605}{0.0076} 
& \msd{0.9047}{0.0169}
& \msd{0.9626}{0.0141} 
& \msd{0.9859}{0.0057} 
& \msd{0.9702}{0.0110} \\

\textbf{funFMC}
& \msd{0.9465}{0.0259} 
& \msd{1.0000}{0.0000} 
& \msd{0.9640}{0.0174}
& \msd{0.9336}{0.0284} 
& \msd{1.0000}{0.0000} 
& \msd{0.9553}{0.0191}
& \msd{1.0000}{0.0000} 
& \msd{1.0000}{0.0000} 
& \msd{1.0000}{0.0000} \\
\bottomrule
\end{tabular}
\end{table}

\begin{table}[ht]
\centering
\caption{Clustering metrics for the three competing methods in Scenario (II). funFMC estimated the true number of clusters correctly in at least 90\% of simulations for each sample size.}
\label{table:compare_kmeans2}
\sisetup{table-format=1.4}
\begin{tabular}{lccccccccc}
\toprule
&
\multicolumn{3}{c}{$n=50$} &
\multicolumn{3}{c}{$n=75$} &
\multicolumn{3}{c}{$n=100$} \\
\cmidrule(lr){2-4}
\cmidrule(lr){5-7}
\cmidrule(lr){8-10}
Method &
{SP} & {SN} & {RI} &
{SP} & {SN} & {RI} &
{SP} & {SN} & {RI} \\
\midrule
$k$-means
& \msd{1.0000}{0.0000} 
& \msd{1.0000}{0.0000} 
& \msd{1.0000}{0.0000}
& \msd{1.0000}{0.0000} 
& \msd{1.0000}{0.0000} 
& \msd{1.0000}{0.0000}
& \msd{0.9892}{0.0108} 
& \msd{0.9904}{0.0096} 
& \msd{0.9898}{0.0102} \\

funHDDC
& \msd{0.9960}{0.0040} 
& \msd{0.9967}{0.0033} 
& \msd{0.9963}{0.0037}
& \msd{1.0000}{0.0000} 
& \msd{1.0000}{0.0000} 
& \msd{1.0000}{0.0000}
& \msd{1.0000}{0.0000} 
& \msd{1.0000}{0.0000} 
& \msd{1.0000}{0.0000} \\

\textbf{funFMC}
& \msd{0.8977}{0.0403} 
& \msd{1.0000}{0.0000} 
& \msd{0.9478}{0.0206}
& \msd{0.9944}{0.0026} 
& \msd{1.0000}{0.0000} 
& \msd{0.9971}{0.0013}
& \msd{1.0000}{0.0000} 
& \msd{1.0000}{0.0000} 
& \msd{1.0000}{0.0000} \\
\bottomrule
\end{tabular}
\end{table}


\subsection{Overlapping Clusters} \label{sec:overlap}

In this section, we assess the accuracy of funFMC in a scenario with overlapping clusters. 
For overlapping clusters, we say that the function $X_i$ belongs to the cluster $k$ if $|\widehat  A_{ik}| > 0.10$.
With this modified notion for $i \in G_a$, the same metrics in Section~\ref{sec:eval-metric} are used for evaluation.

We generate data for this simulation using Model \eqref{eqn:model}. 
We set $K=3$ and draw the latent factors independently from zero-mean Gaussian processes with Matern covariance kernels, defined in \eqref{eqn:matern}. 
Similar to Scenario (I) in Section \ref{sec:non-overlap}, $\nu =1.5$ for all latent factors, but they differ in their scale parameters, which are set to $5,1$ and $0.1$. 
To specify $A_I$, we randomly select $30$ out of $p$ rows to correspond to pure variables.
Out of these $30$ rows, $10$ are randomly assigned to each cluster.
For a particular cluster, half of its pure variables have sign $+1$, while the other half have sign $-1$.
For each row $j \notin I$, we first choose the number of non-zero elements in the row as $\mathrm{card}_j$, randomly from $ \{2,3\}$. 
We then select $\mathrm{card}_j$-many column indices randomly and set those entries of the $j$th row to $\mathrm{sign}_j/ \mathrm{card}_j$, where $\mathrm{sign}_j$ is drawn uniformly from $\{-1,1\}$.
We present results averaged across $50$ simulations for $p\in \{60, 120\}$ and varying $n$ in Table \ref{table:overlap-matern}. 

\begin{table}[ht]
\centering
\caption{Clustering metrics of funFMC for overlapping clusters.}
\label{table:overlap-matern}
\sisetup{table-format=1.4}
\begin{tabular}{lccccccccc}
\toprule
&
\multicolumn{3}{c}{$n=50$} &
\multicolumn{3}{c}{$n=100$} &
\multicolumn{3}{c}{$n=250$} \\
\cmidrule(lr){2-4}
\cmidrule(lr){5-7}
\cmidrule(lr){8-10}
$p$ &
{SP} & {SN} & {RI} &
{SP} & {SN} & {RI} &
{SP} & {SN} & {RI} \\
\midrule

60
& \msd{0.9028}{0.0251}
& \msd{0.9155}{0.0147}
& \msd{0.9126}{0.0125}
& \msd{0.9834}{0.0117}
& \msd{0.9892}{0.0066}
& \msd{0.9875}{0.0061}
& \msd{0.9935}{0.0057}
& \msd{0.9842}{0.0082}
& \msd{0.9865}{0.0067}
\\

120
& \msd{0.8508}{0.0292} 
& \msd{0.9015}{0.0157}
& \msd{0.8963}{0.0152}
& \msd{0.9808}{0.0135} 
& \msd{0.9990}{0.0007} 
& \msd{0.9971}{0.0021}
& \msd{0.9928}{0.0072} 
& \msd{0.9987}{0.0013} 
& \msd{0.9981}{0.0019} \\
\bottomrule
\end{tabular}
\end{table}

The proposed method, funFMC, shows strong performance for $n=100$ and $n=250$.
For $n=50$, the estimation of $K$ is less reliable, which has an adverse effect on the evaluation metrics.
For the larger sample sizes, however, funFMC recovers the true number of clusters $K$ correctly in at least $90\%$ of the simulation replicates. 
As is common in soft clustering procedures, the results depend on the choice of threshold $\alpha$. 
Smaller values of $\alpha$ lead to an increase in false positives, whereas larger values result in a higher rate of false negatives, reflecting the tradeoff in membership assignment.


\subsection{Validation of the Limiting Covariance Structure of $A_J$} \label{sec:num-exp-aj}
In this section, we empirically validate the covariance structure of the limiting random variable of $\sqrt{n}(\widehat A_J^\top - A_J^\top)$, as described in Theorem~\ref{thm:asymp_aj}.
For different values of $n$, we perform Monte-Carlo simulations to estimate $\widehat A_J$ under the assumption that $A_I$ is known.
We plot the standard error of the $(s_1j_1, s_2j_2)$th element of $\sqrt{n}(\widehat A_J^\top - A_J^\top)$ against the theoretical value of covariance, given by $[V^{-1}\mathscr{T}V^{-1}]_{s_1j_1, s_2j_2}$ in Theorem~\ref{thm:asymp_aj}. 
We expect the points to fall on the identity line, especially as $n$ increases.
Since the computation of $\mathscr{S}$ and $\widehat A_J$ is expensive for the functional case, we set $\HH=\RR$.

We generate data using Model \eqref{eqn:model}.
Since $\HH=\RR$, $X\in \RR^p$ and $Z \in \RR^K$.
We set $K=5$ and draw latent factors from a zero-mean multivariate normal distribution with covariance matrix $C_{ab} = 0.5^{|a-b|}$ for all $a, b \in [K]$.
To specify $A_I$, we set $2$ pure variables for each cluster.
We set $p=60$ and randomly select $10$ rows to correspond to the pure variables.
For each index $i\in I_a$, $A_{ia}$ is randomly chosen to be $+1$ or $-1$. 
For each row of $A_J$, we first select the support $\text{card}_j$ randomly from $\{2,3,4,5\}$. 
Then we randomly select $\text{card}_j$-many columns and set those values to be $\text{sign}_j/\text{supp}_j$, where $\text{sign}_j$ is drawn uniformly from $\{-1, 1\}$.
The errors are homoskedastic and are drawn from a multivariate normal distribution with zero mean and variance $0.5 \times I_p$. 

\begin{figure}[ht]
    \centering
    \includegraphics[width=0.95\linewidth]{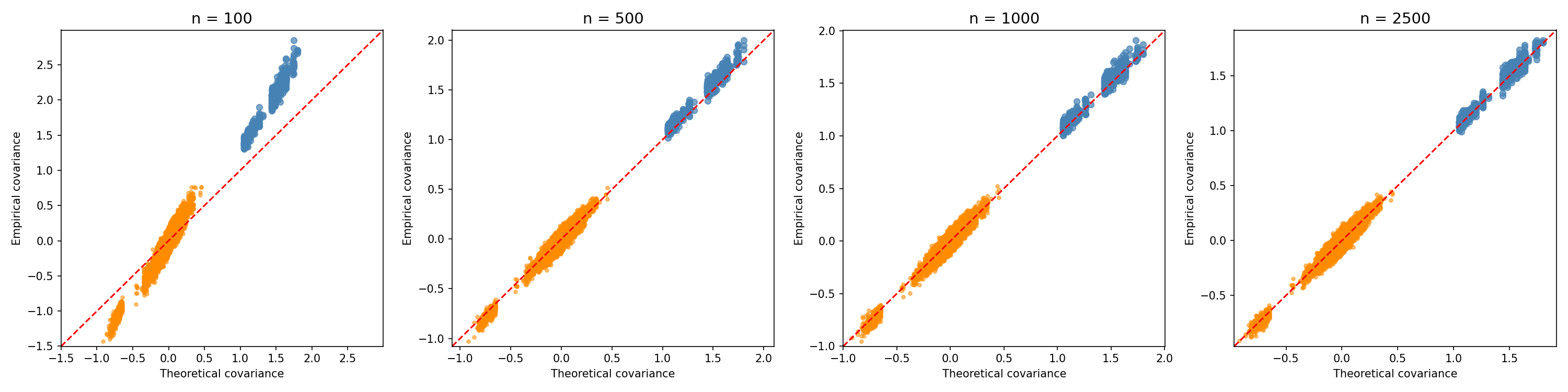}
    \caption{Empirical covariances (computed using Monte Carlo) versus theoretical covariances. Blue points represent variances and orange points represent covariances.}
    \label{fig:clt_verify1}
\end{figure}

The empirical versus theoretical covariances for varying values of $n$, using $2000$ Mote-Carlo replicates, are shown in Figure~\ref{fig:clt_verify1}.
As $n$ increases, the points follow the identity line more closely.
Additional diagnostic plots assessing the normality of the distribution of [$\sqrt{n}(\widehat A_J - A_J)]_{js}$ and the coverage of the $95\%$ confidence intervals can be found in Section~\ref{appx:clt_simulations}  in the supplementary material.


\section{fMRI Data Analysis} \label{sec:fMRI}
Functional magnetic resonance imaging is a neuroimaging technique that measures brain activity indirectly through changes in blood oxygenation level dependent (BOLD) signals.
For each subject, the data take the form of a multivariate time series, where each component corresponds to the BOLD signal recorded over a brain region (parcellation) over time. 
Understanding the similarity of these time series across parcellations is an important question, as parcellations with similar temporal dynamics may indicate related cognitive functions \citep{Yeo2011CortexConnectivity}. 
Prior work has shown that brain parcellations do not cleanly partition into disjoint groups \citep{LaBar1999, Karahanoglu2015, Castanho2022},
motivating methods that can identify overlapping clusters.

We analyze movie-watching fMRI, which has emerged as a popular paradigm for studying brain function under naturalistic conditions, as it may elicit richer and more ecologically valid neural responses than traditional task or resting-state designs \citep{vanderwal2017individual, Sonkusare2019Naturalistic, Kringelbach2023}. 
However, the majority of such studies have focused on datasets featuring narrative movie clips, such as the Human Connectome Project \citep{VanEssen2013}, where high-level cognitive processes are engaged.
The ID1000 dataset in the Amsterdam Open MRI Collection \citep{Snoek2021} is distinctive in this respect: the subjects watched naturalistic scenes from a movie that was designed to lack a narrative.
Subjects watched clips from the film \textit{Koyaanisqatsi}, consisting of $22$ natural scenes selected to vary multiple visual parameters, accompanied by a continuous musical score \citep{Snoek2021}. 
While this design is not intended for studying semantic processes, it is well-suited to studying low-level visual and attentional processing in a naturalistic setting. 
We apply funFMC to BOLD time series extracted from $n=878$ subjects, preprocessed using a standard pipeline comprising principal component analysis with $12$ components followed by confounding regression. 
Time series were extracted using the Schaefer atlas \citep{Schaefer2018} with Yeo $17$-network alignment \citep{Yeo2011CortexConnectivity}, yielding $p=200$ parcellations and $m=290$ time points per subject.

\begin{figure}[ht]
    \centering
    \includegraphics[width=0.9\linewidth]{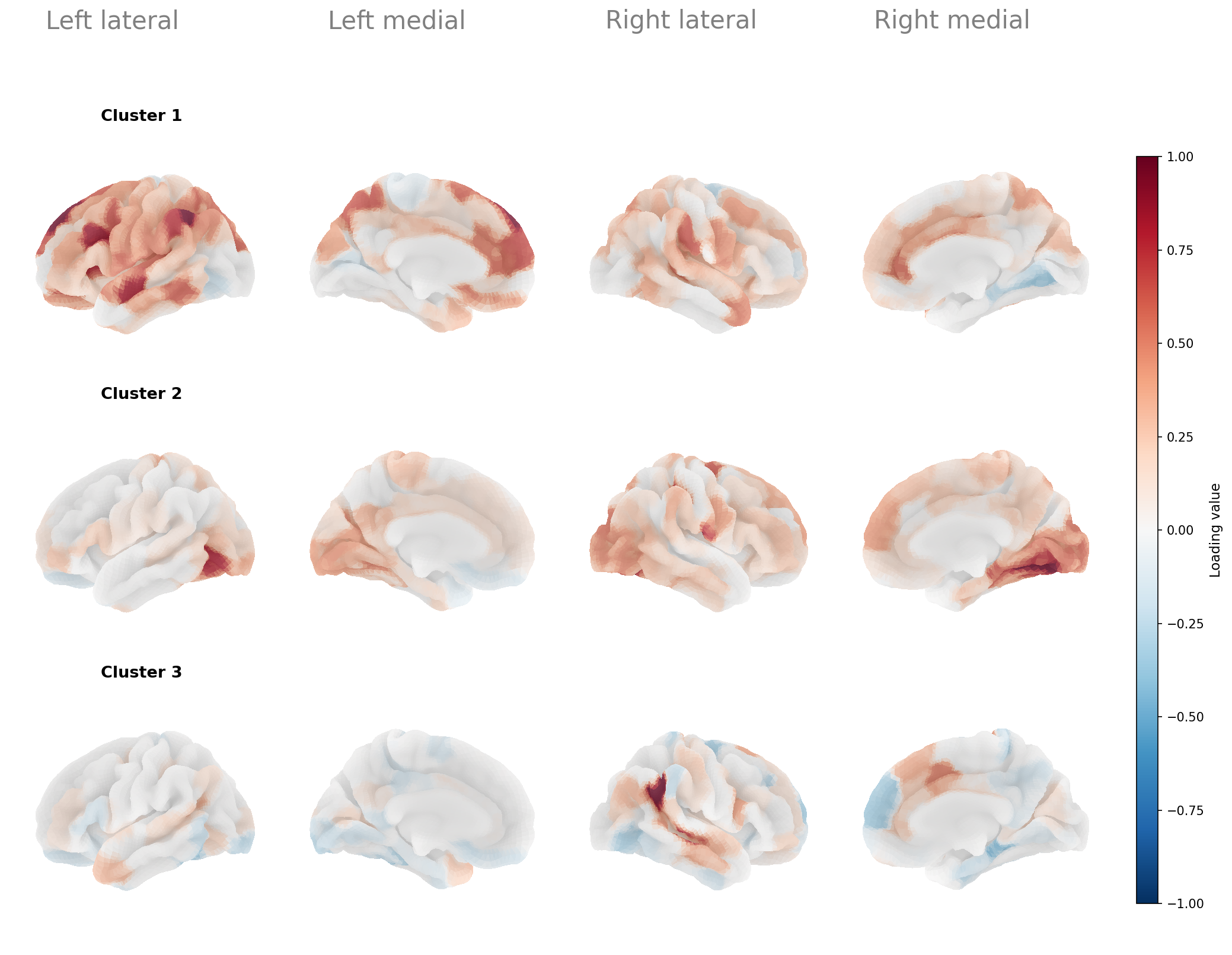}
    \caption{Dependence of parcellations on the estimated clusters}
    \label{fig:fmri_clus}
\end{figure}

\begin{figure}[ht]
    \centering
    \includegraphics[width=0.9\linewidth]{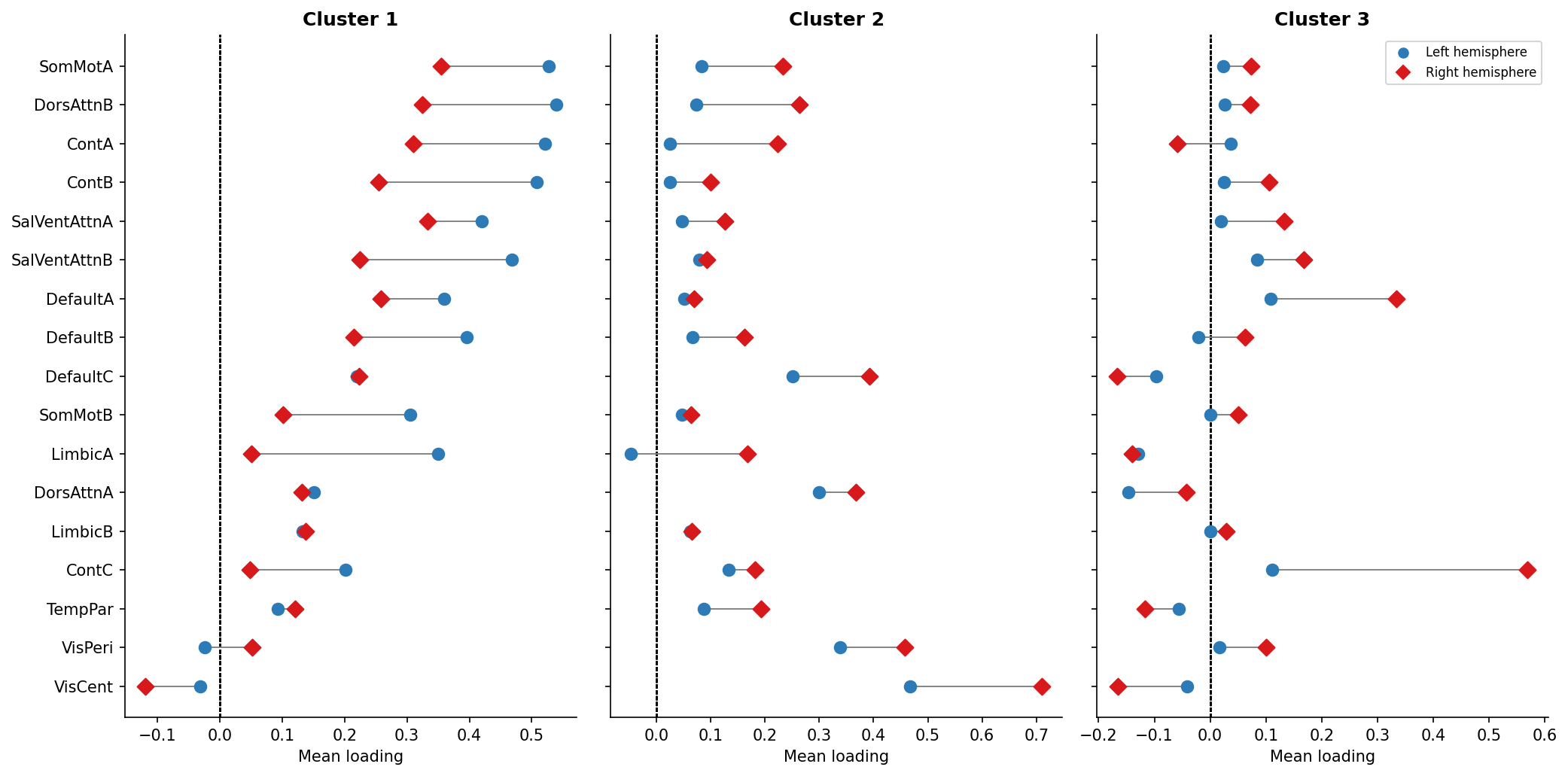}
    \caption{Mean cluster loadings by Yeo-17 network and hemisphere}
    \label{fig:mean_loadings}
\end{figure}

We apply funFMC to assess whether brain parcellations exhibit overlapping cluster structure, where a cluster is interpreted as a set of parcellations with similar BOLD signals. 
The hyperparameter $\delta$ was selected as the median over $10$ cross-validations, and funFMC estimates $\widehat K =3$ clusters.
Figure~\ref{fig:fmri_clus} displays the estimated loading matrix $\widehat A$, thresholded by setting all entries with absolute value below $0.10$ to zero, projected onto the cortical surface. Figure~\ref{fig:mean_loadings} reports the mean loading for each cluster across the $17$ canonical networks of \cite{Yeo2011CortexConnectivity}, separated by hemisphere.

The three clusters admit interpretable descriptions in terms of the task structure. 
Cluster $1$ shows broad loadings across multiple networks, including ContA, ContB, DefaultA, DefaultB, DorsAttnB, LimbicA, both SalVentAttn networks and both SomMot networks.
A continuously-changing stimulus with no narrative structure would be expected to engage sustained attention without a single dominant process. 
Cluster $2$ is strongly and selectively associated with the visual networks (VisCent and VisPeri), with negligible loadings elsewhere. 
This clean separation arises due to the presence of naturalistic visual stimulus and the modulation of visual parameters \citep{Lu2016BOLDMovie}.
Cluster $3$ is characterized by a strongly right-lateralized loading on the ContC network, with additional positive loadings on DefaultA. 
This pattern is consistent with the role of the right temporoparietal junction in attentional reorienting \citep{Corbetta2002, Corbetta2008}, which would be engaged by the unpredictable scene transitions in the stimulus.
We note that the auditory cortex, represented primarily in the SomMotB and TempPar networks, does not emerge as a distinct cluster in our analysis. 
This is possibly due to the design of the stimulus, in which the musical score by Philip Glass is continuous and largely invariant across scenes \citep{Snoek2021}, and would therefore produce auditory activation that is continuous over time and independent of the visual content.

The overlaps among the networks in Figure~\ref{fig:mean_loadings} suggest directions for further investigation.
For example, Cluster $2$ has moderate loadings on DefaultC and DorsAttnA, whose relation to the visual networks may be investigated further.
Cluster $3$ shows negative loadings on DefaultC and LimbicA.
This opposing direction of dependence with respect to ContC and DefaultA can be studied more closely.

\section{Discussion}
In this work, we proposed a novel method for identifying overlapping cluster structure in multivariate functional data using a latent factor model, where the loading matrix encodes the dependence of variables on latent clusters.  
By leveraging operator norms and Hilbert-Schmidt inner products, we established identifiability of the loading matrix (cluster memberships) up to signed permutations.  
We also developed theoretical guarantees for recovery of the number of clusters, characterized false positives in pure variable estimation, and established the asymptotic distribution of the loadings corresponding to non-pure variables.
 
The current framework assumes the existence of  two pure variables, which serve as observable anchors for recovering the latent structure. While similar assumptions are common in latent factor models, it would be of interest to investigate weaker conditions that allow identification. 
In addition, the model assumes uncorrelated errors and independent observations, which may be restrictive in settings with temporal and spatial dependence, respectively. 
Extending the proposed framework to accommodate more general dependence structures would broaden its scope and applicability.
Finally, our theoretical analysis assumes that the underlying functional observations are fully available, whereas in practice functions are observed on discrete grids and require smoothing or basis approximations. 
A more complete treatment of the additional bias introduced by discretization and function reconstruction would be  practically relevant.

\bibliography{references}
\newpage


\appendix
\renewcommand{\theequation}{S.\arabic{equation}}
\setcounter{equation}{0}
\renewcommand{\thefigure}{S.\arabic{figure}}
\setcounter{figure}{0}
\section*{Supplementary Material}

All accompanying code for the paper and this supplementary file can be found at \url{https://github.com/abhiti23/funFMC}.

The supplementary material in this document is arranged as follows: in Section~\ref{appx:model-proofs}, we present the proofs of Theorems~\ref{thm2:existence}--\ref{thm:op_to_matrix}, which deal with identifiability and estimation of the loading matrix $A$. 
In Section~\ref{appx:section4-proofs}, we present proofs for Theorem~\ref{thm:pureindexguaran} and Propositions~\ref{prop:event}, which establish statistical guarantees for the estimation of pure variables. 
Section~\ref{appx:proofs-asymp-normality} contains the proof of Theorem~\ref{thm:asymp_aj}, which states an asymptotic normality result for $\widehat A_J$. 
The proofs of all the Lemmas are written chronologically in Section~\ref{appx:proofs-of-lemmas}. 
We provide details for the numerical experiments in Section~\ref{appx:num-methods}.

\section{Proof of Theorems~\ref{thm2:existence}--\ref{thm:op_to_matrix}}
\label{appx:model-proofs}

\subsection{Proof of Theorem~\ref{thm:pure_var}}

We start with the proof of Theorem~\ref{thm:pure_var} since it facilitates that of Theorems~\ref{thm2:existence}. 
To do so, we first present two lemmas that are fundamental for establishing the identifiability of our model. 

\begin{lemma} \label{lem:1}
Under Model \eqref{eqn:model} and Assumptions \ref{assump1}--\ref{assump3}, for any latent factor $Z_{a}$ and $i \in I_{a}$, we have 
\begin{enumerate}[label=(\alph*)]
\item $\|\Sigma_{ij}\|  \le \|C_{aa}\|$ for all $j\not=i$, where the equality holds if and only if $j\in I_a\backslash\{i\}$.

\item $S_i \cup \{i\} = I_a$ and $M_i = \|C_{aa}\|$.
\end{enumerate}
\end{lemma}

This lemma shows that if $i \in I_a$, then the operator norms of the block operators in the $i$th row of $\Sigma$ are at most $\|C_{aa}\|$ (not considering the operator $\Sigma_{ii}$). 
The equality is attained if and only if $j \in I_a$.  
From \eqref{eqn:Si}, recall that $S_i$ consists of all column indices $j$ for which $\|\Sigma_{ij}\|$ is the maximum value in the $i$th row.
Then Lemma \ref{lem:1} allows us to compute $I_a = S_i \cup \{i\} $.

\begin{lemma} \label{lem:M_i}
    Under Model \eqref{eqn:model} and Assumptions \ref{assump1}--\ref{assump3}, $S_i \cap I \neq \varnothing$ for all $i \in [p]$.
\end{lemma} 

While Lemma \ref{lem:1} characterizes the set $S_i$ when $i \in I_a$, it is not applicable for indices $i \notin I$.
Lemma \ref{lem:M_i} derives a weaker condition stating that for any index $i\in [p]$, at least one index in $S_i$ corresponds to a pure variable. 
We need this condition in the proof below to establish a criterion for identifying pure variables.

We are now ready to prove Theorem \ref{thm:pure_var}. Recall that in part (a), a necessary and sufficient condition is given to identify pure variables, and 
in part (b), we state that this condition can be used to identify the entire set of pure variables $I$ and its partition $\cI$.

To prove part (a), we first show that for an index $i \in [p]$, if $M_i = M_j$ for all $j \in S_i$, then $i \in I$. 
From Lemma \ref{lem:M_i}, we know that at least one index in the set $S_i$ corresponds to a pure variable. 
Let this index be $j^* \in I \cap S_i$. Suppose $X_{j^*}$ is a pure variable for the latent factor $Z_a$, that is, $j^* \in I_a$. 
If we can show that $\|\Sigma_{j^*i}\| = \|C_{aa}\|$, then we can use Lemma \ref{lem:1} to conclude that $i \in I_a$. 
We prove this in the following paragraph.

Since $j^* \in I_a$, from Lemma \ref{lem:1} we know that the maximum operator norm in the $j^*$th row of $\Sigma$ is equal to $\|C_{aa}\|$. 
Since $M_{j^*} = \|C_{aa}\|$ and we have assumed that $M_i = M_{j^*}$, we obtain the equality $M_i = \|C_{aa}\|$. 
Since $j^* \in S_i$, the maximum $M_i$ is achieved for $\|\Sigma_{ij^*}\|$. Hence, $\|\Sigma_{ij^*}\| = \|C_{aa}\|$. 
Since $\Sigma_{j^*i}$ is the adjoint of the operator $\Sigma_{ij^*}$, we obtain $\|\Sigma_{j^*i}\| =\|\Sigma_{ij^*}\|= \|C_{aa}\|$. 
We conclude that $i \in I_a \subseteq I$. 

We now prove that if $i \in I$, then $M_i = M_j$ for all $j \in S_i$. 
Let $X_i$ be a pure variable corresponding to the latent factor $Z_a$, that is, $i \in I_a$. 
From Lemma \ref{lem:1}, we know that the maximum operator norm $M_i = \|C_{aa}\| $. 
We choose an index $j \in S_i$. 
Using the definition of $S_i$, the maximum is achieved at $\|\Sigma_{ij}\|  = \|C_{aa}\| $. 
Since $i$ is the index for a pure variable, Lemma \ref{lem:1} part (a) states that $j \in I_a$. 
Again, since $j \in I_a$, applying Lemma \ref{lem:1} to the $j$th index gives us that the maximum norm in the $j$th row is $M_j = \|C_{aa}\| $. 
This proves that $M_i = M_j$.

For the purpose of proving part (b) of Theorem \ref{thm:pure_var}, we present a procedure to construct $I$ and $\cI$ using the condition in part (a). 
For each variable $i \in [p]$, initialize the empty set $I^i$. 
Then compute $M_i= \max_{\ell \neq i}\|\Sigma_{i\ell}\|$ and define $S_i$ as the set of column indices at which this maximum is attained.
For each $j \in S_i$, compute $M_j = \max_{\ell \neq j}\|\Sigma_{j\ell}\|$.
If $M_j = M_i$ for every index $j \in S_i$, set $I^i = S_i ~\cup~\{i\}$; otherwise leave the set $I^i$ empty.
Once we have iterated through all $i \in [p]$, collect all sets $I^i$ into a family $\cI$. 
From Lemma~\ref{lem:1}, any two sets in $\cI$ are either disjoint or identical.
Remove any duplicates sets in $\cI$.
Set $K$ to be the number of distinct sets in $\cI$ and relabel the sets as $\cI = \{I^1, \ldots, I^K\}$.

Let $\tilde I$ be the set of pure indices obtained from the procedure described above. 
Since the labels are not known, there exists a permutation $\pi$ of $\{1, \ldots, K\}$ such that $\tilde I_a = I_{\pi(a)}$. 
Hence, $I$ and $\cI$ can be determined up to a permutation of the labels.


\subsection{Proof of Theorem~\ref{thm2:existence}}
Recall that Theorem \ref{thm2:existence} states that, under Assumptions \ref{assump1}--\ref{assump3}, we can identify the loading matrix $A$ and covariance matrix $C$ up to a signed permutation matrix based on knowledge of $\Sigma$.
This proof is structured as follows: (i) identifiability of $A_I$ up to signed permutation; 
(ii) construction of $A_J$ given $A_I$ and $\Sigma$; (iii) identifiability of $A_J$ up to signed permutation. 
\vskip.3cm
\textbf{Identifiability of $A_I$ up to signed permutation:} 
Theorem \ref{thm:pure_var} shows that, given $\Sigma$, we can identify the number of latent variables, $K$, and the partition of indices of pure variables, $\cI = \{I_1, \ldots, I_K\}$, up to a permutation of labels, namely, for another possible partition $\tilde \cI = \{\tilde I_1, \ldots, \tilde I_K\}$, we must have $\tilde I_a = I_{\pi(a)}$ for some permutation $\pi$ of $\{1, \ldots, K\}$. Thus, if $i\in I_a$ for some $a\in [K]$, then 
$$
|A_{i.}| = e_a \quad\mbox{and}\quad
|\tilde A_{i.}| = e_{\pi(a)},
$$
where $e_a$ is the canonical basis vector in $\RR^K$.
Consequently, $\tilde A_I = A_I Q$ where $Q$ is a signed permutation matrix, i.e., a square matrix that has exactly one nonzero entry in each row and each column, where each nonzero entry is either $+1$ or $-1$.

\textbf{Construction of $A_J$ given $A_I$ and $\Sigma$:}
Recall from Section~\ref{sec:identify-a} that the block structure of $\Sigma$ allows \eqref{eqn:cov_decomp} to be decomposed as 
\begin{equation} \label{eqn:appx-compos-main}
    \Sigma_{II} = A_ICA_I^\top + \Gamma_{II},\qquad \Sigma_{IJ} = A_ICA_J^\top,\qquad \Sigma_{JJ} = A_JCA_J^\top + \Gamma_{JJ}.
\end{equation}
By~\eqref{eqn:appx-compos-main}, it can be shown that $\Sigma_{i_1i_2} = A_{i_1a}A_{i_2a}C_{aa}$ for all $i_1,i_2 \in I_a$ such that $i_1 \neq i_2$.
Averaging over all such indices yields 
\begin{equation} \label{eqn:appx-c_aa}
    C_{aa} = \frac{1}{|I_a|(|I_a|-1)} \sum_{i_1,i_2 \in I_a, i_1 \neq i_2} A_{i_1a}A_{i_2a}\Sigma_{i_1i_2},
\end{equation}
since $|A_{ia}| =1 $ for all $i \in I_a$. Let $W =A_I^\top A_I$, which is a diagonal matrix with entries $W_{aa} = |I_a|$ for all $1\leq a \leq K$.
To estimate $C_{ab}$ for $a \neq b$, it can be shown by~\eqref{eqn:appx-compos-main} that
\begin{equation} \label{eqn:appx-I-rows}
    A_I^\top \Sigma_{II} A_I = WCW + A_I^\top \Gamma_{II} A_I.
\end{equation}
 When $a \neq b$, the columns of $A$ at indices $a$ and $b$ do not have any overlapping non-zero entries. 
 Since $\Gamma_{II}$ is a diagonal matrix, $(A_I^\top \Gamma_{II} A_I)_{ab}$ $ = 0$. 
 The $(a,b)$th element of the matrices in \eqref{eqn:appx-I-rows} yields the equality
$|I_a\|I_b|C_{ab} = \sum_{i_1,i_2 \in I} A_{i_1a}\Sigma_{i_1i_2}A_{i_2b}.$ 
The elements $\{A_{ia}:i \in I\}$ are non-zero if and only if $i \in I_a$. Therefore,
\begin{equation}\label{eqn:appx-c_ab}
    C_{ab} = \frac{1}{|I_a\|I_b|} \sum_{i_1 \in I_a,i_2 \in I_b} A_{i_1a}A_{i_2b}\Sigma_{i_1i_2}.
\end{equation}

Recall the expression for $\Sigma_{IJ}$ from \eqref{eqn:appx-compos-main} and pre-multiply both sides by $W^{-1}A_I^\top$ to derive
\begin{equation} \label{eqn:appx-aj-main} 
    W^{-1} A_I^\top \Sigma_{IJ} = C A_J^\top,
\end{equation}  
where all terms other than $A_J$ have been computed. 
Hence, \eqref{eqn:appx-aj-main} can be used to solve for $A_J$.

\textbf{Identifiability of $A_J$ up to signed permutation:}
We now show that if $\tilde A_J$ is the matrix constructed from $\tilde A_I$ using the procedure in (i), then $\tilde A_J = A_J Q$. 
We will show that if $\widetilde C$ is the block operator constructed using $\widetilde A_I$ and \eqref{eqn:appx-c_aa} and \eqref{eqn:appx-c_ab}, then $\widetilde C = Q^\top CQ$. 
Therefore, the pair $\widetilde A_I$ and $\widetilde C$ satisfy \eqref{eqn:appx-compos-main}.

Let $Q = DL$, where $D$ is a diagonal matrix with each entry $+1$ or $-1$, and $L$ corresponds to the permutation $\pi$. 
Each column of $A_I$ can be written as $\tilde A_{.k} = (AQ)_{.k} = d_k A_{.\pi(k)}$, where $d_k$ is the $k$th entry of the matrix $D$.
Using \eqref{eqn:appx-c_aa}, the diagonal elements of $\widetilde C$ are given by
    \begin{align} \label{eqn:appx-c-aa-tilde}
        \tilde C_{aa} &= \frac{1}{|\tilde I_a|(| \tilde I_a|-1)} \sum_{i_1,i_2 \in \tilde I_a, i_1 \neq i_2} \tilde A_{i_1a} \tilde A_{i_2a}\Sigma_{i_1i_2}.
    \end{align}
We know that $\tilde I_a = I_{\pi(a)}$ and for $i_1 \in \tilde I_a$, $\tilde A_{i_1a} = A_{i'_1 \pi(a)}$ for some $i'_1 \in I_{\pi(a)}$.
Hence, we can write \eqref{eqn:appx-c-aa-tilde} as 
    \begin{align} \label{eqn:appx-cpiapia}
    \begin{split}
        \tilde C_{aa} &= \frac{1}{| I_{\pi(a)}|(| I_{\pi(a)}|-1)} \sum_{i_1,i_2 \in I_{\pi(a)}, i_1 \neq i_2} (d_a A_{i_1 \pi(a)}) (d_a A_{i_2 \pi(a)})\Sigma_{i_1i_2} \\
        &= \frac{1}{| I_{\pi(a)}|(| I_{\pi(a)}|-1)} \sum_{i_1,i_2 \in I_{\pi(a)}, i_1 \neq i_2}  A_{i_1 \pi(a)} A_{i_2 \pi(a)}\Sigma_{i_1i_2} \\
        &= C_{\pi(a)\pi(a)},
        \end{split}
    \end{align}
where the second equality follows from the fact that $d_a \in \{-1, +1\}$ and the third equality follows from \eqref{eqn:appx-c_aa}.

Similarly, for the operator $\tilde C_{ab}$ where $a \neq b$, we use \eqref{eqn:appx-c_ab} to derive
    \begin{align}\label{eqn:appx-cpiapib}
    \begin{split}
        \tilde C_{ab} &= \frac{1}{|\tilde I_a\|\tilde I_b|} \sum_{i_1 \in \tilde I_a,i_2 \in \tilde I_b} \tilde A_{i_1a} \tilde A_{i_2b}\Sigma_{i_1i_2}  \\
        &= \frac{1}{|I_{\pi(a)}\|I_{\pi(b)}|} \sum_{i_1 \in I_{\pi(a)},i_2 \in I_{\pi(b)}} (d_a A_{i_1\pi(a)})(d_bA_{i_2\pi(b)})\Sigma_{i_1i_2}  \\
        &= d_ad_b \frac{1}{|I_{\pi(a)}\|I_{\pi(b)}|} \sum_{i_1 \in I_{\pi(a)},i_2 \in I_{\pi(b)}}  A_{i_1\pi(a)}A_{i_2\pi(b)}\Sigma_{i_1i_2} \\
        &= d_a d_b C_{\pi(a)\pi(b)}.
        \end{split}
    \end{align}
Since \eqref{eqn:appx-cpiapia} and \eqref{eqn:appx-cpiapib} hold for all $a, b \in [K]$ and $Q=DL$, where $L$ is the permutation matrix corresponding to $\pi$, we derive
    \begin{equation}
        \tilde C = Q^\top C Q.
    \end{equation}
Once $\tilde C$ is constructed, we compute $\tilde A_J$ using \eqref{eqn:appx-aj-main}
    $$\tilde C \tilde A_J^\top = \tilde W^{-1} \tilde A_I^\top \Sigma_{IJ},$$
where $\tilde W = \tilde A_I^\top \tilde A_I$. Using (ii), $\tilde W = Q^\top A_I^\top A_I Q = Q^\top W Q$. 
Then,
    \begin{align*}
        \tilde C \tilde A_J^\top &= (Q^\top W Q)^{-1}  \tilde A_I^\top \Sigma_{IJ} \\
         Q^\top CQ \tilde A_J^\top &= Q^\top W^{-1} Q Q^\top A_I^\top \Sigma_{IJ} \\
         CQ \tilde A_J^\top &= W^{-1} A_I^\top \Sigma_{IJ},
    \end{align*}
where the second and third equality use the fact that $Q^\top = Q^{-1}$. 
By Assumption~\ref{assump3}, the operator $C$ has independent columns and comparing with \eqref{eqn:appx-aj-main} leads to
$$\tilde A_J = A_JQ.$$
We have shown that the entire loading matrix $\tilde A = AQ$, where $Q$ is a signed permutation matrix.


\subsection{Proof of Theorem~\ref{thm:op_to_matrix}}
Recall from \eqref{eqn:optim-aj} that $A_J$ is derived as the solution of the following minimization problem in the Hilbert-Schmidt norm:
\begin{equation} \label{eqn:appx-optim-aj}
    A_J = \argmin_{Y \in \mathbb{R}^{|J| \times K} } \left\|W^{-1} A_I^\top \Sigma_{IJ} - CY^\top \right\|_{\TT_{|J|,K}}.
\end{equation}
In Theorem \ref{thm:op_to_matrix}, we use the inner product on $\TT_{1,1}$ to demonstrate that the solution to \eqref{eqn:appx-optim-aj} admits a representation involving only matrices with scalar entries. 

We first set up this problem abstractly. 
Let $\CU$ be an operator from $\HH^{|J|}$ to $\HH^K$. 
Without loss of generality, we can express $\CU$ as
$(\CU_{k,j}, k\in [K], j\in [|J|]),$
where each $\CU_{k,j}$ is an operator from $\HH$ to $\HH$.
Then $\CU$ acts on any element $(y_1,\ldots, y_{|J|})\in\HH^{|J|}$ as follows:
$$
\CU (y_1,\ldots,y_{|J|}) = \left(\sum_{j\in [|J|]} \CU_{k,j}y_j, k\in [K]\right).
$$
For the rest of the proof, we assume that $\CU \in \TT_{|J|,K}$. Consequently, $\CU_{k,j} \in \TT_{1,1}$ for all $1\le k \le K, 1\le j\le |J|$.

Let $\CCC=(\CCC_{k_1,k_2}, k_1,k_2\in [K]))  \in \TT_{K,K}$ and $Y$ be a real matrix lying in $\RR^{|J|\times K}$.
Let $\Psi$ be the mapping
$$\Psi: Y\mapsto \CCC Y^\top = \left(\sum_{s=1}^K Y_{j,s} \CCC_{k,s}, k \in [K], j\in [|J|]\right), \ \RR^{|J|\times K} \mapsto \TT_{|J|,K}.$$
For $\CU \in\TT_{|J|,K}$, the goal is to solve the following minimization problem:
\begin{align} \label{e:opt}
\argmin_{Y\in\RR^{|J|\times K}} \|\CU -\Psi Y\|_{\TT_{|J|,K}},
\end{align}
where $\|\cdot\|_{\TT_{|J|,K}}$ is the Hilbert-Schmidt norm on $\TT_{|J|,K}$. 
We note that this abstract problem corresponds to \eqref{eqn:appx-optim-aj} if we set $\CU = W^{-1}A_I^\top \Sigma_{IJ}$ and $\CCC = C$.

We now derive the solution to \eqref{e:opt}. 
Under the assumption that $\CU \in \mathrm{Dom}(\Psi^\dagger)$, the solution to \eqref{e:opt} is given by
$$ Y^{\mathrm{optim}} = (\Psi^*\Psi)^{\dagger} \Psi^* \CU,$$
where $M^{\dagger}$ denotes the Moore-Penrose inverse of $M$ (cf. Theorem 6.2.2 of \cite{fdabook}).
In the case that $\Psi^*\Psi$ is invertible, the Moore-Penrose inverse coincides with $(\Psi^*\Psi)^{-1}$.
We now derive an explicit expression for the adjoint map 
 $\Psi^* :\TT_{|J|,K} \mapsto \RR^{|J|\times K}$. 
By the definition of an adjoint operator, the following holds for any $Y\in\RR^{|J|\times K}$ and $ \CW\in\TT_{|J|,K}$:
\begin{equation} \label{e:adj}
    \langle \Psi Y, \CW\rangle_{\TT_{|J|,K}} = \langle Y, \Psi^*\CW\rangle_{\RR^{|J|\times K}}.
\end{equation}
The left hand side of this equation has the form
\begin{align*}
\langle \Psi Y, \CW\rangle_{\TT_{|J|,K}} 
&=  \sum_{j=1}^{|J|} \sum_{k=1}^K \left\langle \sum_{s=1}^K Y_{j,s} \CCC_{k,s}, \CW_{k,j}\right\rangle_{\TT_{1,1}} \\
&= \sum_{j=1}^{|J|} \sum_{s=1}^K Y_{j,s} \sum_{k=1}^K \langle \CCC_{k,s}, \CW_{k,j}\rangle_{\TT_{1,1}},
\end{align*}
where the first equality follows from the definition of the inner product on $\TT_{|J|,K}$, and the second equality uses the fact that $Y_{j,s} \in \RR$.
Comparing this to the right hand side of \eqref{e:adj} leads to \begin{align} \label{e:B*}
\Psi^*\CW = \left\{\sum_{k=1}^K \langle \CCC_{k,s}, \CW_{k,j}\rangle_{\TT_{1,1}}\right\}_{j \in [|J|], s \in [K]}.
\end{align}
Since $\CW$ was arbitrary, we have specified the map $\Psi^*$.

We now specify the map $\Psi^*\Psi: \RR^{|J| \times K} \mapsto \RR^{|J| \times K}$, which acts on $Y \in \RR^{|J| \times K}$ as follows:
\begin{align*}
\Psi^*\Psi Y & = \left\{\sum_{k=1}^K \left\langle \CCC_{k,s}, \sum_{s'=1}^K Y_{j,s'} \CCC_{k,s'}\right\rangle_{\TT_{1,1}}\right\}_{j \in [|J|], s \in [K]} \\
& = \left\{\sum_{s'=1}^K Y_{j,s'} \sum_{k=1}^K\left\langle \CCC_{k,s},  \CCC_{k,s'}\right\rangle_{\TT_{1,1}}\right\}_{j \in [|J|], s \in [K]} \\
& = \left\{\sum_{k=1}^K\left\langle \CCC_{k,s},  \CCC_{k,s'}\right\rangle_{\TT_{1,1}}\right\}_{s,s' \in [K]} Y^\top
\end{align*}
Consequently, by \eqref{e:B*},
\begin{align*}
 Y^{\mathrm{optim}} & = (\Psi^*\Psi)^{\dagger} \Psi^*\CU \\
&= \left\{\sum_{k=1}^K\left\langle \CCC_{k,s},  \CCC_{k,s'}\right\rangle_{\TT_{1,1}}\right\}_{s,s'\in [K]} ^{\dagger} \left\{\sum_{k=1}^K \langle \CCC_{k,s}, \CU_{k,j}\rangle_{\TT_{1,1}}\right\}_{s\in [K], j\in [|J|]}.
\end{align*}
Both the matrices in the above equation have elements in $\RR$.


\section{Proofs of Theorem~\ref{thm:pureindexguaran} and Proposition~\ref{prop:event}} \label{appx:section4-proofs}

In this section, we establish statistical guarantees for the estimation of pure variables.

\subsection{Proof of Proposition~\ref{prop:event}}
In this proof, we show that for suitable choices of $\delta$, the event $\cE(\delta)$ in \eqref{eqn:eventepsilon} holds with probability at least $1-p^2e^{-t}$.

\textbf{Part $(a)$:} 
We will use Theorem 9 in \cite{conc}, which we state now. 
Let $Y$ be a centered, subgaussian  random variable in $\HH$ with covariance operator $\Sigma_Y$. 
There exists a constant $c'>0$ such that, for all $t \ge 1$, with probability at least $1-e^{-t}$,
\begin{equation} \label{e:kol_lun_conc}
    \|\widehat \Sigma_Y -  \Sigma_Y \| \le c' \|\Sigma_Y\| \left( \sqrt{\frac{r(\Sigma_Y)}{n}},\frac{r(\Sigma_Y)}{n}, \sqrt{\frac{t}{n}}, \frac{t}{n}  \right).
\end{equation}

To derive a similar result for the covariance operator $\Sigma \in \HH^p$, we first consider the diagonal block operators $\Sigma_{ii}$.
For each index $i \in [p]$, we define the event
$$\mathcal{E}_{ii}(\delta_{ii})= \{ \|\widehat{\Sigma}_{ii} - \Sigma_{ii}\| \leq \delta_{ii} \}.$$
A direct application of \eqref{e:kol_lun_conc} shows that there exists a constant $c_{ii}$ such that for all $t\ge 1$, 
\begin{equation}\label{e:epsilon-ii}
    \PP[\mathcal{E}_{ii}(\delta_{ii})] \geq 1 - e^{-t},
\end{equation}
where $\delta_{ii} = c_{ii} \|\Sigma_{ii}\| \left( \sqrt{\frac{r(\Sigma_{ii})}{n}} \bigvee \frac{r(\Sigma_{ii})}{n} \bigvee \sqrt{ \frac{t}{n}} \bigvee \frac{t}{n}\right)$. 
This holds true for all $i \in [p]$, but the constant $c_{ii}$ depends on $i$.

Next, we consider the off-diagonal block operators of $\Sigma$.
Pick any two indices $i,j \in [p]$ such that $i \neq j$. 
We consider the random variable $(X_i, X_j) \in \mathbb{H}^2$ with covariance operator $\Psi_{ij} \in \cL(\HH^2)$. 
Let $\widehat \Psi_{ij}$ denote its sample covariance operator.
Both operators have a block operator structure as follows:
$$\Psi_{ij} = \begin{bmatrix}
    \Sigma_{ii} & \Sigma_{ij} \\
    \Sigma_{ji} & \Sigma_{jj}
\end{bmatrix}, \qquad\hat{\Psi}_{ij} = \begin{bmatrix}
    \hat{\Sigma}_{ii} & \hat{\Sigma}_{ij} \\
    \hat{\Sigma}_{ji} & \hat{\Sigma}_{jj}
\end{bmatrix}.$$
We exploit this structure to derive a bound for $\|\hat{\Sigma}_{ij} - \Sigma_{ij}\|$ in terms of $r(\Psi_{ij})$.
Let $I, \mathbf{0} \in\cL(\HH)$ denote the identity operator and the zero operator, respectively.
Then
\begin{align*}
    \|\widehat{\Sigma}_{ij} - \Sigma_{ij}\| &= \left\| \begin{bmatrix}
        I & \mathbf{0}
    \end{bmatrix} \widehat{\Psi}_{ij} \begin{bmatrix}
        \mathbf{0} \\ I
    \end{bmatrix} - \begin{bmatrix}
        I & \mathbf{0}
    \end{bmatrix} \Psi_{ij} \begin{bmatrix}
        \mathbf{0} \\ I
    \end{bmatrix} \right\| \\
    &= \left\| \begin{bmatrix}
        I & \mathbf{0}
    \end{bmatrix}( \widehat{\Psi}_{ij} - \Psi_{ij}) \begin{bmatrix}
        \mathbf{0} \\ I
    \end{bmatrix} \right\| \\
    &\leq \left\|\begin{bmatrix}
        I & \mathbf{0}
    \end{bmatrix} \right\| \left\| \widehat{\Psi}_{ij} - \Psi_{ij} \right\| \left\|\begin{bmatrix}
        \mathbf{0} \\ I
    \end{bmatrix} \right\|, 
\end{align*}
where the inequality uses the sub-multiplicative property of the operator norm. 
The block operators $\begin{bmatrix}
        I & \mathbf{0}
    \end{bmatrix}$ and $\begin{bmatrix}
        \mathbf{0} & I
    \end{bmatrix}^\top$ both embed elements of $\HH$ into the space $\HH^2$ since  one of their components is $\mathbf{0}$. 
Hence, their operator norm is at most $1$. 
The computations above simplify to
    $$\|\widehat{\Sigma}_{ij} - \Sigma_{ij}\| \leq \|\hat{\Psi}_{ij} - \Psi_{ij}\|.$$
We define the event
$$\cE_{ij}(\delta_{ij}) = \{ \|\widehat{\Sigma}_{ij} - \Sigma_{ij}\| \leq \delta_{ij} \}.$$
An application of \eqref{e:kol_lun_conc} shows that there exists a constant $c_{ij}$ such that for all $t\ge 1$, 
\begin{equation}\label{e:epsilon-ij}
    \PP[\mathcal{E}_{ij}(\delta_{ij})] \geq 1 - e^{-t},
\end{equation}
where  
$\delta_{ij} = c_{ij} \|\Psi_{ij}\| \left( \sqrt{\frac{r(\Psi_{ij})}{n}} \bigvee \frac{r(\Psi_{ij})}{n} \bigvee \sqrt{ \frac{t}{n}} \bigvee \frac{t}{n} \right)$.

For event $\cE(\delta)$ in \eqref{eqn:eventepsilon} to occur, all events $\cE_{ij}(\delta_{ij})$ where $i,j \in [p]$, must occur simultaneously.
Define
$$\delta_a = c_0 \max_{1 \leq i, j \leq p} \|\Xi_{ij}\| \left( \sqrt{\frac{r(\Xi_{ij})}{n}} \bigvee \frac{r(\Xi_{ij})}{n} \bigvee \sqrt{ \frac{t}{n}} \bigvee \frac{t}{n} \right),$$
where $c_0 = \max_{i, j \in [p]}c_{ij}$ and
\begin{equation*}
\Xi_{ij} = \begin{cases} 
\Sigma_{ii}
    & \text{if } i=j \\
\Psi_{ij} 
    & \text{if } i\neq j
\end{cases}.
\end{equation*}
Then
\begin{align*}
    \PP\left[{ \bigcap_{1\leq i,j \leq p}\cE_{ij}(\delta_a)}\right] &\geq 1-\sum_{1\leq i,j \leq p}(1- \PP[\cE_{ij} (\delta_a)])  \\
    &=\sum_{1\leq i,j \leq p} \PP[\cE_{ij}(\delta_a)] - (p^2 - 1)  \\
    &\geq \sum_{1\leq i,j \leq p} (1 - e^{-t})- (p^2 - 1)  \\
    &= 1 - \sum_{1\leq i,j \leq p}e^{-t} \\
    &= 1 - p^2 e^{-t},
\end{align*}
where the first step uses a basic bound on the probability of intersection of event.
The third step uses the fact that for all $i,j \in [p]$, $\delta_a\ge \delta_{ij}$, hence the event $\cE_{ij}(\delta_a) \supseteq \cE_{ij}(\delta_{ij})$.

\textbf{Part $(b)$:} 
We now express $r(\Psi_{ij})$ and $\|\Psi_{ij}\|$ in terms of the corresponding quantities for $\Sigma_{ii}$ and $\Sigma_{jj}$.
We first establish an upper bound on $r(\Psi_{ij})$.
Since $\|(X_i, X_j)\|_{\HH^2} = \sqrt{\|X_{i}\|_{\HH}^2 + \|X_{j}\|_{\HH}^2} \le \|X_i\|_{\HH} + \|X_j\|_{\HH}$ by the Cauchy-Schwarz inequality, we derive
\begin{align}\label{eqn:r-exp-upper-bound}
    (\EE\|(X_i, X_j)\|_{\HH^2})^2 &\le (\EE \|X_i\|_{\HH} + \EE \|X_j\|_{\HH})^2 \nonumber \\
    &= (\EE\|X_i\|_{\HH})^2 + (\EE\|X_j\|_{\HH})^2 + 2\EE\|X_i\|_{\HH}\EE\|X_j\|_{\HH}\nonumber \\
    &\le 2((\EE\|X_i\|_{\HH})^2 + (\EE\|X_j\|_{\HH})^2 ),
\end{align}
where the last step uses the fact that $2ab \le a^2 +b^2$ for any real numbers $a$ and $b$.

We now derive a lower bound on $\|\Psi_{ij}\|$. 
We use the Cauchy-Interlacing Theorem (Poincare Separation Theorem), which is usually stated for symmetric matrices, but also holds for covariance operators.
Let $B \in \RR^{b \times b}$ be a symmetric matrix. 
Any matrix $B' \in \RR^{b' \times b'}$, where $b' \leq b$, is called a compression of $B$ if there exists an orthogonal projection $P$ onto a subspace of dimension $b'$ such that $PBP^\top = B'$. 
In particular, any sub-matrix of $B$ is a compression. 
The eigenvalues of $B$ are denoted by $\lambda_1(B) \le \ldots \le \lambda_b(B)$, and those of $B'$ are denoted by $\lambda_1(B') \le \ldots \le\lambda_{b'}(B')$.
Then the Cauchy Interlacing Theorem states that for all $j \leq  b'$, $\lambda_j(B) \leq \lambda_j(B') \leq  \lambda_{b-b'+j}(B).$
Let $\lambda_{\mathrm{max}}$ denote the largest eigenvalue.
Using  $j = b'$ results in $\lambda_{\mathrm{max}}(B') \le \lambda_{\max}(B).$

Let $B = \Psi_{ij}$ and $B' = \Sigma_{ii}$.
Then by the Cauchy-Interlacing Theorem,
\begin{equation} \label{e:cauchy-interlacing}
\lambda_{\mathrm{max}}(\Sigma_{ii}) \le \lambda_{\max}(\Psi_{ij}).\end{equation}
The operator norm of a self-adjoint, positive semi-definite operator can be expressed in terms of its largest eigenvalue.
Let $\Sigma \in \cL(\HH)$ be any self-adjoint, positive semi-definite operator.
Then
\begin{align*}
    \|\Sigma\| = \sup_{\|y\|_{\HH}=1} \|\Sigma y\|_{\HH}
    = \sup_{\|y\|_{\HH}=1} \langle \Sigma y,  \Sigma y\rangle_{\HH}^{1/2}
    = \sup_{\|y\|_{\HH}=1} \langle \Sigma^* \Sigma y, y\rangle_{\HH}^{1/2} 
    = \sqrt{\lambda_{\max} (\Sigma^* \Sigma)}
    = \lambda_{\max} (\Sigma),
\end{align*}
where the third step uses the definition of an adjoint operator.
The fourth step follows from the self-adjoint nature of $\Sigma$ and the last step uses the self-adjoint, positive nature of $\Sigma$.
Then using \eqref{e:cauchy-interlacing}, $\|\Sigma_{ii}\|\leq \|\Psi_{ij}\|.$
A similar inequality holds for $\|\Sigma_{jj}\|$. 
Hence, $\|\Psi_{ij}\| \geq \|\Sigma_{ii}\| \vee \|\Sigma_{jj}\|.$
We derive
\begin{align*}
    r(\Psi_{ij}) &= \frac{(\EE\|(X_i, X_j)\|_{\HH^2})^2}{\|\Psi_{ij}\|} \\
    &\le \frac{2((\EE\|X_i\|_{\HH})^2 + (\EE\|X_j\|_{\HH})^2 )}{\|\Sigma_{ii}\| \vee \|\Sigma_{jj}\|} \\
    &\le 2 \left(\frac{(\EE\|X_i\|_{\HH})^2}{\|\Sigma_{ii}\| } + \frac{(\EE\|X_j\|_{\HH})^2}{\|\Sigma_{jj}\| } \right) \\
    &= 2(r(\Sigma_{ii})+r(\Sigma_{jj})).
\end{align*}
Here, the second step uses \eqref{eqn:r-exp-upper-bound} and the lower bound for $\|\Psi_{ij}\|$. 

Based on the expression for $\delta_a$, we now derive an upper bound for $\|\Psi_{ij}\|$. 
Since $\Psi_{ij} = \begin{bmatrix}
    \Sigma_{ii} & \Sigma_{ij}\\ \Sigma_{ji} & \Sigma_{jj}
\end{bmatrix}$ is a positive semi-definite block matrix, 
$\|\Psi_{ij}\| \leq \|\Sigma_{ii}\|+\|\Sigma_{jj}\|$ using Proposition 8.3 in \cite{siam_op_norm_inequality}.

We define
\begin{equation*} 
    \delta_b = c_0 \max_{1 \leq i , j \leq p} (\|\Sigma_{ii}\| + \|\Sigma_{jj}\| ) \left( \sqrt{\frac{2 ( r(\Sigma_{ii})+ r(\Sigma_{jj}))}{n}} \bigvee \frac{2 ( r(\Sigma_{ii})+ r(\Sigma_{jj}))}{n} \bigvee \sqrt{ \frac{t}{n}} \bigvee \frac{t}{n} \right),
\end{equation*}
where $c_0 = \max_{i,j \in [p]}c_{ij}$. 
Let $\delta_{ij} = c_{ij} \|\Xi_{ij}\| \left( \sqrt{\frac{r(\Xi_{ij})}{n}} \bigvee \frac{r(\Xi_{ij})}{n} \bigvee \sqrt{ \frac{t}{n}} \bigvee \frac{t}{n} \right)$ for all $i, j \in [p]$.
The event $\cE_{ij}(\delta_b) \supseteq \cE_{ij}(\delta_{ij})$, and consequently,
$$\PP\left[{ \bigcap_{1\leq i,j \leq p}\cE_{ij}(\delta_b)}\right] \ge \PP\left[{ \bigcap_{1\leq i,j \leq p}\cE_{ij}(\delta_{ij})}\right].$$
The probability bounds derived in part $(a)$ showed that $\PP[\cap_{1\leq i,j \leq p}\cE_{ij}(\delta_{ij})] \ge 1-p^2e^{-t}$.
Therefore, $\PP[\cE(\delta_b)] \ge 1-p^2e^{-t}$.


\subsection{Proof of Theorem~\ref{thm:pureindexguaran}: Parts $(a)$ and $(b)$}
Recall that the maximum operator norm in the $i$th row is denoted by $\widehat M_i = \max_{j \in [p] \backslash \{i\}} \left\|\widehat{\Sigma}_{ij}\right\|$. 
The set $\widehat S_i = \left\{ j\in [p] \backslash \{i\} : \widehat M_i- \left\|\widehat{\Sigma}_{ij}\right\|   \leq 2\delta \right\}$ captures the column indices at which the operator norm is within $2\delta$ distance of $\widehat M_i$. 
Before proving Theorem~\ref{thm:pureindexguaran}, we state Lemma~\ref{lem:sigma_hat_norms}, which characterizes the behavior of the operator norms  of $\widehat \Sigma$.

\begin{lemma}\label{lem:sigma_hat_norms}
    Under the same assumptions as in Theorem~\ref{thm:pureindexguaran}, the following inequalities hold on the event $\cE(\delta)$ for all $i\in I_a$:
    \begin{enumerate}[label=(\alph*)]
        \item $\left| \|\widehat \Sigma_{ij_1}\| - \|\widehat \Sigma_{ij_2}\| \right| \le 2\delta$ for all $j_1, j_2 \in I_a \backslash \{i\}$ and $j_1\neq j_2$.
        \item $\|\widehat \Sigma_{ij_1}\| - \|\widehat \Sigma_{ij_2}\| < 2\delta$ for all $j_1\not\in I_a$ and $j_2 \in I_a \backslash \{i\}$. 
        \item $\|\widehat \Sigma_{ij_2}\| - \|\widehat \Sigma_{ij_1}\| > 2\delta$ for all $j_1 \notin (I_a \cup J_1^a)$ and $j_2 \in I_a \backslash \{i\}$. 
        \item
        $\widehat M_j -  \|\widehat \Sigma_{ij}\| \le 2\delta$ for all $j \in J_1^a$. 
    \end{enumerate}
\end{lemma}

Lemma~\ref{lem:sigma_hat_norms} describes how the results of Lemma~\ref{lem:1} change due to sampling error. 
Part (a) characterizes the behavior of pure variables.
From part (b) and (c), we infer that when $j_1$ is a quasi-pure variable and $j_2$ is a pure variable, then $-2\delta < \|\widehat \Sigma_{ij_1}\| - \|\widehat \Sigma_{ij_2}\| < 2\delta$.
In particular the second inequality shows that pure variables and quasi-pure variables share similar properties. 
Part (c) shows this property is not shared by non-pure variables that are not quasi-pure.
Part (d) indicates that quasi-pure variables may be selected as pure variables in Algorithm~\ref{algo:pure_set}. 
We defer the proof of Lemma~\ref{lem:sigma_hat_norms} to Section~\ref{appx:proofs-of-lemmas}.

We now proceed with the proof of Theorem \ref{thm:pureindexguaran}. 
Recall that in Algorithm~\ref{algo:pure_set}, $X_i$ is selected as a pure variable if 
\begin{equation} \label{eqn:appx-pure-var-algo1}
    \left| \widehat M_j - \left\|\widehat{\Sigma}_{ij}\right\|  \right|  \leq 2 \delta
\end{equation}
for all $j \in \widehat S_i$.
We also construct the potential pure variable set $\widehat I^i = \widehat S_i ~\cup~ \{i\}$ in this case.
However, if $\widehat I^{i} \cap \widehat I^{j} \neq \varnothing$ for some index $j \in [p]$, then we replace $\widehat I^{i}$ and $\widehat I^{j}$ by $\widehat I^{i} \cap \widehat I^{j}$.
We will show that (i) if $i \in I_a$, then $X_i$ is selected as a pure variable and $I_a \subseteq \widehat{I}^{i} \subseteq I_a \cup J_1^a$;
(ii) if $i \in J_1^a$, then $I_a \subseteq \widehat{I}^{i}$;
(iii) and lastly, if $i \in J \backslash J_1$, then $X_i$ is not selected as a pure variable. 

\textbf{If $i \in I_a$, then $X_i$ is selected as a pure variable and $I_a \subseteq \widehat{I}^{i} \subseteq I_a \cup J_1^a$:} This part of the proof shows that if the $i$th index corresponds to a pure variable, then it is selected as a pure variable in Algorithm~\ref{algo:pure_set}. 
The potential pure variable set $\widehat I^i$ contains all of $I_a$, and may contain indices in $J_1^a$.

Let $i \in I_a$.
Lemma~\ref{lem:sigma_hat_norms} characterizes the operator norms in the $i$th row of $\widehat \Sigma$.
From part (c) of Lemma~\ref{lem:sigma_hat_norms}, we infer that $\widehat M_i$ cannot be attained for $\|\widehat \Sigma_{ij_1}\|$ where $j_1 \notin (I_a \cup J_1^a)$.
Hence, $\widehat S_i \subseteq (I_a \cup J_1^a)$.
The maximum $\widehat M_i$ can be attained at $\|\Sigma_{ij}\|$ for either $j \in I_a$ or $j \in J_1^a$.
If $j \in I_a$, then part (a) of Lemma~\ref{lem:sigma_hat_norms} ensures that $I_a \subseteq \widehat S_i$.
Otherwise, part (b) of Lemma~\ref{lem:sigma_hat_norms} ensures that $I_a \subseteq \widehat S_i$.
Therefore, we have shown that $I_a \subseteq \widehat S_i \subseteq (I_a \cup J_1^a)$.

We now show that \eqref{eqn:appx-pure-var-algo1} holds for all $j \in \widehat S_i$.
First, we consider $j \in I_a$.
By applying part (c) of Lemma~\ref{lem:sigma_hat_norms} to the norms in the $j$th row of $\widehat \Sigma$, we conclude that $\widehat M_j$ cannot be attained at $\|\Sigma_{jj_1}\|$ where $j_1 \notin (I_1 \cup J_1^a)$.
If $\widehat M_j = \|\widehat \Sigma_{jj_1}\|$ for $j_1 \in I_a$, then part (a) of Lemma~\ref{lem:sigma_hat_norms} shows that $\left|\widehat M_j - \|\widehat \Sigma_{ji}\| \right| \le 2\delta$.
Then using the fact that $\|\widehat \Sigma_{ji}\| = \|\widehat \Sigma_{ij}\|$, we see that \eqref{eqn:appx-pure-var-algo1} is satisfied.
If  $\widehat M_j = \|\widehat \Sigma_{jj_1}\|$ for $j_1 \in J_1^a$, then part (b) shows that \eqref{eqn:appx-pure-var-algo1} is satisfied.
Next, we consider $j \in J_1^a$.
As shown in Lemma~\ref{lem:sigma_hat_norms}, $\widehat M_j - \|\widehat\Sigma_{ij}\| \le 2\delta$ since we assumed $i \in I_a$.
Hence, \eqref{eqn:appx-pure-var-algo1} is attained for all $j \in \widehat S_i$ since $\widehat S_i$ is a subset of $(I_a \cup J_1^a)$.
This proves that $X_i$ is selected as a pure variable and $I_a \subseteq \widehat{I}^{i} \subseteq I_a \cup J_1^a$ since $\widehat{I}^{i} = \widehat S_i \cup \{i\}$.

\textbf{If $i \in J_1^a$, then $I_a \subseteq \widehat{I}^{i}$:}
This claim immediately follows from Lemma~\ref{lem:sigma_hat_norms}. 
As noted earlier, a quasi-pure variable $X_i$ may be selected as a pure variable.
Hence, it is essential that $\widehat I^i$ contains the entire set $I_a$, otherwise some indices in $I_a$ will be ousted from the pure index set due to the intersection performed in the second step of Algorithm~\ref{algo:pure_set}.

\textbf{If $i \in J \backslash J_1$, then $X_i$ is not selected as a pure variable:}
This part of the proof guarantees that the only non-pure variables estimated as pure variables are the quasi-pure variables.
Let $i \in J \backslash J_1$. 
We will show that there exists an index $j \in \widehat S_i$ such that condition \eqref{eqn:appx-pure-var-algo1} does not hold, hence $X_i$ is not selected as a pure variable.
Recall from Lemma~\ref{lem:M_i} that when there is no sampling error, $S_i \cap I \neq \varnothing$.
We will show that $\widehat S_i \cap I \neq \varnothing$ and for any index $j \in \widehat S_i \cap I$, the condition \eqref{eqn:appx-pure-var-algo1} is violated.

As shown in Lemma~\ref{lem:M_i}, there exists an index $j \in S_i \cap I$.
Then $M_i = \|\Sigma_{ij}\|$ since $j \in S_i$.
On the event $\cE(\delta)$, $M_i -\delta \le \|\widehat \Sigma_{ij} \| \le M_i +\delta$ and $M_i-\delta\le \widehat M_i \le M_i +\delta$.
Combining these inequalities gives us $|\widehat M_i - \|\widehat \Sigma_{ij}\||\le 2\delta$. 
Therefore, $j \in \widehat S_i$.

We now prove that \eqref{eqn:appx-pure-var-algo1} is violated for the index $j \in \widehat S_i \cap I$.
Let $j \in I_b$ for some $b \in [K]$.
Then $\|\Sigma_{ij}\| = \left\|\sum_{k_1 = 1}^K A_{ik_1}C_{k_1b} \right\|$ and using the triangle inequality, $\|\Sigma_{ij}\| \le |A_{ib}|\|C_{bb}\|+\sum_{k_1\neq b} |A_{ik_1}| \|C_{k_1b}\|$.
Since $\nu$ is the minimum separation between norms, $\|C_{k_1b}\| < \|C_{bb}\| - \nu$ for all $k_1 \neq b$.
Moreover, from Assumption~\ref{assump1}, $\sum_{k_1\neq b} |A_{ik_1}| \le 1-|A_{ib}|$.
Then
\begin{align*}
    \|\Sigma_{ij}\| &< |A_{ib}|\|C_{bb}\| + (1-|A_{ib}|)(\|C_{bb}\| - \nu ) \\
    &= \|C_{bb}\| -\nu(1-|A_{ib}|)\\
    &< \|C_{bb}\| - 4\delta, 
\end{align*}
where in the last inequality, we use $|A_{ib}| < 1-4\delta/\nu$ since $i \in J\backslash J_1$.
Moreover, on the event $\cE(\delta)$, $\|\widehat \Sigma_{ij}\| \le \|\Sigma_{ij}\| +\delta< \|C_{bb}\| - 3\delta$.
As discussed earlier, $\widehat M_j \ge M_j -\delta$, and $M_j = \|C_{bb}\|$ since we assumed that $j \in I_b$ (Lemma~\ref{lem:1}).
Then
$$\widehat M_j - \|\Sigma_{ij}\| > 2\delta.$$
Thus, $X_i$ is not selected as a pure variable in Algorithm~\ref{algo:pure_set}.


\subsection{Proof of Theorem~\ref{thm:pureindexguaran}: Part $(c)$}
Throughout this proof, we will work on the event $\cE(\delta)$.
Theorem~\ref{thm:pureindexguaran} parts $(a)$ and $(b)$ show that there exists a permutation $\pi$ of the cluster labels such that $I_{\pi(a)} \subseteq \widehat I_a$.
Recall the procedure for construction of $\widehat A_I$ given $\wh \cI$ in Section~\ref{sec:identify-a}.
In the absence of quasi-pure variables, the procedure aligns $\wh A_I$ with $A_I$ and the required matrix $Q= I$.
This is not true when quasi-pure variables may be estimated as pure variables and we work in this general scenario.

Recall that we used $\wh A_I$ to denote the matrix $\wh A_{\wh I}$.
In this proof, we use $[\wh A]_I$ to denote the rows of $\wh A$ corresponding to the true pure variable set $I$.
Let $P$ be the permutation matrix associated with $\pi$ in Theorem~\ref{thm:pureindexguaran} part $(b)$.
We have shown that $|[\widehat A]_I| = |A_{I}P|$, where $|B|$ denotes the matrix composed of the absolute values of the elements of $B$.
Moreover, for any $i \in \wh I_a \backslash I_a$, $|[AP]_{ia}| > 1-4\delta/\nu$.
This means that $P$ aligns the columns of $\widehat A_I$.
To prove part $(c)$, we have to show that there exists a diagonal matrix $D$ with each entry in the set $\{-1,+1\}$ such that $[\widehat A]_I = A_{I}PD$ and for any $i \in \wh I_a \backslash I_a$, the sign of $\wh A_{ia}$ and $[APD]_{ia}$ are the same.
Then $Q=PD$ will be the required signed permutation matrix.

We first derive a condition equivalent to the existence of $Q$.
Let $\tilde{A}_I=A_{\hat I}P$, which implies that the sign of $\tilde{A}_{ia}$ will be equal to the sign of $A_{i\pi(a)}$ for all $i \in \widehat I_a$.
Let the diagonal elements of $D$ be denoted by $d_a, \ldots, d_K$.
We have to show that $\mathrm{sign}(\hat A_{ia}) = \mathrm{sign}( \tilde{A}_{i\pi(a)})d_a = \mathrm{sign}( A_{i\pi(a)})d_a$ for all $i \in \widehat I_a$ and $a \in K$.
Since $\hat I_a \subseteq I_{\pi(a)} \cup J_1^{\pi(a)}$, we will show the following equivalent condition:
\begin{equation}\label{e:sign-align}
    \frac{\mathrm{sign}(\hat A_{ia})}{\mathrm{sign}( A_{i\pi(a)})} = \frac{\mathrm{sign}(\hat A_{ja})}{\mathrm{sign}( A_{j\pi(a)})}    
\end{equation}
for all pairs $i,j \in I_{\pi(a)} \cup J_1^{\pi(a)}$.
This condition is more amenable because in Section~\ref{sec:est-sample}, we first estimate $I$ using Algorithm~\ref{algo:pure_set}, and then  we designated signs by comparing pairs of indices.
Recall that for each pure set $\widehat I_a$, we chose the smallest index $i_0 \in \hat{I}_a$ and set $\hat{A}_{i_0a}=1$. 
For the remaining indices $j \in \hat{I}_a \backslash \{i_0\}$, we set $A_{ja} = 1$ if
$\left\|\hat\Sigma_{i_0i_0}+\hat\Sigma_{i_0j} \right\| > \left\|\hat\Sigma_{i_0i_0}-\hat\Sigma_{i_0j}\right\|$ and $-1$ otherwise.
This condition is agnostic to the choice of the reference index $i_0$.
In particular, for any $i, j \in \widehat I_a$,  $\mathrm{sign}(\hat A_{i a}) = \mathrm{sign}(\hat A_{j a})$ if and only if 
\begin{equation}\label{e:sign-condition}
    \|\hat \Sigma_{ii} + \hat \Sigma_{ij}\| > \|\hat \Sigma_{ii} - \hat \Sigma_{ij}\|.
\end{equation}
From \eqref{e:sign-align}, it is clear that we have to prove that if $\text{sign}(A_{i\pi(a)})=\text{sign}(A_{j\pi(a)})$, then $\left\|\hat\Sigma_{ii}+\hat\Sigma_{ij} \right\|$ $ > \left\|\hat\Sigma_{ii}-\hat\Sigma_{ij}\right\|$ and if $\text{sign}(A_{i\pi(a)}) \neq \text{sign}(A_{j\pi(a)})$, then $\left\|\hat\Sigma_{ii}+\hat\Sigma_{ij} \right\| \le \left\|\hat\Sigma_{ii}-\hat\Sigma_{ij}\right\|$ for all $i,j\in \widehat I_a$ and $a\in [K]$.

We prove this in three steps: (i) \eqref{e:sign-condition} for $\Sigma$ and pure variables; (ii) \eqref{e:sign-condition} for $\Sigma$ and quasi-pure variables; (iii) \eqref{e:sign-condition} for $\widehat \Sigma$.

\textbf{Condition \eqref{e:sign-condition} for $\Sigma$ and pure variables:}
Since we are working on the event $\cE(\delta)$, where $\widehat \Sigma_{ij}$ is an accurate estimator of $\Sigma_{ij}$ for all $i, j\in[p]$, we will begin by proving \eqref{e:sign-condition} for block operators of $\Sigma$. 
We will consider both pure and quasi-pure variables since $\widehat I_a \subseteq I_{\pi(a)} \cup J_1^{\pi(a)}$.
Lastly, in the third part of this proof, we will incorporate the sampling error.

Firstly, we assume that both $i, j \in I_a$. 
Recall from \eqref{eqn:cov_decomp} that in this case, $\Sigma_{ii} = C_{\pi(a)\pi(a)} + \Gamma_{ii}$ and $\Sigma_{ij} = A_{i\pi(a)}A_{j\pi(a)}C_{\pi(a)\pi(a)}$. 
If $\mathrm{sign}(A_{i\pi(a)}) = \mathrm{sign}(A_{j\pi(a)})$, then $\Sigma_{ii} - \Sigma_{ij} = \Gamma_{ii}$ and $\Sigma_{ii} + \Sigma_{ij} = 2C_{\pi(a)\pi(a)} + \Gamma_{ii}$. 
Therefore,
$\|\Sigma_{ii} + \Sigma_{ij}\| - \|\Sigma_{ii} - \Sigma_{ij}\| = \|2C_{aa} + \Gamma_{ii}\| - \|\Gamma_{ii}\|  > 0,$ where the inequality holds since both $C_{aa}$ and $\Gamma_{ii}$ are self-adjoint and non-negative definite operators.
Similarly, if $\mathrm{sign}(A_{i \pi(a)}) \neq \mathrm{sign}(A_{j \pi(a)})$, then $\Sigma_{ii} + \Sigma_{ij} = \Gamma_{ii}$ and $\Sigma_{ii} - \Sigma_{ij} = 2C_{aa} + \Gamma_{ii}$ and $\|\Sigma_{ii} + \Sigma_{ij}\| - \|\Sigma_{ii} - \Sigma_{ij}\| < 0$.
    
These computations indicate that criterion \eqref{e:sign-condition} satisfies a `transitive property'. 
For this we will show that if $\|\Sigma_{ii} + \Sigma_{ij}\| > \|\Sigma_{ii} - \Sigma_{ij}\|$ and $\|\Sigma_{ii} + \Sigma_{ik}\| > \|\Sigma_{ii} - \Sigma_{ik}\|$, then $\|\Sigma_{jj} + \Sigma_{jk}\| > \|\Sigma_{jj} - \Sigma_{jk}\|$.
Since $A_{ia}A_{ja}=1$ and $A_{ia}A_{ka}=1$, we also have $A_{ja}A_{ka}=1$.
This implies  $\|\Sigma_{jj} + \Sigma_{jk}\| > \|\Sigma_{jj} - \Sigma_{jk}\|$.

\textbf{Condition \eqref{e:sign-condition} for $\Sigma$ and quasi-pure variables:}
Since $\widehat I_a$ may contain quasi-pure variables, we will consider them as well.
Assume that $i, j \in I_a \cup J_1^a$.
We express $\Sigma_{ii}$ and $\Sigma_{ij}$ using \eqref{eqn:cov_decomp} as
\begin{align}\label{e:sigmas}
    \Sigma_{ii} &= C_{\pi(a)\pi(a)} + 2A_{i\pi(a)}\sum_{b \neq \pi(a)} A_{ib}C_{\pi(a)b} + \sum_{b,c \neq \pi(a)} A_{ib}A_{ic}C_{bc} + \Gamma_{ii},\nonumber  \\ 
    \Sigma_{ij}&= C_{\pi(a)\pi(a)} +  A_{j \pi(a)} \sum_{b \neq \pi(a)}A_{ib}C_{\pi(a)b} + A_{i\pi(a)}\sum_{b \neq a}A_{jb}C_{\pi(a)b} + \sum_{b,c \neq \pi(a)}A_{ib}A_{jc}C_{bc}.
\end{align}
In the case that $\mathrm{sign}(A_{i\pi(a)}) = \mathrm{sign}(A_{j\pi(a)})$, 
\begin{align}\label{e:two-op-terms}
\|\Sigma_{ii}+\Sigma_{ij}\| 
&= \left\| 2C_{\pi(a)\pi(a)} + A_{i\pi(a)} \sum_{b \neq \pi(a)}(3A_{ib}+A_{jb}) C_{\pi(a)b} + \sum_{b,c \neq \pi(a)}(A_{ic}+A_{jc})A_{ib}C_{bc} +\Gamma_{ii} \right\|, \nonumber \\
 \|\Sigma_{ii} - \Sigma_{ij}\|&= \left\| A_{i\pi(a)}\sum_{b \neq \pi(a)} (A_{ib}-A_{jb})C_{\pi(a)b} + \sum_{b,c \neq \pi(a)}(A_{ic} -A_{jc})A_{ib}C_{bc}+\Gamma_{ii} \right\|.
\end{align}
Using the triangle inequality,
{\allowdisplaybreaks
\begin{align*}
    \|\Sigma_{ii}+\Sigma_{ij}\| &\ge \| 2C_{\pi(a)\pi(a)} + \Gamma_{ii}\|  -\left \| A_{i\pi(a)} \sum_{b \neq \pi(a)}(3A_{ib}+A_{jb}) C_{\pi(a)b}\right\| -\left\|\sum_{b,c \neq \pi(a)}(A_{ic} -A_{jc})A_{ib}C_{bc}\right\|,\\
    \|\Sigma_{ii}-\Sigma_{ij}\| &\le \|\Gamma_{ii}\| + \left\|A_{i\pi(a)}\sum_{b \neq \pi(a)} (A_{ib}-A_{jb})C_{\pi(a)b} \right\|+\left\| \sum_{b,c \neq \pi(a)}(A_{ic} -A_{jc})A_{ib}C_{bc} \right\|.
\end{align*}}
Hence, the difference $\|\Sigma_{ii}+\Sigma_{ij}\| -  \|\Sigma_{ii} - \Sigma_{ij}\|$ has the following lower bound:
\begin{align}
     &\ge \| 2C_{\pi(a)\pi(a)} + \Gamma_{ii}\| -\|\Gamma_{ii}\| \label{e:gamma_term} \\
    &\quad - \left \| A_{i\pi(a)} \sum_{b \neq \pi(a)}(3A_{ib}+A_{jb}) C_{\pi(a)b}\right\| -\left\|A_{i\pi(a)}\sum_{b \neq \pi(a)} (A_{ib}-A_{jb})C_{\pi(a)b} \right\| \label{e:cpiab_term} \\
    &\quad -\left\|\sum_{b,c \neq \pi(a)}(A_{ic} -A_{jc})A_{ib}C_{bc}\right\| -\left\| \sum_{b,c \neq \pi(a)}(A_{ic} -A_{jc})A_{ib}C_{bc} \right\| \label{e:cbc_term}.
\end{align}
Another application of the triangle inequality yields 
\begin{align*}
    \eqref{e:gamma_term} &\ge 2\|C_{\pi(a)\pi(a)}\|-2\|\Gamma_{ii}\|;\\
    \eqref{e:cpiab_term} &\ge -4 \left\|A_{i\pi(a)}\sum_{b \neq \pi(a)} A_{ib}C_{\pi(a)b} \right\|- 2 \left\|A_{i\pi(a)}\sum_{b \neq \pi(a)} A_{jb}C_{\pi(a)b} \right\|; \\
    \eqref{e:cbc_term} &\ge -2 \left\|\sum_{b,c \neq \pi(a)}A_{ic}A_{ib}C_{bc} \right\|- 2 \left\|\sum_{b,c \neq \pi(a)}A_{jc}A_{ib}C_{bc} \right\|.
\end{align*}

In order to further simplify the lower bound on \eqref{e:cpiab_term}, we observe that
\begin{align*}
    \left\| A_{i\pi(a)}\sum_{b \neq \pi(a)} A_{jb}C_{\pi(a)b} \right\| &\leq |A_{i\pi(a)}|\sum_{b \neq \pi(a)} |A_{jb}| \big\|C_{\pi(a)b}\big\| \\
    &\leq |A_{i\pi(a)}| \left(\big\|C_{\pi(a)\pi(a)} \big\| -\nu \right) \sum_{b \neq \pi(a)} |A_{jb}|\\
    &\leq |A_{i\pi(a)}| \left(1-|A_{j\pi(a)}| \right) \left(\big\|C_{\pi(a)\pi(a)} \big\| -\nu \right) \\
    &\leq \frac{4\delta}{\nu} \left(\big\|C_{\pi(a)\pi(a)} \big\| -\nu \right).
\end{align*}
The first step follows from the triangle inequality and the second step uses the fact that $\nu = \Delta(C)$ is the minimum separation between covariance operators, as described in Assumption~\ref{assump3}.
The third step uses the fact that row sums of the loading matrix are at most one (Assumption~\ref{assump1}).
The last step follows because $|A_{i\pi(a)}| \leq 1$ for any index $i$, and $|A_{j \pi(a)}| \geq 1-\frac{4\delta}{\nu}$ since we had assumed that $j \in I_a \cup J_1^a$ is a pure or a quasi-pure variable. 
A similar argument shows that the other term in \eqref{e:cpiab_term} is also bounded as $\left\|A_{i\pi(a)}\sum_{b \neq \pi(a)} A_{ib}C_{\pi(a)b} \right\| \leq \frac{4\delta}{\nu} \left(\big\|C_{\pi(a)\pi(a)} \big\| -\nu \right)$ since $i \in I_a\cup J_1^a$.

We further simplify the bound on \eqref{e:cbc_term} by first applying the triangle inequality as follows:
$$\left\| \sum_{b,c \neq \pi(a)}A_{ib}A_{jc}C_{bc} \right\| \leq \sum_{b,c \neq \pi(a)}|A_{ib}||A_{jc}|\| C_{bc}\|.$$
Since $i \in I_a\cup J_1^a$, $|A_{i\pi(a)}| \geq 1-\frac{4\delta}{\nu}$, and using Assumption~\ref{assump1}, $\sum_{b \neq \pi(a)}|A_{ib}| \leq \frac{4\delta}{\nu}$. 
Similarly, $\sum_{b \neq \pi(a)}|A_{jb}| \leq \frac{4\delta}{\nu}$.
Recall that $\|C\|_{\infty} = \max_{1\leq b, c\leq K} \|C_{bc}\|.$
Then
$$\sum_{b,c \neq \pi(a)}|A_{ib}||A_{jc}|\| C_{bc}\| \leq \|C\|_{\infty} \sum_{b,c \neq \pi(a)}|A_{ib}||A_{jc}| \le  \frac{16\delta^2}{\nu^2} \|C\|_{\infty}.$$
A similar computation yields $\left\|\sum_{b,c \neq \pi(a)}A_{ic}A_{ib}C_{bc} \right\| \le \frac{16\delta^2}{\nu^2}\|C\|_{\infty}$.

We combine the bounds derived for \eqref{e:gamma_term}, \eqref{e:cpiab_term} and \eqref{e:cbc_term} to get:
\begin{align*}
    &\|\Sigma_{ii}+\Sigma_{ij}\| -  \|\Sigma_{ii} - \Sigma_{ij}\| \\
    &\geq 2\|C_{\pi(a)\pi(a)}\| - 2\|\Gamma_{ii}\|- \frac{24\delta}{\nu}\left(\big\|C_{\pi(a)\pi(a)} \big\| -\nu \right) - \frac{48\delta^2}{\nu^2} \|C\|_{\infty}\\
    &\geq  2\|C_{\pi(a)\pi(a)}\| - 2\|\Gamma_{ii}\|-\frac{24\delta}{\nu}\|C_{\pi(a)\pi(a)}\| + 24\delta -\frac{48\delta^2}{\nu^2} \|C\|_{\infty}.
\end{align*}
Recall the assumption that $\nu > 2 \max (2 \delta, \sqrt{2 \|C\|_{\infty} \delta})$ in Theorem~\ref{thm:pureindexguaran}.
Then
\begin{equation} \label{eqn:subgp}
    \|\Sigma_{ii}+\Sigma_{ij}\| -  \|\Sigma_{ii} - \Sigma_{ij}\| \geq \|C_{\pi(a)\pi(a)}\| - 2\|\Gamma_{ii}\| +12 \delta > 12\delta.
\end{equation}
We have proved that in the case that $\mathrm{sign}(A_{i\pi(a)}) =\mathrm{sign} (A_{j\pi(a)})$, $ \|\Sigma_{ii}+\Sigma_{ij}\| -  \|\Sigma_{ii} - \Sigma_{ij}\| > 12\delta >0$.

In the case that $\mathrm{sign}(A_{i\pi(a)}) \neq \mathrm{sign}(A_{j \pi(a)})$, we observe that the expressions for $\|\Sigma_{ii}+\Sigma_{ij}\|$ and $\|\Sigma_{ii}-\Sigma_{ij}\|$ in \eqref{e:two-op-terms} are reversed.
That is, 
\begin{align*}
\|\Sigma_{ii}-\Sigma_{ij}\| 
&= \left\| 2C_{\pi(a)\pi(a)} + A_{i\pi(a)} \sum_{b \neq \pi(a)}(3A_{ib}+A_{jb}) C_{\pi(a)b} + \sum_{b,c \neq \pi(a)}(A_{ic}+A_{jc})A_{ib}C_{bc} +\Gamma_{ii} \right\|, \nonumber \\
 \|\Sigma_{ii} + \Sigma_{ij}\|&= \left\| A_{i\pi(a)}\sum_{b \neq \pi(a)} (A_{ib}-A_{jb})C_{\pi(a)b} + \sum_{b,c \neq \pi(a)}(A_{ic} -A_{jc})A_{ib}C_{bc}+\Gamma_{ii} \right\|.
\end{align*}
Then following the entire argument for the previous case yields
\begin{equation} \label{eqn:subgrp-neg}
    \|\Sigma_{ii} + \Sigma_{ij}\| -  \|\Sigma_{ii} - \Sigma_{ij}\| < -12\delta <0.
\end{equation}

\textbf{Condition \eqref{e:sign-condition} for $\widehat \Sigma$:}
We now prove \eqref{e:sign-condition} in its original form.
Since we are working on the event $\cE(\delta)$, $\widehat \Sigma$ is an accurate estimator of $\Sigma$. 
We derive a lower bound for \eqref{e:sign-condition} in terms of $\Sigma$:
\begin{align*}
    \|\widehat\Sigma_{ii} + \widehat\Sigma_{ij}\|  &= \|-\widehat\Sigma_{ii} - \widehat\Sigma_{ij}\| \\
    &=\|(\Sigma_{ii} - \widehat\Sigma_{ii})+(\Sigma_{ij} - \widehat\Sigma_{ij}) - (\Sigma_{ii} + \Sigma_{ij})\| \\
    &\ge \|\Sigma_{ii} + \Sigma_{ij}\| - \|\Sigma_{ii} - \widehat\Sigma_{ii}\| - \|\Sigma_{ij} - \widehat\Sigma_{ij}\| \\
    &\ge \|\Sigma_{ii} + \Sigma_{ij}\| -2\delta,
\end{align*}
where the second step uses property \ref{normprop:2} of norms.
The third step follows from the triangle inequality and the fourth step uses the fact that covariance operators are estimated accurately on the event $\cE(\delta)$.
Similarly, 
\begin{align*}
    \|\widehat\Sigma_{ii} - \widehat\Sigma_{ij}\| &=  \|(\Sigma_{ii}-\widehat\Sigma_{ii})+(\widehat\Sigma_{ij} - \Sigma_{ij})+(\Sigma_{ij} - \Sigma_{ii})\| \\ &\le \|\Sigma_{ij} - \Sigma_{ii}\| + \|\Sigma_{ii}-\widehat\Sigma_{ii}\| + \|\widehat\Sigma_{ij}-\Sigma_{ij}\| \\
    &\le \|\Sigma_{ij} - \Sigma_{ii}\| +2\delta.
\end{align*}
Then $$\|\widehat\Sigma_{ii} + \widehat\Sigma_{ij}\| - \|\widehat\Sigma_{ii} - \widehat\Sigma_{ij}\| \ge \|\Sigma_{ij} + \Sigma_{ii}\|- \|\Sigma_{ij} - \Sigma_{ii}\| -4\delta.$$
In the case that $\mathrm{sign}(A_{i \pi(a)}) = \mathrm{sign}(A_{j \pi(a)})$, we use \eqref{eqn:subgp} to derive $$\|\hat\Sigma_{ii} + \hat\Sigma_{ij}\| - \|\hat\Sigma_{ii} - \hat\Sigma_{ij}\| > 12\delta - 4\delta > 0.$$

Next, we establish an upper bound for \eqref{e:sign-condition} using the same techniques.
The term
\begin{align*}
    \|\widehat\Sigma_{ii} + \widehat\Sigma_{ij}\|
    &=\|(\Sigma_{ii} - \widehat\Sigma_{ii})+(\Sigma_{ij} - \widehat\Sigma_{ij}) - (\Sigma_{ii} + \Sigma_{ij})\| \\
    &\le \|\Sigma_{ii} + \Sigma_{ij}\| + \|\Sigma_{ii} - \widehat\Sigma_{ii}\| + \|\Sigma_{ij} - \widehat\Sigma_{ij}\| \\
    &\le \|\Sigma_{ii} + \Sigma_{ij}\| +2\delta,
\end{align*}
where the step step follows from the triangle inequality and the third step uses the definition of $\cE(\delta)$.
Similarly,
\begin{align*}
    \|\widehat\Sigma_{ii} - \widehat\Sigma_{ij}\| &=  \|(\Sigma_{ii}-\widehat\Sigma_{ii})+(\widehat\Sigma_{ij} - \Sigma_{ij})+(\Sigma_{ij} - \Sigma_{ii})\| \\ 
    &\ge \|\Sigma_{ij} - \Sigma_{ii}\| - \|\Sigma_{ii}-\widehat\Sigma_{ii}\| - \|\widehat\Sigma_{ij}-\Sigma_{ij}\| \\
    &\ge \|\Sigma_{ij} - \Sigma_{ii}\| -2\delta.
\end{align*}
In the case that $\mathrm{sign}(A_{i \pi(a)}) \neq \mathrm{sign}(A_{j \pi(a)})$, we use \eqref{eqn:subgrp-neg} to derive
$$\|\widehat\Sigma_{ii} + \widehat\Sigma_{ij}\| - \|\widehat\Sigma_{ii} - \widehat\Sigma_{ij}\| < -12\delta + 4\delta < 0.$$
This completes the proof.


\section{Proof of Theorem~\ref{thm:asymp_aj}}\label{appx:proofs-asymp-normality}

In this proof, we show that $\sqrt{n}(\widehat A_J^\top - A_J^\top)$ converges in distribution to a zero-mean Gaussian random matrix.
The main idea is that $\widehat{A}_J$ is a function of $\widehat{\Sigma}$, whose asymptotics are known.
Specifically, $\sqrt{n}(\widehat \Sigma - \Sigma) \cid \mathfrak{Z}$, and the covariance of the limiting random variable $\mathfrak{Z}$ is denoted by $\mathscr{S} = \mathbb{E}[(X \otimes X-\Sigma) \otimes_{HS}(X \otimes X-\Sigma)]$. 

We recall the assumption $\wh I= I$, which guarantees that $\wh A_I=A_I$ due to the algorithm for construction of $\wh A_I$ given $\wh I$.
Hence, we assume that $\wh A_I = A_I$ for the rest of the proof.
We now state a property of cross-covariance operators that will be used repeatedly.
We refer the reader to \citep[Theorem~7.2.9]{fdabook} for a proof.
Let $Y_1, Y_2 \in \mathbb{H}$ be zero-mean random variables with $\EE\|Y_i\|^2 < \infty$ for $i=1,2$. 
Let $\mathcal{K}_{12}$ denote the cross-covariance operator of $Y_1$ and $Y_2$.
Then for any $f, g \in \mathbb{H}$, 
\begin{equation}\label{eqn:appx-cross-cov}
    \langle \mathcal{K}_{12} f, g\rangle = \EE\left[ \langle Y_1, g \rangle \langle Y_2, f\rangle \right].
\end{equation}
In particular, the right hand side of \eqref{eqn:appx-cross-cov} is the covariance of the real-valued random variables $\langle Y_1, g \rangle$ and $\langle Y_2, f \rangle$ because their means are zero.

The outline for the rest of the proof is as follows: (i) decomposition of $\widehat A_J^\top - A_J^\top$ into two linear terms and a non-linear term; (ii) (auto)covariance contributions from each term; and (iii) cross-covariance contributions from each term.

\textbf{Decomposition of $\widehat A_J^\top - A_J^\top$ into linear and non-linear terms:}
We decompose $\widehat A_J$ into various terms, some of which are linear functions of $\widehat \Sigma$ and others which are non-linear functions.
Under the invertibility assumptions of Theorem~\ref{thm:asymp_aj}, from \eqref{eqn:optim_scalar} and Algorithm~\ref{algo:2}, we can write 
\begin{align*}
    A_J^\top &=  \left\{\sum_{k=1}^K\left\langle C_{k,s},  C_{k,s'}\right\rangle_{\TT_{1,1}}\right\}_{s,s'\in [\widehat K]} ^{-1} \left\{\sum_{k=1}^K \langle C_{k,s}, \CU_{k,j}\rangle_{\TT_{1,1}}\right\}_{s\in [ K], j\in J}, \\
    \widehat A_J^\top &=  \left\{\sum_{k=1}^K\left\langle \widehat{C}_{k,s},  \widehat{C}_{k,s'}\right\rangle_{\TT_{1,1}}\right\}_{s,s'\in [\widehat K]} ^{-1} \left\{\sum_{k=1}^K \langle \widehat{C}_{k,s}, \hat{\CU}_{k,j}\rangle_{\TT_{1,1}}\right\}_{s\in [\widehat K], j\in \widehat J},
\end{align*}
where $\CU = W^{-1}A_I^\top \Sigma_{IJ}$ and $W = A_I^\top A_I$, and $\widehat \CU$ and $\widehat W$ are the corresponding sample versions. 
Since we have assumed that $\wh A_I = A_I$, $\wh W = W$ and the matrices $\widehat A_J$ and $A_J$ have the same dimensions.
Recall the matrices $$V = \left\{\sum_{k=1}^K\left\langle C_{k,s},  C_{k,s'}\right\rangle_{\TT_{1,1}}\right\}_{s,s'\in [ K]}, \qquad \widehat V = \left\{\sum_{k=1}^K\left\langle \widehat{C}_{k,s},  \widehat{C}_{k,s'}\right\rangle_{\TT_{1,1}}\right\}_{s,s'\in [\widehat K]}.$$ 
Since $\widehat K=K$ on $\cE(\delta_n)$, $\widehat V$ and $V$ are both matrices in $\RR^{K \times K}$.

To understand the dependence on $\widehat \Sigma$ and $\Sigma$, we define the following maps. 
Let $\cF: \mathbb{T}_{p,p} \to \mathbb{T}_{|J|,K}$ be defined as $\cF(B) = W^{-1}A_I^\top B_{IJ}$, where $B_{IJ} = \{B_{ij}\}_{i \in I, j \in J}$ denotes the $(I,J)$ block of $B$.
In particular, $\cF(\Sigma) = \CU$ and since $\wh A_I=A_I$, $\cF(\widehat \Sigma) = \widehat \CU$.
Let $\mathcal{T}: \mathbb{T}_{p,p} \to \mathbb{T}_{K,K}$ be defined as 
\begin{align}
[\cT (B)]_{aa} &= \frac{1}{|I_a|(|I_a|-1)} \sum_{i_1,i_2 \in I_a, i_1 \neq i_2} A_{i_1a}A_{i_2a}B_{i_1i_2}, \quad \text{for all } a\in [K], \label{eqn:appx-taa} \\
[\cT (B)]_{ab} &= \frac{1}{|I_a||I_b|} \sum_{i_1 \in I_a, i_2 \in I_b} A_{i_1a}A_{i_2b}B_{i_1i_2}, \quad \text{for all } a, b\in [K] \text{ such that } a\neq b. \label{eqn:appx-tab}
\end{align}
In particular, from \eqref{eqn:c_aa} and \eqref{eqn:c_ab}, $\mathcal{T}(\Sigma) = C$.
From Algorithm~\ref{algo:2}, $\mathcal{T}(\widehat \Sigma) = \widehat C$.
We also define the following maps used to express $\widehat{V}^{-1}$.
Let $g_1: \mathbb{T}_{K,K} \to \mathbb{R}^{K\times K}$ be defined by
\begin{equation}\label{eqn:appx-g1}
    g_1(B) = \left\{\sum_{k=1}^K
  \langle B_{ks}, B_{ks'}\rangle_{\mathbb{T}_{1,1}}\right\}_{s,s'\in[K]}.
\end{equation}
The map $g_2$ is defined on the space of all non-singular $K \times K$ matrices such that any matrix is mapped to its inverse.
The composition $\mathcal{H} = g_2 \circ g_1 \circ \mathcal{T}:
\mathbb{T}_{p,p} \to \mathbb{R}^{K\times K}$ satisfies
$\mathcal{H}(\Sigma) = V^{-1}$ and $\mathcal{H}(\widehat \Sigma) = \widehat V^{-1}$.

We can decompose the estimator $\widehat A_J$ as
\begin{align} \label{eqn:appx-aj_minus_hataj}
\widehat A_J^\top - A_J^\top & = \widehat V^{-1}\widehat V\widehat A_J^\top - V^{-1} VA_J^\top \nonumber \\
& = \widehat V^{-1}(\widehat V\widehat A_J^\top - VA_J^\top) + (\widehat V^{-1}-V^{-1}) V A_J^\top.
\end{align}
Under a $\sqrt{n}$ scaling, the first term fo \eqref{eqn:appx-aj_minus_hataj} is a linear function of $\widehat \Sigma - \Sigma$ and the second term is non-linear. 
To tackle the first term, we further decompose
\begin{align} \label{eqn:appx-asymp_linear_term}
    \widehat V \widehat A_J^\top - VA_J^\top &=  \left\{\sum_{k=1}^K \langle \widehat{C}_{k,s}, \hat{\CU}_{k,j}\rangle_{\TT_{1,1}}\right\}_{s\in [\widehat K], j\in [\widehat J]} - \left\{\sum_{k=1}^K \langle C_{k,s}, \CU_{k,j}\rangle_{\TT_{1,1}}\right\}_{s\in [K], j\in J} \nonumber \\
    &= \left\{ \sum_{k=1}^K \left(\langle C_{ks}, \widehat \CU_{kj} - \CU_{kj} \rangle  + \langle \widehat C_{ks}- C_{ks}, \CU_{kj} \rangle + \langle \widehat C_{ks} -C_{ks}, \widehat \CU_{kj} - \CU_{kj} \rangle \right)\right\}_{s\in [K], j\in J}.
\end{align}

The first term of \eqref{eqn:appx-asymp_linear_term} is a real matrix whose $(s,j)$th element can be written as 
$\sum_{k=1}^K \langle C_{ks}, \widehat \CU_{kj} - \CU_{kj}\rangle = \sum_{k=1}^K \langle C_{ks}, [\cF(\widehat \Sigma)]_{kj} - [\cF(\Sigma)]_{kj} \rangle.$
Since $\cF(B) = (A_I^\top A_I)^{-1}A_I^\top B_{IJ}$, it is a linear map.
Let $\mathcal{G}_1: \mathbb{T}_{p,p} \to \mathbb{R}^{K \times |J|}$ be the map such that for $s \in [K]$ and $j \in J$,
$[\mathcal{G}_1(B)]_{sj} = \sum_{k=1}^K \langle C_{ks},\, \cF(B)_{kj}\rangle_{\mathbb{T}_{1,1}}.$
Since $\cF$ is a linear map and inner products are linear in each component, $\mathcal{G}_1$ is also a linear map.
Then 
\begin{align*}
    \left\{ \sum_{k=1}^K \langle C_{ks}, \widehat \CU_{kj} - \CU_{kj} \rangle\right\}_{s\in [K], j\in J} &= \left\{ \sum_{k=1}^K \langle C_{ks}, [\cF(\widehat \Sigma-\Sigma)]_{kj} \rangle \right\}_{s\in [K], j\in J} \\
    &=  \mathcal{G}_1(\widehat \Sigma -\Sigma),
\end{align*}
where the first step follows from the linearity of $\cF$ and the second step uses the definition of $\mathcal{G}_1$.

The second term of \eqref{eqn:appx-asymp_linear_term} is also a linear map $\widehat \Sigma - \Sigma$. 
To see this, we note that its $(s,j)$th element can be written as
$\langle \widehat C_{ks}- C_{ks}, \CU_{kj} \rangle = \langle [\cT(\widehat \Sigma)]_{ks} - [\cT(\Sigma)]_{ks}, \CU_{kj} \rangle.$
The map $\cT$ is linear because for any scalar $t \in \RR$ and $\Sigma, \Sigma' \in \TT_{p, p}$,
\begin{align*}
    [\cT(t\Sigma + \Sigma')]_{aa} = \frac{1}{|I_a|(|I_a|-1)} \sum_{i_1,i_2 \in I_a, i_1 \neq i_2} A_{i_1a}A_{i_2a} (t \Sigma_{i_1i_2}+\Sigma'_{i_1i_2}) =  [t\cT(\Sigma) + \cT(\Sigma')]_{aa}. 
\end{align*}
A similar calculation shows that $[\cT(t\Sigma + \Sigma')]_{ab} = [t\cT(\Sigma) + \cT(\Sigma')]_{ab}$ for all $a \neq b$.
Let $\mathcal{G}_2: \mathbb{T}_{p,p} \to \mathbb{R}^{K \times |J|}$ be the map such that for $s \in [K]$ and $j \in J$, $[\mathcal{G}_2(B)]_{sj} = \sum_{k=1}^K \langle \cT(B)_{ks},\, \CU_{kj}\rangle_{\mathbb{T}_{1,1}}.$
Since $\cT$ is linear, so is $\mathcal{G}_2$.
Then 
\begin{align*}
    \left\{ \sum_{k=1}^K \langle \widehat C_{ks}- C_{ks}, \CU_{kj} \rangle\right\}_{s\in [K], j\in J} &= \left\{ \sum_{k=1}^K \langle [\cT(\widehat \Sigma)]_{ks} - [\cT(\Sigma)]_{ks}, \CU_{kj} \rangle \right\}_{s\in [K], j\in J} \\
    &=  \mathcal{G}_2(\widehat \Sigma -\Sigma),
\end{align*}
where the first step follows from the linearity of $\cT$ and the second step uses the definition of $\mathcal{G}_2$.

The last term of \eqref{eqn:appx-asymp_linear_term} vanishes asymptotically under a $\sqrt{n}$ scaling.
From the previous computations, $\widehat C-C=\cT(\widehat \Sigma - \Sigma) $ and $\widehat \CU -\CU =\cF(\widehat \Sigma - \Sigma) $ where both $\cT$ and $\cF$ are linear maps. 
Then
\begin{align*}
    \sqrt{n} \langle \widehat C_{ks} - C_{ks}, \widehat \CU_{kj} - \CU_{kj}\rangle &= \sqrt{n} \langle [\cT(\widehat \Sigma - \Sigma)]_{ks}, [\cF(\widehat \Sigma - \Sigma)]_{kj} \rangle \\
    &= \langle [\cT(\sqrt{n}(\widehat \Sigma - \Sigma))]_{ks}, [\cF(\widehat \Sigma - \Sigma)]_{kj} \rangle,
\end{align*}
where the second step uses the linearity of the inner product in its first component and the linearity of $\cT$.
As $n \to \infty$, $\sqrt{n}(\widehat \Sigma - \Sigma)$ has a non-degenerate limit, but $\widehat \Sigma - \Sigma \to \mathbf{0}$. 
Therefore, $[\cF(\widehat \Sigma - \Sigma)]_{kj} \to 0$ and $\sqrt{n} \langle \widehat C_{ks} - C_{ks}, \widehat \CU_{kj} - \CU_{kj}\rangle \to 0$.

We have shown that we only need to understand the limiting behavior of the first two terms of \eqref{eqn:appx-asymp_linear_term}. 
We now address the role of $\widehat V^{-1}$ in the first term of \eqref{eqn:appx-aj_minus_hataj}.
Recall the map $\mathcal{H} = g_2 \circ g_1 \circ \cT :\mathbb{T}_{p,p} \to \mathbb{R}^{K\times K}$ satisfying $\mathcal{H}(\Sigma) = V^{-1}$ and $\mathcal{H}(\widehat \Sigma) = \widehat V^{-1}$.
From previous discussion, $\cT$ is linear, hence it is continuous. 
The map $g_1$ is continuous since inner products are bilinear maps.
Under the assumptions of Theorem~\ref{thm:asymp_aj}, 
$g_2$ is continuous.
As the composition of continuous maps, $\cH$ is also continuous.
By the Continuous Mapping Theorem, $\widehat V^{-1} \cip V^{-1}$.
Then
\begin{align}\label{eqn:appx-cg1pluscg2}
    \sqrt{n}\widehat V^{-1} (\widehat V \widehat A_J^\top - VA_J^\top) &= \widehat V^{-1}(\sqrt{n}(\mathcal{G}_1(\widehat \Sigma -\Sigma)+\mathcal{G}_2(\widehat \Sigma -\Sigma)) ) \nonumber\\
    &=\widehat V^{-1}(\mathcal{G}_1(\sqrt{n}(\widehat \Sigma -\Sigma))+\mathcal{G}_2(\sqrt{n}(\widehat \Sigma -\Sigma)) ) \nonumber \\
    &\cid V^{-1} (\mathcal{G}_1(\mathfrak{Z})+ \mathcal{G}_2(\mathfrak{Z})).
\end{align}
The second step uses the linearity of $\cG_1$ and $\cG_2$. The third step uses Slutsky's Theorem.
This characterizes the limiting behavior of the first term of \eqref{eqn:appx-aj_minus_hataj}.

Next, we analyze the second term of \eqref{eqn:appx-aj_minus_hataj}. 
We write $\widehat V^{-1}-V^{-1} = -\widehat V^{-1}(\wh V-V) V^{-1}$ and simplify $\wh V-V$ as follows.
\begin{align}
\wh V - V &= \left\{\sum_{k=1}^K\left\langle \widehat{C}_{ks},  \widehat{C}_{ks'}\right\rangle_{\TT_{1,1}}
- \left\langle C_{ks},  C_{ks'}\right\rangle_{\TT_{1,1}} \right\}_{s,s'\in [K]}\nonumber  \\
&= \left\{\sum_{k=1}^K\left\langle \widehat{C}_{ks},  \widehat{C}_{ks'}-C_{ks'} \right\rangle_{\TT_{1,1}} 
\right\}_{s,s'\in [K]} 
+ \left\{\sum_{k=1}^K\left\langle \widehat{C}_{ks}-C_{ks},  {C}_{ks'}\right\rangle_{\TT_{1,1}}
\right\}_{s,s'\in [K]}\nonumber  \\
&= \left\{\sum_{k=1}^K\left\langle C_{ks},  \wh C_{ks'}-C_{ks'}
\right\rangle_{\TT_{1,1}} \right\}_{s,s'\in [K]} 
+ \left\{\sum_{k=1}^K\left\langle \widehat{C}_{ks}-C_{ks},  {C}_{ks'}\right\rangle_{\TT_{1,1}} \right\}_{s,s'\in [K]}\nonumber  \\
&\quad + 
\left\{\sum_{k=1}^K\left\langle \widehat{C}_{ks}-C_{ks},  \widehat{C}_{ks'}-C_{ks'}
\right\rangle_{\TT_{1,1}} \right\}_{s,s'\in [K]}\nonumber  \\
& =  \left\{\sum_{k=1}^K\left\langle  {C}_{ks}, [\mathcal{T}(\wh\Sigma-\Sigma)]_{ks'}\right\rangle_{\TT_{1,1}} \right\}_{s,s'\in [K]}+ \left\{\sum_{k=1}^K\left\langle [\mathcal{T}(\wh\Sigma-\Sigma)]_{ks}, {C}_{ks'}\right\rangle_{\TT_{1,1}} \right\}_{s,s'\in [K]} \nonumber \\
&\quad + \left\{\sum_{k=1}^K\left\langle[\mathcal{T}(\wh\Sigma-\Sigma)]_{ks},  [\mathcal{T}(\wh\Sigma-\Sigma)]_{ks'}\right\rangle_{\TT_{1,1}} \right\}_{s,s'\in [K]}. \label{eqn:appx-whv-v} \end{align}
The second and third steps require the addition and subtraction of appropriate terms.
The fourth step follows from the definition of $\cT$ in \eqref{eqn:appx-taa}--\eqref{eqn:appx-tab}.
The last term of \eqref{eqn:appx-whv-v} vanishes under a $\sqrt{n}$ scaling, similar to the last term of \eqref{eqn:appx-asymp_linear_term}.
To capture the non-degenerate terms, we define $\cG_3 :\TT_{p,p} \to \RR^{K \times K}$ such that 
$$\cG_3(B)_{ss'} = \sum_{k=1}^K \langle \cT(B)_{ks}, C_{ks'} \rangle + \langle C_{ks}, \cT(B)_{ks'} \rangle.$$
Therefore, 
\begin{align*}
    \sqrt{n}(\wh V^{-1} - V^{-1})VA_J^\top &= -\sqrt{n} \wh V^{-1}(\wh V - V)V^{-1}VA_J^\top \\
    &= -\sqrt{n} \wh V^{-1} \cG_3(\wh \Sigma- \Sigma) A_J^\top \\
    &= -\wh V^{-1} \cG_3(\sqrt{n}(\wh \Sigma- \Sigma)) A_J^\top \\
    &\cid -V^{-1} \cG_3(\mathfrak{Z})A_J^\top,
\end{align*}
where the third step follows from the linearity of $\cG_3$ and the last step follows from the Continuous Mapping Theorem.
Combining this with \eqref{eqn:appx-cg1pluscg2}, as $n\to\infty$,
\begin{equation}\label{eqn:appx-limiting-rv}
    \sqrt{n}(\widehat A_J^\top -A_J^\top) \cid V^{-1}\bigl[\mathcal{G}_1(\mathfrak{Z}) + \mathcal{G}_2(\mathfrak{Z})
    - \mathcal{G}_3(\mathfrak{Z})A_J^\top\bigr].
\end{equation}

\textbf{Covariance Contributions:} We now compute the covariance of the limiting random variable in \eqref{eqn:appx-limiting-rv}.
Contributions to the covariance arise from the (auto)covariance of each term in \eqref{eqn:appx-limiting-rv} and their cross-covariances.
We begin by computing the (auto)covariances.

Recall that $\cG_1(\mathfrak{Z}), \cG_2(\mathfrak{Z})$ and $\cG_3(\mathfrak{Z})A_J^\top \in \RR^{K \times |J|}$.
Their covariance is a symmetric positive semidefinite matrix in $\RR^{K|J| \times K|J|}$.
We also specify the convention for operators. 
For example, $\cF\mathscr{S}\cF^*$ is a bounded operator on $\mathbb{T}_{|J|,K}$, and elements of $\mathbb{T}_{|J|,K}$ decompose into $K \times |J|$ blocks indexed by $(k,j)$ with each block in $\mathbb{T}_{1,1}$.
We write $[\cF\mathscr{S}\cF^*]_{k_1j_1, k_2j_2} \in \mathbb{T}_{1,1}$ for the block corresponding to coordinates $(k_1,j_1)$ and $(k_2,j_2)$.
We use analogous notation for $[\mathcal{T}\mathscr{S}\mathcal{T}^*]_{k_1s_1,k_2s_2}$ and the cross-operator $[\cF\mathscr{S}\mathcal{T}^*]_{k_1j_1,k_2s_2}$ for $k_1, k_2, s_1, s_2 \in [K]$ and $j_1 \in J$.

The covariance between $[\cG_1(\mathfrak{Z})]_{s_1j_1}$ and $[\cG_1(\mathfrak{Z})]_{s_2 j_2}$ is derived below.
\begin{align*}
    \mathrm{Cov}\left([\cG_1(\mathfrak{Z})]_{s_1 j_1}, [\cG_1(\mathfrak{Z})]_{s_2 j_2}\right) &= \mathrm{Cov}\left( \sum_{k=1}^K \langle C_{ks_1}, \cF(\mathfrak{Z})_{kj_1} \rangle_{\mathbb{T}_{1,1}}, \sum_{k=1}^K \langle C_{ks_2}, \cF(\mathfrak{Z})_{kj_2} \rangle_{\mathbb{T}_{1,1}} \right) \\
    &= \sum_{k_1=1}^K \sum_{k_2=1}^K \mathrm{Cov}\left( \langle C_{k_1s_1}, \cF(\mathfrak{Z})_{k_1j_1} \rangle, \langle C_{k_2s_2}, \cF(\mathfrak{Z})_{k_2j_2} \rangle \right) \\
    &= \sum_{k_1=1}^K \sum_{k_2=1}^K  \langle [\cF \mathscr{S}\cF^*]_{k_1j_1, k_2j_2} C_{k_2s_2}, C_{k_1s_1} \rangle, 
\end{align*}
where the second step follows from the bi-linearity of covariance and the third step follows from \eqref{eqn:appx-cross-cov}.

Next, we derive the covariance between $[\cG_2(\mathfrak{Z})]_{s_1 j_1}$ and $[\cG_2(\mathfrak{Z})]_{s_2 j_2}$ below.
\begin{align*}
    \mathrm{Cov}\left([\cG_2(\mathfrak{Z})]_{s_1 j_1}, [\cG_2(\mathfrak{Z})]_{s_2 j_2}\right) &= \mathrm{Cov}\left( \sum_{k=1}^K \langle \cT(\mathfrak{Z})_{ks_1}, \CU_{kj_1} \rangle_{\mathbb{T}_{1,1}}, \sum_{k=1}^K \langle \cT(\mathfrak{Z})_{ks_2}, \CU_{kj_2} \rangle_{\mathbb{T}_{1,1}} \right) \\
    &= \sum_{k_1=1}^K \sum_{k_2=1}^K \mathrm{Cov}\left( \langle \cT(\mathfrak{Z})_{k_1s_1}, \CU_{k_1j_1} \rangle, \langle \cT(\mathfrak{Z})_{k_2s_2}, \CU_{k_2j_2} \rangle \right) \\
    &= \sum_{k_1=1}^K \sum_{k_2=1}^K  \langle [\cT \mathscr{S}\cT^*]_{k_1s_1, k_2s_2} \CU_{k_2j_2}, \CU_{k_1j_1} \rangle, 
\end{align*}
where the second step follows from the bi-linearity of covariance and the third step follows from \eqref{eqn:appx-cross-cov}.

Lastly, we compute the covariance of the third term in \eqref{eqn:appx-limiting-rv}.
For any $s_1, s_2 \in [K]$ and $j_1, j_2 \in J$, 
{\allowdisplaybreaks
\begin{align*}
    &\mathrm{Cov}\left( [\cG_3(\mathfrak{Z})A_J^\top]_{s_1j_1}, [\cG_3(\mathfrak{Z})A_J^\top]_{s_2j_2}  \right) \\
    &= \mathrm{Cov}\left( \sum_{k=1}^K [\cG_3(\mathfrak{Z})]_{s_1k}[A_J]_{j_1k}, \sum_{k=1}^K [\cG_3(\mathfrak{Z})]_{s_2k}[A_J]_{j_2k} \right) \\
    &= \sum_{k_1=1}^K \sum_{k_2=1}^K [A_J]_{j_1k_1}[A_J]_{j_2k_2} \mathrm{Cov}\left( [\cG_3(\mathfrak{Z})]_{s_1k_1}, [\cG_3(\mathfrak{Z})]_{s_2k_2}\right)  \\
    &= \sum_{k_1=1}^K \sum_{k_2=1}^K [A_J]_{j_1k_1}[A_J]_{j_2k_2} \mathrm{Cov}\Bigg( \sum_{k=1}^K \langle \cT(\mathfrak{Z})_{ks_1}, C_{kk_1} \rangle + \langle C_{ks_1}, \cT(\mathfrak{Z})_{kk_1} \rangle,\\
    &\qquad \qquad \qquad \qquad \qquad \qquad \qquad \qquad\sum_{k=1}^K \langle \cT(\mathfrak{Z})_{ks_2}, C_{kk_2} \rangle + \langle C_{ks_2}, \cT(\mathfrak{Z})_{kk_2} \rangle  \Bigg)\\
    &= \sum_{k_1=1}^K \sum_{k_2=1}^K [A_J]_{j_1k_1}[A_J]_{j_2k_2}\sum_{k_3,k_4\in[K]}\Bigl(\langle[\mathcal{T}\mathscr{S}\mathcal{T}^*]_{k_3s_1,k_4s_2}\,C_{k_4k_2},\,C_{k_3k_1}\rangle \\
    &\qquad\qquad\qquad\qquad\qquad +\langle[\mathcal{T}\mathscr{S}\mathcal{T}^*]_{k_3s_1,k_4k_2}\,
       C_{k_4s_2},\,C_{k_3k_1}\rangle
       + \langle[\mathcal{T}\mathscr{S}\mathcal{T}^*]_{k_3k_1,k_4s_2}\,
       C_{k_4k_2},\,C_{k_3s_1}\rangle\\
      &\qquad\qquad\qquad\qquad\qquad +\,\langle[\mathcal{T}\mathscr{S}\mathcal{T}^*]_{k_3k_1,k_4k_2}\,
       C_{k_4s_2},\,C_{k_3s_1}\rangle \Bigr).
\end{align*}}
The first equality uses matrix multiplication and the second equality follows from the bi-linearity of the inner product.
The third equality uses the explicit form of $\cG_3$ and the last equality follows from \eqref{eqn:appx-cross-cov}.
All inner products are in $\TT_{1,1}$.

\textbf{Cross-Covariance Contributions:} We now compute the cross-covariance among the three terms in \eqref{eqn:appx-limiting-rv}. 
First, we compute the cross-covariance between $\cG_1(\mathfrak{Z})$ and $\cG_2(\mathfrak{Z})$.
For any indices $s_1, s_2 \in [K]$ and $j_1, j_2 \in J$,
\begin{align*}
    \mathrm{Cov}\left([\cG_1(\mathfrak{Z})]_{s_1 j_1}, [\cG_2(\mathfrak{Z})]_{s_2 j_2}\right) &= \mathrm{Cov}\left( \sum_{k=1}^K \langle C_{ks_1}, \cF(\mathfrak{Z})_{kj_1} \rangle_{\mathbb{T}_{1,1}}, \sum_{k=1}^K \langle \cT(\mathfrak{Z})_{ks_2}, \CU_{kj_2} \rangle_{\mathbb{T}_{1,1}} \right) \\
    &= \sum_{k_1=1}^K \sum_{k_2=1}^K \mathrm{Cov}\left( \langle \cF(\mathfrak{Z})_{k_1j_1}, C_{k_1s_1} \rangle, \langle \cT(\mathfrak{Z})_{k_2s_2}, \CU_{k_2j_2} \rangle \right) \\
    &= \sum_{k_1=1}^K \sum_{k_2=1}^K  \langle [\cF \mathscr{S}\cT^*]_{k_1j_1, k_2s_2} \CU_{k_2j_2}, C_{k_1s_1} \rangle, 
\end{align*}
where the second step uses the bi-linearity of covariance as well as the symmetric of the inner product $\langle \cdot, \cdot \rangle_{\TT_{1,1}}$.
The last step follows from \eqref{eqn:appx-cross-cov}.

We derive the cross-covariance between $\cG_1(\mathfrak{Z})$ and $\cG_3(\mathfrak{Z})A_J^\top$ below.
For any indices $s_1, s_2 \in [K]$ and $j_1, j_2 \in J$,
\begin{align*}  &\mathrm{Cov}\left([\cG_1(\mathfrak{Z})]_{s_1 j_1}, [\cG_3(\mathfrak{Z})A_J^\top]_{s_2 j_2}\right)\\
    &= \mathrm{Cov}\left( \sum_{k=1}^K \langle C_{ks_1}, \cF(\mathfrak{Z})_{kj_1} \rangle_{\mathbb{T}_{1,1}}, \sum_{k=1}^K [\cG_3(\mathfrak{Z})]_{s_2k}[A_J]_{j_2k} \right)  \\
    &= \sum_{k_2=1}^K [A_J]_{j_2k_2}\sum_{k_1=1}^K \mathrm{Cov}\left( \langle \cF(\mathfrak{Z})_{k_1j_1}, C_{k_1s_1} \rangle, [\cG_3(\mathfrak{Z})]_{s_2k_2} \right) \\
    &= \sum_{k_2=1}^K [A_J]_{j_2k_2}\sum_{k_1=1}^K \mathrm{Cov}\left( \langle \cF(\mathfrak{Z})_{k_1j_1}, C_{k_1s_1} \rangle, \sum_{k_3=1}^K \langle \cT(\mathfrak{Z})_{k_3s_2}, C_{k_3k_2} \rangle + \langle C_{k_3s_2},\cT(\mathfrak{Z})_{k_3k_2} \rangle\right)\\
    &= \sum_{k_2=1}^K [A_J]_{j_2k_2}\sum_{k_1=1}^K \sum_{k_3=1}^K \langle[\cF \mathscr{S}\cT^*]_{k_1j_1, k_3s_2} C_{k_3k_2}, C_{k_1s_1} \rangle + \langle [\cF \mathscr{S}\cT^*]_{k_1j_1, k_3k_2} C_{k_3s_2}, C_{k_1s_1} \rangle.
\end{align*}

Lastly, we derive the cross-covariance between $\cG_2(\mathfrak{Z})$ and $\cG_3(\mathfrak{Z})A_J^\top$ below.
For any indices $s_1, s_2 \in [K]$ and $j_1, j_2 \in J$,
{\allowdisplaybreaks
\begin{align*}
    &\mathrm{Cov}\left([\cG_2(\mathfrak{Z})]_{s_1 j_1}, [\cG_3(\mathfrak{Z})A_J^\top]_{s_2 j_2}\right)\\
    &= \mathrm{Cov}\left( \sum_{k=1}^K \langle \cT(\mathfrak{Z})_{ks_1}, \CU_{kj_1} \rangle, \sum_{k=1}^K [\cG_3(\mathfrak{Z})]_{s_2k}[A_J]_{j_2k} \right)\\
    &= \sum_{k_2=1}^K [A_J]_{j_2k_2}\sum_{k_1=1}^K \mathrm{Cov}\left( \langle \cT(\mathfrak{Z})_{k_1s_1}, \CU_{k_1j_1} \rangle, [\cG_3(\mathfrak{Z})]_{s_2k_2} \right) \\
    &= \sum_{k_2=1}^K [A_J]_{j_2k_2}\sum_{k_1=1}^K \mathrm{Cov}\left( \langle \cT(\mathfrak{Z})_{k_1s_1}, \CU_{k_1j_1} \rangle, \sum_{k_3=1}^K \langle \cT(\mathfrak{Z})_{k_3s_2}, C_{k_3k_2} \rangle + \langle C_{k_3s_2},\cT(\mathfrak{Z})_{k_3k_2} \rangle\right)\\
    &= \sum_{k_2=1}^K [A_J]_{j_2k_2}\sum_{k_1=1}^K \sum_{k_3=1}^K \langle[\cT \mathscr{S}\cT^*]_{k_1s_1, k_3s_2} C_{k_3k_2}, \CU_{k_1j_1} \rangle + \langle [\cT \mathscr{S}\cT^*]_{k_1s_1, k_3k_2} C_{k_3s_2}, \CU_{k_1j_1} \rangle.
\end{align*}}

Then as $n \to \infty$,
\[\sqrt{n}(\widehat{A}_J^\top - A_J^\top)\xrightarrow{d} \cG(\mathfrak{Z})\quad \text{in } \mathbb{R}^{K\times |J|},\]
where $\cG(\mathfrak{Z})$ is a zero-mean Gaussian random matrix with covariance
\begin{equation}\label{e:cov_full}
\mathrm{Cov}\bigl[\cG(\mathfrak{Z})_{s_1,j_1},\,\cG(\mathfrak{Z})_{s_2,j_2}\bigr]= [V^{-1}\,\mathscr{T}\, V^{-1}]_{s_1j_1,\,s_2j_2},\end{equation}
where $\mathscr{T} \in \mathbb{R}^{K|J| \times K|J|}$ is the covariance matrix of $\mathfrak{Z}$. 
Using \eqref{eqn:appx-limiting-rv}, for $s_1,s_2 \in [K]$ and $j_1,j_2 \in J$, $\mathscr{T}_{s_1j_1,\,s_2j_2}$ is defined as
{\allowdisplaybreaks
\begin{align}
    & \underbrace{\sum_{k_1,k_2 \in [K]}
       \langle (\cF\mathscr{S}\cF^*)_{k_1j_1, k_2j_2}\,
       C_{k_2s_2},\, C_{k_1s_1}
       \rangle}_{\text{(I): from }\mathcal{G}_1}
    \quad + \underbrace{\sum_{k_1,k_2 \in [K]}
       \langle (\mathcal{T}\mathscr{S}\mathcal{T}^*)_{k_1s_1,k_2s_2}\,
       \CU_{k_2j_2},\, \CU_{k_1j_1}
       \rangle}_{\text{(II): from }\mathcal{G}_2}
    \nonumber  \\
    & + \underbrace{\sum_{k_1,k_2 \in [K]}
       \Bigl(
       \langle (\cF\mathscr{S}\mathcal{T}^*)_{k_1j_1,k_2s_2}\,
       \CU_{k_2j_2},\, C_{k_1s_1}\rangle +
       \langle (\cF\mathscr{S}\mathcal{T}^*)_{k_1j_2,k_2s_1}\,
       \CU_{k_2j_1},\, C_{k_1s_2}\rangle
       \Bigr)}_{\text{(III): cross covariance }\mathcal{G}_1,\mathcal{G}_2}
    \nonumber \\
    & - \underbrace{2\sum_{s\in[K]}(A_J)_{j_2s}
       \sum_{k_1,k_2\in[K]}
       \Bigl(
       \langle(\cF\mathscr{S}\mathcal{T}^*)_{k_1j_1,k_2s_2}\,
       C_{k_2s},\,C_{k_1s_1}\rangle +
       \langle(\cF\mathscr{S}\mathcal{T}^*)_{k_1j_1,k_2s}\,
       C_{k_2s_2},\,C_{k_1s_1}\rangle
       \Bigr)}_{\text{(IV): cross term }\mathcal{G}_1,\mathcal{G}_3A_J^\top}
    \nonumber \\
    & - \underbrace{2\sum_{s\in[K]}(A_J)_{j_2s}
       \sum_{k_1,k_2\in[K]}
       \Bigl(
       \langle(\mathcal{T}\mathscr{S}\mathcal{T}^*)_{k_1s_1,k_2s_2}\,
       \CU_{k_2s},\,\CU_{k_1j_1}\rangle
       +
       \langle(\mathcal{T}\mathscr{S}\mathcal{T}^*)_{k_1s_1,k_2s}\,
       \CU_{k_2s_2},\,\CU_{k_1j_1}\rangle \Bigr)}_{\text{(V): cross term }\mathcal{G}_2,\mathcal{G}_3A_J^\top}
    \nonumber \\
    & + \underbrace{\sum_{s,s'\in[K]}(A_J)_{j_1s}(A_J)_{j_2s'}
       \sum_{k_1,k_2\in[K]}
       \Bigl(
       \langle(\mathcal{T}\mathscr{S}\mathcal{T}^*)_{k_1s_1,k_2s_2}\,
       C_{k_2s'},\,C_{k_1s}\rangle +\langle(\mathcal{T}\mathscr{S}\mathcal{T}^*)_{k_1s_1,k_2s'}\,
       C_{k_2s_2},\,C_{k_1s}\rangle}_{\text{(VI): from }\mathcal{G}_3A_J^\top}
    \nonumber \\
    &\qquad\qquad\qquad \qquad \qquad \quad \underbrace{
       +\, \langle(\mathcal{T}\mathscr{S}\mathcal{T}^*)_{k_1s,k_2s_2}\,
       C_{k_2s'},\,C_{k_1s_1}\rangle +\,\langle(\mathcal{T}\mathscr{S}\mathcal{T}^*)_{k_1s,k_2s'}\,
       C_{k_2s_2},\,C_{k_1s_1}\rangle
       \Bigr).}_{\text{(VI) cont.}}\nonumber
\end{align}}


\section{Proof of Lemmas~\ref{lem:1}--\ref{lem:sigma_hat_norms}} \label{appx:proofs-of-lemmas}

\subsection{Proof of Lemma~\ref{lem:1}}
The proofs of Lemma~\ref{lem:1} and \ref{lem:M_i} repeatedly use basic properties of a norm. We state these here.

Let $\cX$ be a vector space over $\RR$.
Then a norm $\|\cdot\|$ is a function from $\cX$ to $\RR$ satisfying the following properties:
\begin{property} \label{normprop:1}
    $\|x\| \geq 0$ for all $x \in \cX$. The equality holds if and only if $x=0$.
\end{property}
\begin{property} \label{normprop:2}
    $\|rx\| = |r|\|x\|$ for all scalars $r \in \RR$.
\end{property}
\begin{property} \label{normprop:3}
    $\|x+y\| \le \|x\|+\|y\|$ for all $x,y \in \cX$. This is called the triangle inequality.
\end{property}

We now begin the proof of Lemma~\ref{lem:1}. We first prove part (a) of Lemma~\ref{lem:1}, which states that if $i \in I_a$, then the operator norms of the block operators in the row $\Sigma_{i.}$ are at most $\|C_{aa}\|$ (not considering the operator $\Sigma_{ii}$). 
Moreover, the equality is attained if and only if $j \in I_a$. 

Throughout this proof, we assume that $i \in I_a$. From \eqref{eqn:cov_decomp}, for any index $j \neq i$, the block operator structure of the covariance operators allows us to write 
\begin{equation} \label{e:model-ineqj}
    \Sigma_{ij} = \sum_{k_1=1}^K \sum_{k_2=1}^K A_{ik_1}C_{k_1k_2}A_{jk_2}.
\end{equation}
Since $i \in I_a$, $|A_{ia}|=1$ and all other entries in the $i$th row of $A$ are zero. 
Similarly, if $j \in I_a$, then $|A_{ja}|=1$ and $A_{jb}=0$ for all indices $b \neq a$.
Then \eqref{e:model-ineqj} simplifies to $\Sigma_{ij} = \mathrm{sign}(A_{ia}A_{ja})C_{aa}$. 
Since $\mathrm{sign}(A_{ia}A_{ja}) \in \{-1, +1\}$,  $\|\Sigma_{ij} \| = \| C_{aa}\|$ by the property of norms.

We now show that for an index $j \notin I_a$, $\|\Sigma_{ij}\|$ is strictly less than $\|C_{aa}\|$. 
Since $X_i$ is a pure variable, \eqref{e:model-ineqj} simplifies to
$\Sigma_{ij} =  A_{ia}\sum_{k_2=1}^K A_{jk_2}C_{ak_2}$, where $A_{ia} \in \{-1, +1\}$. 
Then
\begin{equation} \label{e:sigmaij-ineq}
    \|\Sigma_{ij}\|   \leq \sum_{k_2=1}^K \|A_{jk_2}C_{ak_2}\| =\sum_{k_2=1}^K |A_{jk_2}| \|C_{ak_2}\|,
\end{equation}
where the first step uses the triangle inequality and the subsequent equality uses Property \ref{normprop:2} of norms.
From Assumption \ref{assump1}, we know that $\sum_{k_2 = 1}^K |A_{jk_2}| \le 1$.
Suppose that $\sum_{k_2 = 1}^K |A_{jk_2}| < 1$.
Then \ref{e:sigmaij-ineq} has the following upper bound:
$$\sum_{k_2=1}^K |A_{jk_2}| \|C_{ak_2}\|\leq \max_{k_2}\|C_{ak_2}\|  \sum_{k_2=1}^K |A_{jk_2}| < \|C_{aa}\|,$$
where the last inequality follows from the fact that $\|C_{ak_2}\| < \|C_{aa}\|$ for all $k_2 \neq a$  (Assumption \ref{assump3}). 

Next, we suppose that $\sum_{k_2 = 1}^K |A_{jk_2}| = 1$. 
Since the index $j \notin I_a$,  $|A_{ja}| \neq 1$. 
There exists an index $b \neq a$ such that $A_{jb} \neq 0$. 
We derive an upper bound for \ref{e:sigmaij-ineq} in this case as follows:
\begin{align*}
    \sum_{k_2=1}^K |A_{jk_2}| \|C_{ak_2}\| &= \sum_{k_2 \neq b } |A_{jk_2}| \|C_{ak_2}\| + |A_{jb}| \|C_{ab}\| \\&< \sum_{k_2 \neq b } |A_{jk_2}| \|C_{ak_2}\| + |A_{jb}| \|C_{aa}\| \\
    &\leq  \sum_{k_2 \neq b } |A_{jk_2}| \|C_{aa}\| + |A_{jb}| \|C_{aa}\| \\
    &= \|C_{aa}\| \sum_{k_2 =1 }^K |A_{jk_2}|
    = \|C_{aa}\|.
\end{align*}
In the second step, the term $\|C_{ab}\|$ is upper bounded by $\|C_{aa}\|$ using Assumption~\ref{assump3}.
In the third step, we use the same assumption to obtain $\|C_{ak_2}\| \le \|C_{aa}\|$. The last step uses our assumption that $\sum_{k_2 = 1}^K |A_{jk_2}| = 1$. We have shown that if $j \notin I_a$, then $\|\Sigma_{ij}\|< \|C_{aa}\|$. This completes the proof of part (a).

It directly follows that the maximum operator norm in the $i$th row is $M_i = \|C_{aa}\|$.
Moreover, this upper bound is only achieved by $\|\Sigma_{ij}\|$, where $j \in I_a$. 
The set $S_i$ consists of all column indices at which the maximum norm is achieved.
Therefore, $I_a = S_i ~\cup~ \{i\}$.


\subsection{Proof of Lemma \ref{lem:M_i}}
In Lemma~\ref{lem:M_i}, we show that even if $X_i$ is a non-pure variable, the set $S_i$ contains at least one index corresponding to a pure variable. 
To this end, we first establish an upper bound on $M_i$ and then show that it is achieved by $\|\Sigma_{ij}\|$ for an index $j \in I$.

Let $i$ be any index in $[p]$. 
We establish an upper bound on $M_i$ as follows:
\begin{align} \label{e:upper-bound-M_i}
    M_i &= \max_{j \neq i} \left\| \sum_{k_1=1}^K \sum_{k_2=1}^K A_{ik_1}C_{k_1k_2}A_{jk_2} \right\|
    \leq \max_{j \neq i} \sum_{k_2=1}^K \left|A_{jk_2}\right|  \left\| \sum_{k_1=1}^K A_{ik_1}C_{k_1k_2}\right\|, 
\end{align}
where the second step uses the triangle inequality and Property \ref{normprop:2} of norms. 
For any $k_2\in [K]$, let the operator $N_{k_2} =   \sum_{k_1=1}^K A_{ik_1}C_{k_1k_2}$. 
Then \eqref{e:upper-bound-M_i} shows that $M_i \le \max_{j \neq i} \sum_{k_2=1}^K \left|A_{jk_2}\right| \|N_{k_2}\|$. 
In the sum on the right hand side, if $A_{jk} = 0$ for some index $k$, then the $k$th term vanishes in the summation. 
We only consider indices where $A_{jk_2}$ is non-zero and define the index
$$k_2^* = \argmax_{k_2 \in [K], A_{jk_2} \neq 0} \left\| N_{k_2}\right\|.$$ 
We note that $k_2^*$ depends on $j$ and denote it as $k_2^*(j)$ to emphasize this. 
We further simplify \eqref{e:upper-bound-M_i} as
\begin{align*}
    M_i &\le \max_{j \neq i} \sum_{k_2=1}^K \left|A_{jk_2}\right| \|N_{k_2}\| 
    \le \max_{j \neq i} \|N_{k_2^*(j)}\| \sum_{k_2=1}^K \left|A_{jk_2}\right| 
    \le \max_{j \neq i} \|N_{k_2^*(j)}\|,
\end{align*}
where the last step follows from Assumption~\ref{assump1}. 
Let the maximum be achieved for the index $j^* \neq i$. 
Then, we establish the following upper bound 
\begin{equation} \label{eqn:appx-mi-upper-bound}
    M_i \le \|N_{k_2^*(j^*)}\| = \left\|\sum_{k_1=1}^K A_{ik_1}C_{k_1k_2^*(j^*)}\right\|.
\end{equation}

Suppose that the index $i_2\in I_{k_2^*(j^*)}$ and $i_2 \neq i$. 
This is always possible since there are at least two pure variables corresponding to each latent factor (Assumption~\ref{assump2}). 
Since $X_{i_2}$ is a pure variable,
$$ \|\Sigma_{ii_2}\|   = \left\|A_{i_2k_2^*(j^*)} \sum_{k_1=1}^K  A_{ik_1}C_{k_1k_2^*(j^*)} \right\|  =   \left\| \sum_{k_1=1}^K A_{ik_1}C_{k_1k_2^*(j^*)}\right\|,$$
where the last equality uses $|A_{i_2k_2^*(j^*)}|=1$.
By \eqref{eqn:appx-mi-upper-bound}, this shows that $M_i$ is achieved at any index $i_2 \in I_{k_2^*(j^*)}$. 
Then $ I_{k_2^*(j^*)} \subset S_i$ and the set $I_{k_2^*(j^*)}$ is non-empty by Assumption \ref{assump2}. 
Hence, we have proved that there exists at least one index corresponding to a pure variable in $S_i$.


\subsection{Proof of Lemma \ref{lem:sigma_hat_norms}}
On the event $\cE(\delta)$, for any indices $i, j\in [p]$, 
\begin{equation}\label{eqn:appx-epsilon-delta}
    \|\Sigma_{ij}\|-\delta \le \|\widehat \Sigma_{ij}\| \le \|\Sigma_{ij}\|+\delta.
\end{equation}
Recall from Lemma~\ref{lem:1} that $\|\Sigma_{ij_1}\| = \|\Sigma_{ij_2}\| = \|C_{aa}\|$ for all $i \in I_a$ and $j_1, j_2 \in I_a \backslash\{i\}$ and $j_1\neq j_2$. 
On the event $\cE(\delta)$, since $\widehat \Sigma_{ij_1}$ and $\widehat \Sigma_{ij_2}$ are at most $\delta$ distance away from $\Sigma_{ij_1}$ and $\Sigma_{ij_2}$ respectively, the triangle inequality leads to 
\begin{equation} \label{e:est-ai-2delta}
   \left| \|\widehat \Sigma_{ij_1}\|-\|\widehat \Sigma_{ij_2}\| \right| \leq 2\delta.
\end{equation}
This proves part (a). 

We next prove (b) and (c). For any index $j_1 \notin I_a$, using the triangle inequality results in
\begin{align*}
    \|\Sigma_{ij_1}\| = \left\| \sum_{k=1}^K C_{ak}A_{j_1k}\right\| 
    \le |A_{j_1a}|\|C_{aa}\| + \left\|\sum_{k\neq a} A_{j_1k}C_{ak}\right\|.
\end{align*}
By the definition of $\nu$, we know that $\|C_{ak}\| \le \|C_{aa}\|-\nu$ for all $k\neq a$. Then using Assumption~\ref{assump1},
\begin{align} \label{eqn:appx-sigma-nu}
\begin{split}
    \|\Sigma_{ij_1}\| & \le |A_{ja}|\|C_{aa}\|+(\|C_{aa}\|-\nu)(1-|A_{j_1a}|) \\
    & = \|C_{aa}\| -\nu(1-|A_{j_1a}|) \\
    & \le \|C_{aa}\| = \|\Sigma_{ij_2}\|
    \end{split}
\end{align}
for any $j_2 \in I_a \backslash \{i\}$, which implies that
$$
\|\Sigma_{ij_2}\|-\|\Sigma_{ij_1}\|\ge \nu(1-|A_{j_1a}|).
$$
Applying \eqref{eqn:appx-epsilon-delta}, we have
$$
\|\wh\Sigma_{ij_2}\|-\|\wh\Sigma_{ij_1}\|\ge \nu(1-|A_{j_1a}|)-2\delta.
$$
Since $\nu(1-|A_{j_1a}|)>0$, (b) follows readily. If, furthermore, $j_1\not\in J_1^a$, then $\nu(1-|A_{j_1a}|)>4\delta$ and (c) follows. 

We now show (d). In view of \eqref{eqn:appx-epsilon-delta}, it suffices to establish
\begin{equation} \label{eqn:appx-lem5-diff2}
M_j \le \|\Sigma_{ji}\| \mbox{\ for any $j \in J_1^a$ and $i\in I_a$.}
\end{equation}
To do so, recall that from \eqref{eqn:appx-mi-upper-bound} we have the upper bound
\begin{equation}\label{eqn:appx-mi-bnd}
    M_j \leq \left\|\sum_{k_1=1}^K A_{jk_1}C_{k_1k_2^*} \right\|
\end{equation}
for some index $k_2^* \in [K]$. 
By the triangle inequality, it follows that
\begin{align}\label{eqn:appx-k2-star}
\begin{split}
    \left\|\sum_{k_1=1}^K A_{jk_1}C_{k_1k_2^*} \right\| &\le |A_{ja}| \|C_{ak_2^*} \| + \sum_{k_1\neq a} |A_{jk_1}| \left\|  C_{k_1k_2^*} \right\| \\
    &\leq |A_{ja}| \|C_{ak_2^*} \| + (1-|A_{ja}|) \|C_{k_2^*k_2^*}\| \\
    &< |A_{ja}| (\|C_{aa} \| - \nu) + (1-|A_{ja}|) \|C_{k_2^*k_2^*}\|,
    \end{split}
\end{align}
where the second inequality follows from the fact that $\|C_{k_2^*k_2^*}\| \ge \|C_{k_1k_2^*}\|$ for all $k_1 \in [K]$ and from Assumption~\ref{assump1}, and 
the third inequality uses $\|C_{aa}\| - \|C_{ak_2^*} \| > \nu$ from Assumption~\ref{assump3}.
Next, we lower bound $\|\Sigma_{ji}\| = \left\|\sum_{k_1=1}^K A_{jk_1}C_{k_1a} \right\|$ as follows. 
By the triangle inequality,
\begin{align}\label{eqn:appx-sigmaij-lower-bound}
\begin{split}
    \left\|\sum_{k_1=1}^K A_{jk_1}C_{k_1a} \right\| &\geq |A_{ja}| \|C_{aa} \| - \left\| \sum_{k_1 \neq a} A_{jk_1}C_{k_1a}\right\| \\
    &\ge |A_{ja}| \|C_{aa} \| - (\| C_{aa} \| - \nu) \left\| \sum_{k_1 \neq a} A_{jk_1}\right\| \\
    &\geq |A_{ja}| \|C_{aa} \| - (1-|A_{ja}|)(\| C_{aa} \| - \nu),
\end{split}
\end{align}
where the second step invokes the fact that $\|C_{k_1a}\| \le (\|C_{aa}\|-\nu)$, and
the third step follows from Assumption~\ref{assump1}.
By\eqref{eqn:appx-k2-star} and \eqref{eqn:appx-sigmaij-lower-bound}, we have
\begin{align}\label{eqn:appx-lemma5-diff}
\begin{split}
M_j - \|\Sigma_{ji}\| 
& \le \left\|\sum_{k_1=1}^K A_{jk_1}C_{k_1k_2^*} \right\| - \left\|\sum_{k_1=1}^K A_{jk_1}C_{k_1a} \right\| \\
&\le (1-|A_{ja}|)(\| C_{aa} \| + \left\|C_{k_2^*k_2^*}\right\|) -\nu.
    \end{split}
\end{align}
Since we assumed that $j\in J_1^a$, $1-|A_{ja}| \le 4\delta/\nu$. 
Moreover, Theorem~\ref{thm:pureindexguaran} assumes that $\nu > 2\sqrt{2 \|C\|_{\infty} \delta}.$
Equivalently, $\|C\|_{\infty} < \nu^2/8\delta$.
Since both $\| C_{aa} \|$ and $\left\|C_{k_2^*k_2^*}\right\|$ are bounded above by $\|C\|_{\infty}$,
$$(1-|A_{ja}|)(\| C_{aa} \| + \|C_{k_2^*k_2^*})\| \leq \frac{4\delta}{\nu } \frac{2\nu^2}{8\delta} =\nu.
$$
Substituting this into \eqref{eqn:appx-lemma5-diff} leads to $M_j - \|\Sigma_{ji}\| \le 0$, which is the same inequality \eqref{eqn:appx-lem5-diff2} that we set out to establish.
This concludes the proof of Lemma~\ref{lem:sigma_hat_norms}.


\section{Details of Numerical Experiments} \label{appx:num-methods}
\subsection{Approximation of Norms of the Covariance Operator}
In this section, we discuss the details for approximating the operator norm and the Hilbert-Schmidt norm of the covariance operator using its covariance function.
Recall that we are working in the Hilbert space $\HH=L^2[0,1]$.
Any covariance operator in $\cL(\HH)$ is an integral operator.
Let $\Psi \in \cL(\HH)$ be a covariance operator defined as
$\Psi f (\cdot) = \int_{0}^1 \psi(s,\cdot) f(s) \,ds$ for all $f\in L^2[0,1]$. 
The bivariate, symmetric function $\psi$ is called the covariance function associated with the operator $\Psi$. 

To approximate the action of $\Psi$, we first construct a discrete representation of the covariance function $\psi$.
Let $M \in \RR^{m \times m}$ be the matrix whose $(s,t)$th element is $\psi\left(\frac{s}{(m-1)},\frac{t}{(m-1)}\right)$, for any $0 \le s,t \le m-1$.
Let $f \in L^2[0,1]$, and let $\tilde{f} \in \RR^{m}$ such that its $s$th element is $f(s/(m-1))$ for $0\le s\le m-1$.
The action of $\Psi$ on the function $f$ can be approximated at the point $t_0/(m-1)$ as follows:
\begin{align}\label{eqn:reimann-approx}
    \Psi f \left(\frac{t_0}{(m-1)} \right)&= \int_0^1 \psi\left(s, \frac{t_0}{(m-1)}\right)f\left(s \right)\,ds \nonumber \\
    &\approx \frac{1}{m-1}\sum_{s=0}^{m-1} \psi\left(\frac{s}{(m-1)}, \frac{t_0}{(m-1)}\right)f\left(\frac{s}{(m-1)} \right) \nonumber \\
    &= \frac{1}{m-1}[M\tilde f]_{t_0},
\end{align}
where $[M\tilde f]_{t_0}$ denotes the element of the vector $M\tilde f$ at position $t_0$.
In the computations above, the second step follows from Riemann sums and the third step follows from the symmetry of the matrix $M$.

Recall that the operator norm of $\Psi$ is defined as 
$$\|\Psi \| = \sup_{f:\|f\|_{\HH}=1} \|\Psi f\|_{\HH}.$$
Let $\tilde f \in \RR^{m}$ denote the discretization of $f$.
Since
\begin{align*}
    \|f\|^2_{\HH} &= \int_0^1 f(s)^2\,ds \\
    &\approx \frac{1}{m-1}\sum_{s=0}^{m-1}f\left(\frac{s}{m-1} \right)^2 \\
    &= \frac{1}{m-1} \|\tilde f\|_2^2.
\end{align*}
Similarly, 
\begin{align*}
    \|\Psi f\|^2_{\HH} &= \int_0^1 (\Psi f(s))^2\,ds \\
    &\approx \frac{1}{m-1}\sum_{s=0}^{m-1} \left(\Psi f \left(\frac{s}{m-1}\right)\right)^2 \\
    &= \frac{1}{(m-1)^3}\|M\tilde f\|_2^2,
\end{align*}
where the second step uses Riemann sums and the third step invokes \eqref{eqn:reimann-approx}.
Let $\tilde g = \frac{1}{\sqrt{m-1}} \tilde f$. 
Then the operator norm can be approximated as
\begin{align*}
    \|\Psi \| &\approx \sup_{\tilde f: \|\tilde f\|_2 / (\sqrt{m-1})=1} \frac{1}{(m-1)^{3/2}} \|M\tilde f\|_2 \\
    &= \sup_{\|\tilde g\|_2 = 1} \frac{1}{(m-1)^{3/2}}\sqrt{m-1}\left\|M\tilde g \right\|_2 \\
    &= \frac{1}{m-1} \sup_{\|\tilde g\|_2 = 1} \left\|M\tilde g \right\|_2 \\
    &=\frac{1}{m-1} \|M\|_{op}.
\end{align*}
The notation $\|\cdot\|_{op}$ denotes the operator norm or spectral norm for matrices.
Therefore, the operator norm of any covariance operator can be approximated using discretizations of the corresponding covariance function.

We now consider approximation of the Hilbert-Schmidt norm.
For $\HH=L^2[0,1]$, the Hilbert Schmidt norm of $\Psi$ is expressed as
$$\|\Psi\|_{\TT_{1,1}} = \int_0^1 \int_0^1 \psi(s,t)^2\,ds \,dt. $$
We approximate the integral using Riemann sums as follows:
\begin{align*}
    \|\Psi\|_{\TT_{1,1}}^2 &\approx \frac{1}{(m-1)^2}\sum_{t=0}^{m-1} \sum_{s=0}^{m-1} \psi\left(\frac{s}{m-1}, \frac{t}{m-1} \right)^2 \\
    &=\frac{1}{(m-1)^2} \|M\|^2_F,
\end{align*}
where $\|\cdot\|_F$ denotes the Frobenius norm of a matrix.
Therefore, $\|\Psi\|_{\TT_{1,1}} \approx \|M\|_F/(m-1)$.

\subsection{Empirical Validation of Theorem~\ref{thm:asymp_aj}} \label{appx:clt_simulations}
In this section, we present additional simulations for validating Theorem~\ref{thm:asymp_aj}.
Recall that in Section~\ref{sec:num-exp-aj}, we used Monte-Carlo simulations to estimate $\widehat A_J$ under the assumption that $A_I$ is known.
Under the same simulation setting as Section~\ref{sec:num-exp-aj}, we check the distribution of elements of $\sqrt{n}(\widehat A_J - A_J)$, and check coverage of the confidence intervals computed using the asymptotic distribution.
We perform all three diagnostic checks for another simulation setting.

We first motivate the additional diagnostic checks.
From Theorem~\ref{thm:asymp_aj}, as $n \to \infty$, $[\sqrt{n}(\widehat A_J - A_J)]_{js}/\sqrt{\text{Cov}([\cG(\mathfrak{Z})]_{sj}, [\cG(\mathfrak{Z})]_{sj})}$ converges in distribution to a standard normal random variable.
A Q-Q plot of $\frac{[\sqrt{n}(\widehat A_J - A_J)]_{js}}{\wh \sigma_{js}}$ is used to assess normality, where $\widehat \sigma^2_{js}$ is the $(js, js)$th element of the estimated $\widehat V^{-1} \widehat{\mathscr{T}} \widehat V^{-1}$.
Lastly, we check the coverage of the $95\%$ confidence intervals obtained from the asymptotic distribution. 
This is equivalent to checking if $$\frac{|[\widehat A_J - A_J]_{js}|}{\sqrt{\text{Cov}([\cG(\mathfrak{Z})]_{sj}, [\cG(\mathfrak{Z})]_{sj})/n}} \le 1.96.$$
We also check the coverage when the plug-in estimate $\widehat \sigma^2_{js}$ is used instead of the true covariance $\text{Cov}([\cG(\mathfrak{Z})]_{sj}, [\cG(\mathfrak{Z})]_{sj})$.

We now describe Scenario (IV), which resembles the simulation setting in \cite{bing}.
We set $K=5$ and draw the latent factors from a multivariate normal distribution with covariance $C$ such that the diagonal elements of $C$ are $C_{aa} = 2+(a-1)/4 $. 
The non-diagonal elements are $C_{ab} = (-1)^{(a+b)}0.3^{|a-b|} \min(C_{aa}, C_{bb})$.
To specify $A_I$, we set $5$ pure variables for each cluster.
We set $p=100$ and randomly select $25$ rows to correspond to the pure variables.
For each index $i\in I_a$, $A_{ia}$ is randomly chosen to be $+1$ or $-1$. 
For each row of $A_J$, we first select the support $\text{card}_j$ randomly from $\{2,3,4,5\}$. 
Then we randomly select $\text{card}_j$-many columns and set those values to be $\text{sign}_j/\text{supp}_j$, where $\text{sign}_j$ is drawn uniformly from $\{-1, 1\}$.
The errors are heteroskedastic and for each $i\in [p]$, $E_i$ is drawn from zero-mean normal distribution with variance $\sigma_i^2$, where $\sigma_i^2 \sim \text{Uniform}[1,3]$.

\begin{figure}
    \centering
    \includegraphics[width=0.9\linewidth]{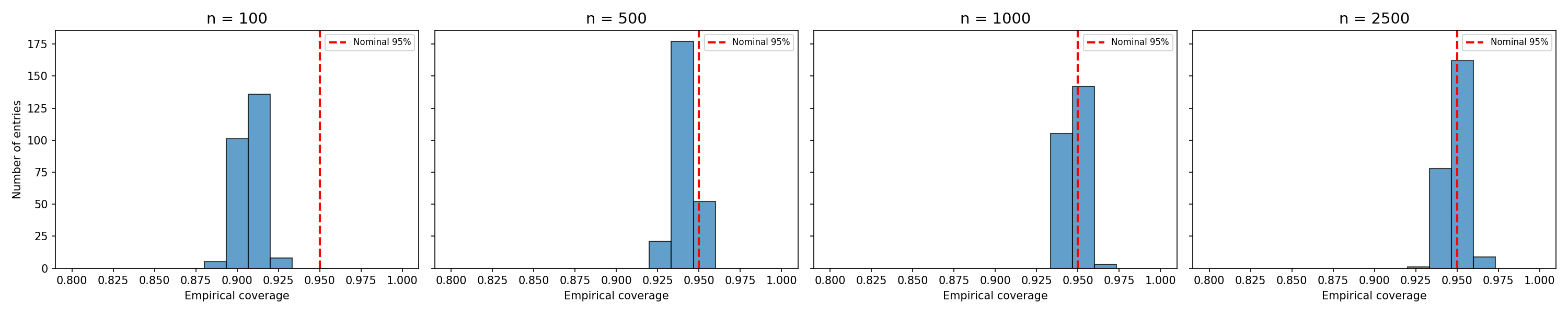}
    \caption{Coverage of 95\% intervals computed using the true covariance; Scenario (III)}
    \label{fig:scenario3cov}
\end{figure}

\begin{figure}
    \centering
    \includegraphics[width=0.9\linewidth]{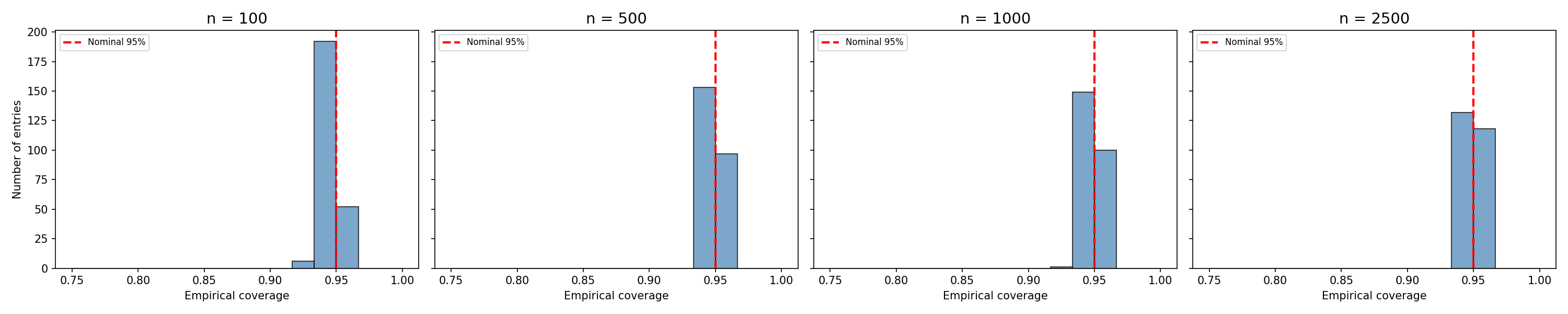}
    \caption{Coverage of 95\% intervals computed using plug-in estimates of the covariance; Scenario (III)}
    \label{fig:scenario3covplugin}
\end{figure}

\begin{figure}
    \centering
    \includegraphics[width=0.9\linewidth]{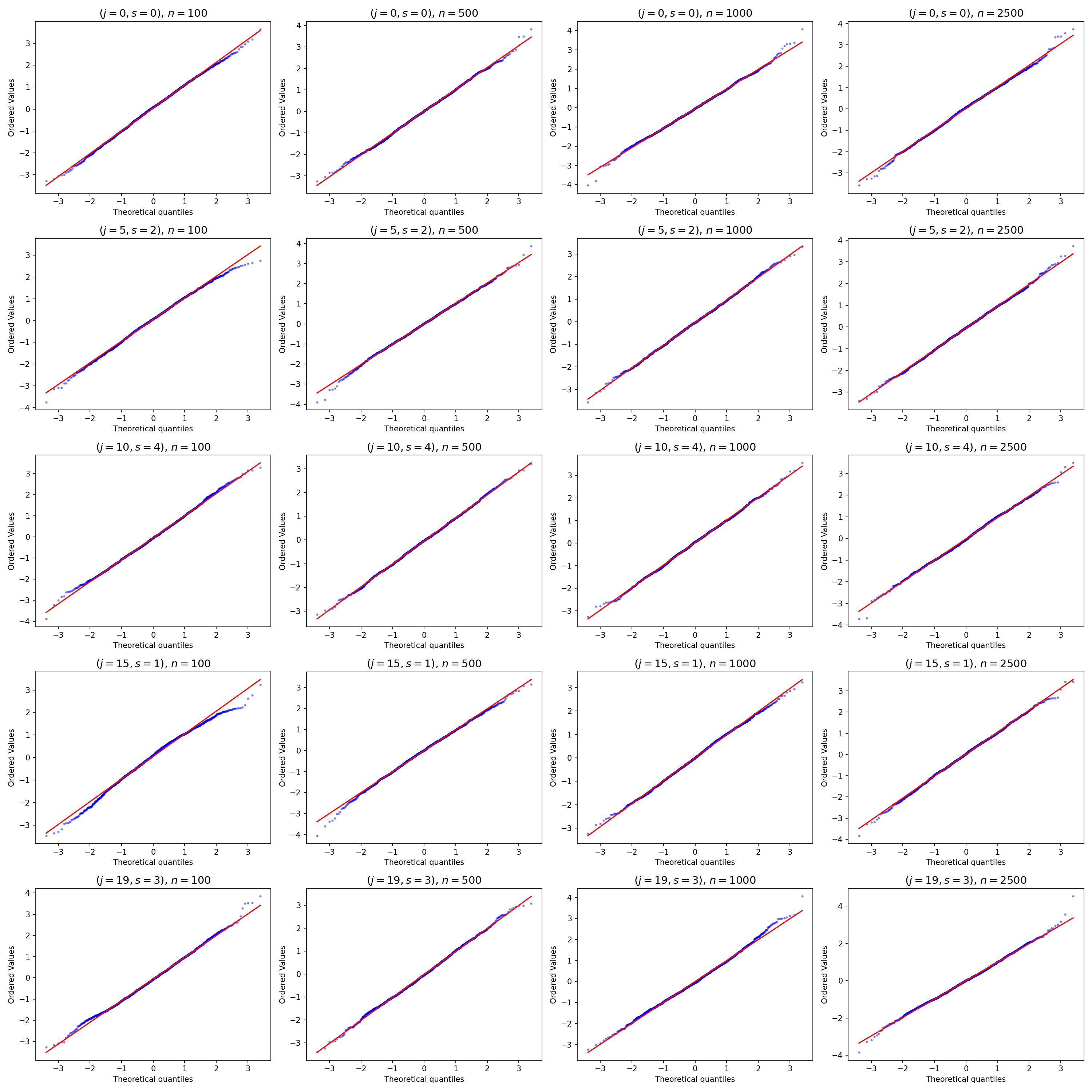}
    \caption{Q-Q plots of $\sqrt{n}[\widehat A_J-A_J]_{js}/\widehat\sigma_{js}$ for a subset of indices $j \in J, s\in [K]$; Scenario (III).}
    \label{fig:scenario3qqplot}
\end{figure}

\begin{figure}
    \centering
    \includegraphics[width=0.9\linewidth]{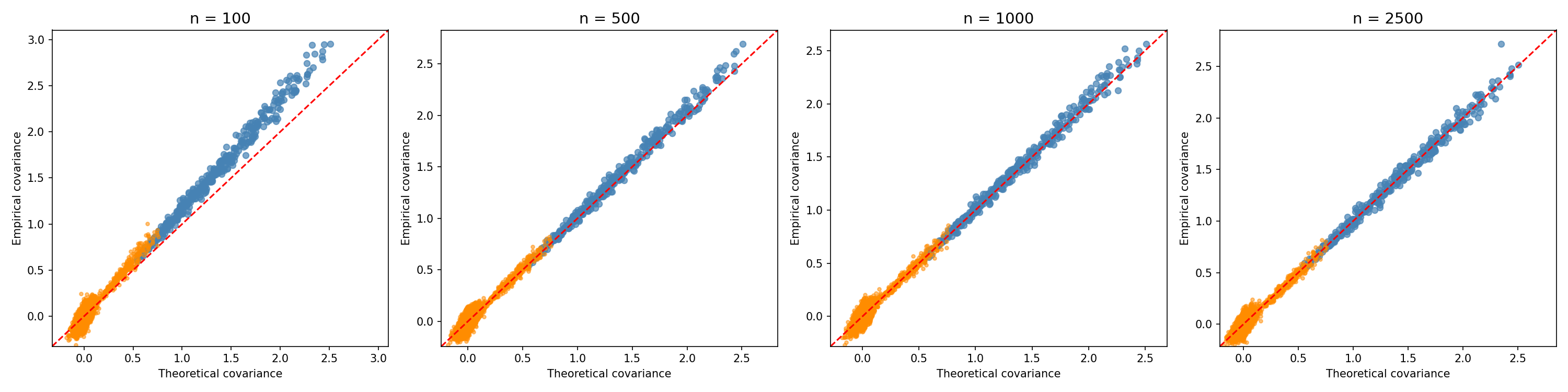}
    \caption{Empirical covariances (computed using Monte Carlo) versus theoretical covariances for Scenario (IV). Blue points represent variances and orange points represent covariances.}
    \label{fig:scenario4checkcov}
\end{figure}

\begin{figure}
    \centering
    \includegraphics[width=0.9\linewidth]{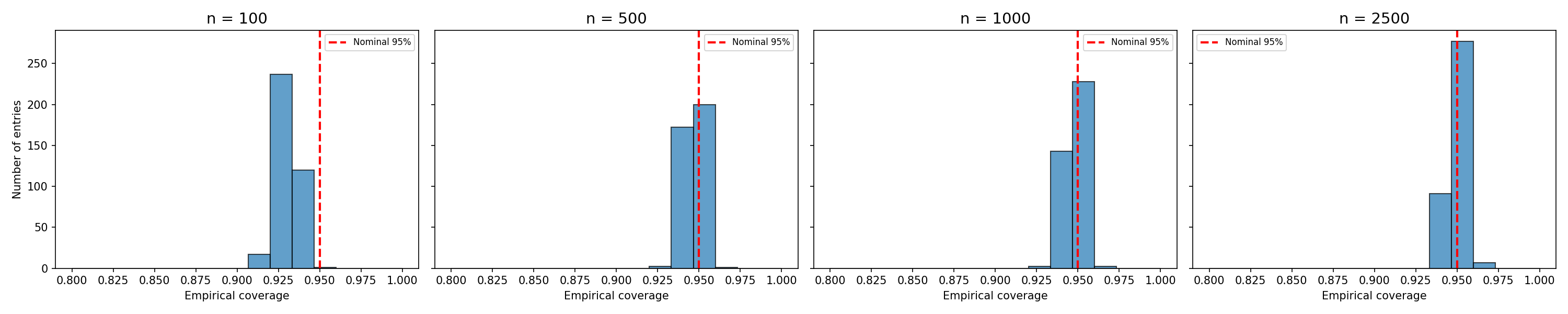}
    \caption{Coverage of 95\% intervals computed using the true covariance; Scenario (IV)}
    \label{fig:scenario4cov}
\end{figure}

\begin{figure}
    \centering
    \includegraphics[width=0.9\linewidth]{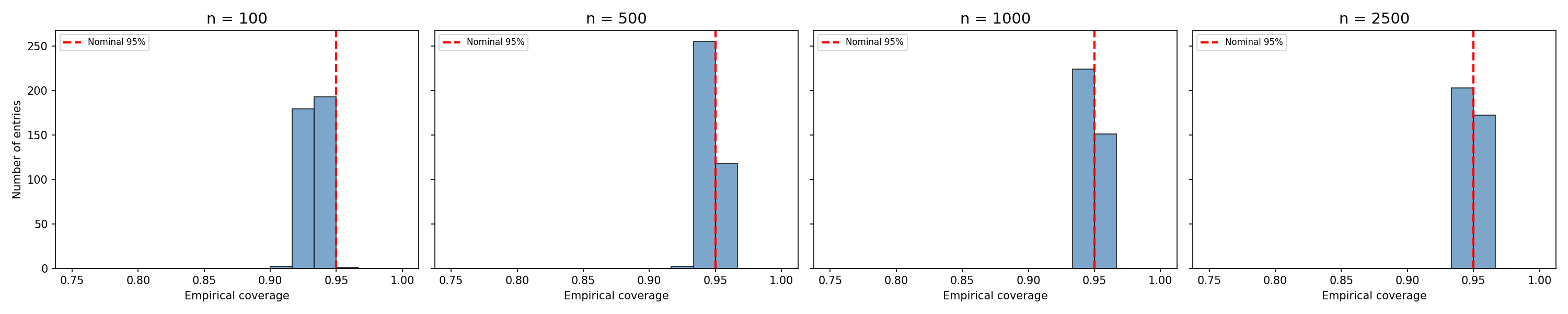}
    \caption{Coverage of 95\% intervals computed using plug-in estimates of the covariance; Scenario (IV)}
    \label{fig:scenario4covplugin}
\end{figure}

\begin{figure}
    \centering
    \includegraphics[width=0.9\linewidth]{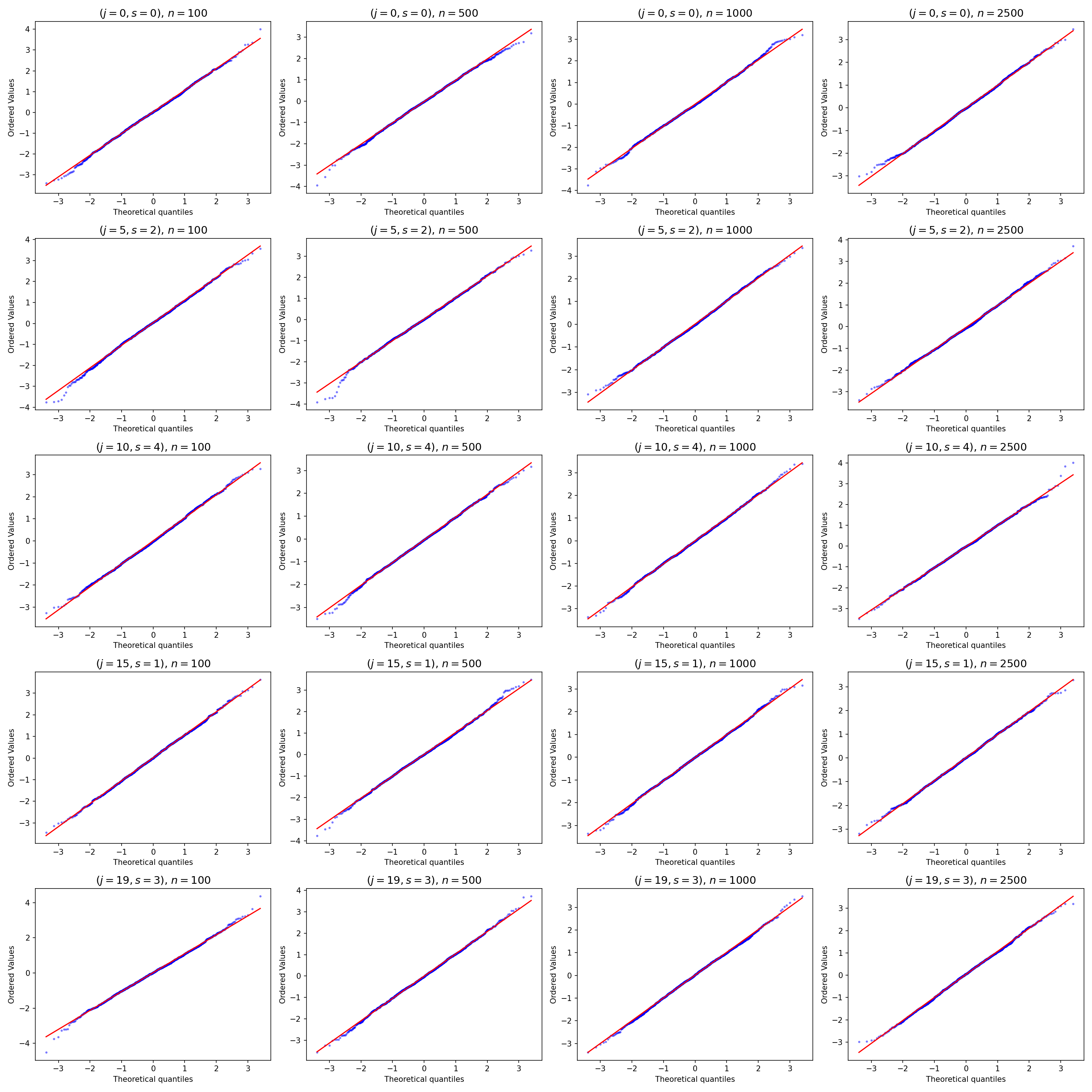}
    \caption{Q-Q plots of $\sqrt{n}[\widehat A_J-A_J]_{js}/\widehat\sigma_{js}$ for a subset of indices $j \in J, s\in [K]$; Scenario (IV)}
    \label{fig:scenario4qqplot}
\end{figure}

Figure~\ref{fig:scenario3cov}--\ref{fig:scenario3qqplot} show the diagnostic plots for the setting described in Section~\ref{sec:num-exp-aj}, which we refer to as Scenario (III).
The diagnostic plots for Scenario (IV) are shown in Figure~\ref{fig:scenario4checkcov}--\ref{fig:scenario4qqplot}.
The results were obtained from $2000$ Monte-Carlo simulations.
In both the Q-Q plots (Figure~\ref{fig:scenario3qqplot} and \ref{fig:scenario4qqplot}), the points generally follow the identity line with slight deviation at the tails.
Adherence to the identity line becomes better as $n$ increases.
Figure~\ref{fig:scenario4checkcov} plots the empirical covariances versus the theoretical covariances and they fall closer to the identity line as $n$ increases.
In all four histograms showing the coverage of the $95\%$ confidence intervals, the coverage gets better as $n$ increases.
The coverage obtained by using plug-in estimates of the variance (Figure~\ref{fig:scenario3covplugin} and \ref{fig:scenario4covplugin}) is better than that obtained using the true variance (Figure~\ref{fig:scenario3cov} and \ref{fig:scenario4cov}) in both scenarios.
This is attributable to the fact that the theoretical variances are only valid asymptotically, but the estimated variances better capture finite-sample properties.

\end{document}